\documentclass[aps,prd,reprint,twocolumn,superscriptaddress,letterpaper,longbibliography]{revtex4-1}

\usepackage[caption=false]{subfig}
\usepackage{amsmath}
\usepackage{amsfonts}
\usepackage{amssymb}
\usepackage{bbm}
\usepackage[nolist]{acronym}
\usepackage[normalem]{ulem}
\usepackage{graphicx}
\usepackage{epsfig}
\usepackage{color}
\usepackage{multirow}
\usepackage{hyperref}
\usepackage{cleveref}
\usepackage{physics}

\def\bea{\begin{eqnarray}}
\def\eea{\end{eqnarray}}
\def\bef{\begin{flalign}}
\def\eef{\end{flalign}}
\def\nn{\nonumber\\}                    % No number and new line for align and eqnarray
\def\d{\mathrm{d}}                      % Differential "d", can be done also with \dd using "physics" package
\newcommand{\id}{\hat{\mathbbm{I}}}
\def\R{\mathbb{R}}                      % Reals
\def\ipic{\text{I}}                     % Interaction picture subscript
\def\intac{\text{int}}                  % Interaction subscript
\newcommand{\eff}{\text{eff}}           % effective subscript
\def\ct{\text{ct}}                      % Counterterm subscript
\def\dagg{^{\dagger}}                   % Dagger
\def\conj{^{*}}                         % Conjugate
\def\pr{^{\prime}}                      % Prime
\def\HC{\text{H.c.}}                    % Hermitian conjugate terms
\def\Sy{\text{S}}                       % System super/sub-script identifier
\def\E{\text{E}}                        % Environment super/sub-script identifier
\def\r{\text{R}}                        % Environment super/sub-script identifier
\newcommand{\CG}{\mathcal{G}}           % Calligraphic G

\def\al{\alpha}
\def\be{\beta}
\def\la{\lambda}

\def\om{\omega}
\newcommand{\omS}{\omega_{\text{S}}}     
\def\Om{\Omega}
\def\ep{\epsilon}

\def\vk{\vec{k}}
\def\vx{\vec{x}}
\def\vy{\vec{y}}
\def\vp{\vec{p}}

\def\hsig{\hat{\sigma}}                 % Composite system density operator
\def\hsigi{\hat{\sigma}_{\ipic}}        % Composite system density operator in interaction picture
\def\hrho{\hat{\rho}}                   % Reduced system density operator 
\def\hrhoi{\hat{\rho}_{\ipic}}          % Reduced system density operator in interaction picture

\def\hphi{\hat{\phi}}
\def\hchi{\hat{\chi}}
\def\hpi{\hat{\pi}}

\def\hU{\hat{U}}
\def\hH{\hat{H}}

\newcommand{\hO}{\hat{\mathcal{O}}}
\newcommand{\hCP}{\hat{\mathcal{P}}}
\newcommand{\hCQ}{\hat{\mathcal{Q}}}

\newcommand{\ha}[1]{\hat{a}^{\vphantom{\dagger}}_{#1}}
\newcommand{\hadagg}[1]{\hat{a}^{\dagger}_{#1}}
\newcommand{\hb}[1]{\hat{b}^{\vphantom{\dagger}}_{#1}}
\newcommand{\hbdagg}[1]{\hat{b}^{\dagger}_{#1}}
\newcommand{\hL}[1]{\hat{L}^{\vphantom{\dagger}}_{#1}}
\newcommand{\hLdagg}[1]{\hat{L}^{\dagger}_{#1}}

\newcommand{\Fk}[1]{F_{k, #1}}
\newcommand{\Gamk}[1]{\Gamma_{k, #1}}
\newcommand{\gamk}[1]{\gamma_{k, #1}}
\newcommand{\Sk}[1]{S_{k, #1}}
\newcommand{\gamd}[1]{\gamma_{#1}}
\newcommand{\Sd}[1]{S_{#1}}

\begin{document}
%%%%%%%%%%%%%%%%%%%%%%%%%%%%%%%%%%%%%%%%%%%%%%%%%%

\title{Open system dynamics in interacting quantum field theories}

\author{Brenden Bowen}
\email{brenden\_bowen@student.uml.edu}
\affiliation{Department of Physics and Applied Physics, University of Massachusetts, Lowell, Massachusetts 01854, USA}

\author{Nishant Agarwal}
\email{nishant\_agarwal@uml.edu}
\affiliation{Department of Physics and Applied Physics, University of Massachusetts, Lowell, Massachusetts 01854, USA}

\author{Archana Kamal}
\email{archana.kamal@northwestern.edu}
\affiliation{Department of Physics and Applied Physics, University of Massachusetts, Lowell, Massachusetts 01854, USA}
\affiliation{Department of Physics and Astronomy, Northwestern University, Evanston, Illinois 60208, USA}

\date{\today}

\begin{abstract}
A quantum system that interacts with an environment generally undergoes nonunitary evolution described by a non-Markovian or Markovian master equation. In this paper, we construct the non-Markovian Redfield master equation for a quantum scalar field that interacts with a second field through a bilinear or nonlinear interaction on a Minkowski background. We use the resulting master equation to set up coupled differential equations that can be solved to obtain the equal-time two-point function of the system field. We show how the equations simplify under various approximations including the Markovian limit and argue that the Redfield equation-based solution provides a perturbative resummation to the standard second-order Dyson series result. For the bilinear interaction, we explicitly show that the Redfield solution is closer to the exact solution compared to the perturbation theory-based one. Further, the environment correlation function is oscillatory and nondecaying in this case, making the Markovian master equation a poor approximation. For the nonlinear interaction, on the other hand, the environment correlation function is sharply peaked and the Redfield solution matches that obtained using a Markovian master equation in the late-time limit.
\end{abstract}

\maketitle

%%%%%%%%%%%%%%%%%%%%%%%%%%%%%%%%%%%%%%%%%%%%%%%%%%

%--------------------------
\section{Introduction}
\label{sec:intro}
%--------------------------

Most physical systems that we encounter, classical or quantum, are {\it open}, in the sense that they interact with some environment that is either not of interest or unknown to us. In the case of quantum systems, the open system approach is natural when observables of interest act only on a subspace of the full Hilbert space. Equations of motion for reduced system objects, such as the density operator, can be obtained by systematically {\it tracing out} environment degrees of freedom, converting environment operators into correlation functions. A common technique that is used to describe the nonunitary/non-Hamiltonian dynamics of an open quantum system is the master equation approach. In particular, for open systems where the environment consists of many degrees of freedom, the reduced dynamics are often well described by a quantum Markov process. For such systems, the time evolution of the reduced density operator is described by the celebrated Gorini-Kossakowski-Lindblad-Sudarshan (GKLS) master equation \cite{Gorini:1975nb,Lindblad:1975ef},
\begin{align}
	\frac{
		\dd
		\hrho
	}{
		\dd t
	}
&=
	-
	i
	\big[
		\hH
		,
		\hrho
	\big]
	+
	\sum_{\al}
	\gamma_{\al}
	\bigg[
		\hat{L}_{\al}
		\hrho
		\hat{L}_{\al}\dagg
		-
		\frac{
			1
		}{
			2
		}
		\big\{
			\hat{L}_{\al}\dagg
			\hat{L}_{\al}
			,
			\hrho
		\big\}
	\bigg]
	\,,
\label{eq:GKLSME}
\end{align}
where $ \hrho = \hrho(t) $ is the reduced density operator in the Schr\"odinger picture. The operators $ \hat{L}_{\al} $ are referred to as Lindblad or jump operators and the coefficients $ \gamma_{\al} $ are relaxation rates for different decay channels of the system. More generally, one can carry out a microscopic derivation of the master equation, starting from the von Neumann equation for a composite system and explicitly tracing out environment degrees of freedom under a series of approximations. This approach has the advantage that it leads to an intermediate {\it non-Markovian} master equation called the Redfield equation~\cite{Redfield1957}~\footnote{The Redfield equation is also referred to as the Born-Redfield equation \cite{Keefe:2024cia}.}, which will be our prime focus in this work.

The Redfield equation is similar to the GKLS equation in that it is time local but differs from it as the $ \gamma_{\al} $ coefficients are time-dependent. This is important when coherent feedback from the environment is non-negligible and induces {\it memory} in the reduced system. It is worth noting, however, that the Redfield equation is not guaranteed to generate a completely positive map, unlike the GKLS equation. In fact, it does not necessarily generate a completely positive map even when environment correlation functions decay rapidly enough to justify taking a Markovian limit, wherein time-dependent rates are replaced by their late-time values. Instead, one has to additionally make the rotating wave approximation (RWA), which drops rapidly oscillating terms in the Redfield equation, to obtain the completely positive Davies equation \cite{Davies_1974,Davies_1976} that can then be put in GKLS form. It is also worth noting that there have been many significant efforts to restore positivity in the Redfield equation \cite{Gaspard:1999,Whitney:2008,Schaller:2008,Majenz:2013,Farina:2019,Mozgunov_2020,Davidovic:2020,Nathan:2020,Trushechkin2021}, but each solution comes with limited validity. Some of us recently constructed a frequency-domain master equation that goes beyond the Redfield description by retaining full memory of the quantum state and preserved positivity \cite{Keefe:2024cia}. Despite its limitations, as highlighted by these works, the Redfield master equation remains a valuable tool to describe non-Markovian dynamics in a wide variety of open quantum systems, especially in studies of quantum transport \cite{Levy2014}.

There has been much recent interest in applying the open system approach to relativistic settings to gain insights into, for example, interacting quantum field theories \cite{Lombardo:1995fg,Koksma:2009wa,Koksma:2011dy,Agon:2014uxa,Boyanovsky:2015xoa,Agon:2017oia,Boyanovsky:2018fxl,Burrage:2018pyg,Banerjee:2021lqu,Kading:2022jjl,Kading:2025cwg}, primordial perturbation theory \cite{Lombardo:2005iz,Boyanovsky:2015tba,Burgess:2015ajz,Hollowood:2017bil,Shandera:2017qkg,Boyanovsky:2018soy,Akhtar:2019qdn,Gong:2019yyz,Brahma:2020zpk,Banerjee:2020ljo,Brahma:2021mng,Hsiang:2021kgh,Colas:2022hlq,DaddiHammou:2022itk,Burgess:2022nwu,Colas:2022kfu,Burgess:2024eng,Salcedo:2024smn,Brahma:2024ycc,Lopez:2025arw,Li:2025azq}, dark matter \cite{Cao:2022kjn,Kading:2023mdk}, gravitational decoherence \cite{Bassi:2017szd,Sharifian:2023jem}, gravitational waves \cite{Zarei:2021dpb}, construction of particle detectors in curved spacetime \cite{Kaplanek:2019dqu,Kaplanek:2019vzj,Kaplanek:2022xrr}, and holography \cite{Jana:2020vyx,Loganayagam:2020eue,Loganayagam:2022zmq,Pelliconi:2023ojb}. Motivated by primordial perturbation theory, our interest in this paper is to revisit the open quantum field theory (QFT) paradigm and to use it to calculate equal-time correlation functions of a system field of interest. Specifically, we demonstrate how the open system approach allows for approximations that prioritize the dynamics of the reduced system, which can lead to more accurate solutions than standard loop corrections or Dyson series perturbation theory results.

First, we provide a careful derivation of the Redfield equation for two interacting scalar QFTs on a Minkowski background. We consider two forms of interaction: one, bilinear in the system and environment fields, and another where the system field remains linear but the environment field is nonlinear. While the first interaction is exactly solvable and hence provides a useful benchmark, the general form of the second interaction is familiar from primordial perturbation theory. We then use the resulting Redfield equation to calculate the equal-time two-point function of the system field and compare the result to a standard loop correction. For the bilinear interaction, we find that the Redfield equation-based solution is closer to the exact solution than a loop correction, suggesting that the Redfield equation provides a {\it perturbative resummation} to the standard loop correction. We also find that both the Markovian limit and RWA fail to capture the exact dynamics in this case, as the environment correlation function that enters the Redfield equation is oscillatory and non-decaying. For the nonlinear interaction, on the other hand, the environment correlation function is sharply peaked, and the late-time behavior of the Redfield solution matches that obtained in the Markovian limit.

The remainder of this paper is organized as follows. In \cref{sec:setup}, we introduce the Lagrangian for a composite system of two scalar fields and write the quantized Hamiltonian in terms of their Fourier modes. We next provide a comprehensive derivation of the well-known Redfield equation in \cref{sec:me}, making a special effort to explicate the various standard approximations and how their regimes of validity manifest in the context of QFTs. In \cref{sec:TwoPointCorrelator}, we review a method to calculate the equal-time two-point function from the Redfield equation. We also discuss how to obtain limiting cases such as the Markovian limit and the standard loop correction, relegating details of the solution in the Markovian limit to \Cref{app:MarkovianLimitSolution}. In \cref{sec:phichi}, we apply this method to calculate the two-point function in the case of a bilinear interaction and compare it to the exact result. We present the details of the exact solution in \Cref{app:exactsoln} and show the various coefficients that enter the Redfield equation in \Cref{app:phichicoeff}. We next apply the Redfield equation-based method to the case of a nonlinear environment in \cref{sec:phichi2} and show the various coefficients that enter the Redfield equation in \Cref{app:phichi2coeff}. We finally discuss a few subtleties of the calculation that deserve further understanding but are outside the scope of this work in \cref{sec:URU} and end with a discussion in \cref{sec:disc}.

%%%%%%%%%%%%%%%%%%%%%%%%%%%%%%%%%%%%%%%%%%%%%%%%%%

%--------------------------
\section{Setup}
\label{sec:setup}
%--------------------------

We consider a composite system of two real, interacting Klein-Gordon fields $ \phi(\vx,t) $ and $ \chi(\vx,t) $, with masses $ m $ and $ M $, respectively, in $(3+1)$D. We restrict our interest to observables of the $ \phi $ field, treating the $ \chi $ field as an environment, and specialize to a single interaction term that is linear in $ \phi $ and either linear or quadratic in $ \chi $. The Lagrangian density that we consider is then
\begin{align}
	\mathcal{L}
&=
	-
	\frac{1}{2}
	(\partial \phi)^{2}
	-
	\frac{1}{2}
	m^{2}
	\phi^{2}
	-
	\frac{1}{2}
	(\partial \chi)^{2}
	-
	\frac{1}{2}
	M^{2}
	\chi^{2}
	-
	\frac{\la}{n!}
	\phi
	\chi^{n}
	,
\label{eq:Phi_Chi_Action}
\end{align}
where $(\partial\phi)^2 = -\dot{\phi}^2 + \big( \vec{\nabla}\phi \big)^2$, the dot denoting a derivative with time, and $ \la $ is a coupling constant with mass dimension $ 3 - n $, where $ n \in \{ 1,2 \} $. We have adopted the mostly plus metric signature $ (-,+,+,+) $ and work in units where $ \hbar = c = 1 $. Identifying the canonically conjugate momenta $ \pi_{\phi} = \dot{\phi} $ and $ \pi_{\chi} = \dot{\chi} $, we perform the usual Legendre transformation and integrate over $ \R^{3} $ to obtain the Hamiltonian. We then impose that the only nonvanishing field commutation relations are $ \big[ \hphi(\vx) , \hpi_{\phi}(\vy) \big] = \big[ \hchi(\vx) , \hpi_{\chi}(\vy) \big] = i \delta^{3} \left( \vx - \vy \right) $ and write the quantized Hamiltonian in the Schr\"{o}dinger picture as
\begin{align}
	\hH
&=
	\hH_{0, \Sy}
	+
	\hH_{0, \E}
	+
	\hH_{\intac}
	\,,
\label{eq:fullhamiltonian}
\end{align}
where
\begin{align}
	\hH_{0, \Sy}
&=
	\frac{1}{2}
	\int \dd^{3}x
	\left[
		\hpi_{\phi}^{2}
		+
		\big(
			\vec{\nabla} 
			\hphi
		\big)^{2}
		+
		m^{2}
		\hphi^{2}
	\right]
	,
\\
	\hH_{0, \E}
&=
	\frac{1}{2}
	\int \dd^{3}x
	\left[
		\hpi_{\chi}^{2}
		+
		\big(
			\vec{\nabla} 
			\hchi
		\big)^{2}
		+
		M^{2}
		\hchi^{2}
	\right]
	,
\end{align}
are the system and environment free Hamiltonians, respectively, and
\begin{align}
\label{eq:Interaction_Hamiltonian}
    \hH_{\intac}
&=
    \frac{\la}{n!}
    \int \dd^{3}x
    \hphi
    \hchi^{n}
\end{align}
is the interaction Hamiltonian. We remark here that this interaction is not bounded from below due to the odd power of $ \hphi $. However, this instability in the system dynamics can be easily removed by including suitable self-interactions for $ \hphi $ and $ \hchi $. We ignore such additional terms for simplicity and also note that the instability does not cause issues in a perturbative treatment for $ m \neq 0 $, $ M \neq 0 $, and with the $\hphi$ field initialized in the vacuum~\cite{Srednicki:2007qs}.

Since it will be convenient to write the master equation in terms of eigenoperators of the system's free Hamiltonian, $ \hH_{0, \Sy} $, we choose to work in momentum space where the creation and annihilation operators are eigenoperators of the free Hamiltonian. We use the Fourier expansions
\begin{align}
	\hphi(\vx)
&=
	\int
	\frac{
		\dd^{3} k
	}{
		(2 \pi)^{3}
	}
	\hphi_{\vk}
	e^{
		i 
		\vk 
		\cdot 
		\vx
	}
	\,,
\label{eq:Fourier_phi} \\
	\hchi(\vx)
&=
	\int
	\frac{
		\dd^{3} p
	}{
		(2 \pi)^{3}
	}
	\hchi_{\vp}
	e^{
		i 
		\vp 
		\cdot 
		\vx
	}
	\,,
\label{eq:Fourier_chi}
\end{align}
where we have indicated the arguments of the Fourier modes with subscripts such that $ \hphi_{\vec{k}} \equiv \hphi(\vec{k}) $ and $ \hchi_{\vec{p}} \equiv \hchi(\vec{p}) $. We further write the field operators $ \hphi_{\vk} $ and $ \hchi_{\vp} $ in terms of their creation and annihilation operators,
\begin{align}
	\hphi_{\vk}
&=
	\frac{
		1
	}{
		\sqrt{
			2 
			\om_{k}
		}
	}
	\left(
		\ha{\vk}
		+
		\hadagg{-\vk}
	\right)
	,
\label{eq:Schrodinger_picture_Fourier_phi} \\
	\hchi_{\vp}
&=
	\frac{
		1
	}{
		\sqrt{
			2 
			\Om_{p}
		}
	}
	\left(
		\hb{\vp}
		+
		\hbdagg{-\vp}
	\right)
	,
\label{eq:Schrodinger_picture_Fourier_chi}
\end{align}
where we have defined the frequencies $ \om_{k}^{2} \equiv k^{2} + m^{2} $ and $ \Om_{p}^{2} \equiv p^{2} + M^{2} $, which inherit the commutation relations
\begin{align}
	\left[
		\ha{\vk}
		,
		\hadagg{\vk\pr}
	\right]
&=
	(2 \pi)^{3}
	\delta^{3}
	\big(
		\vk 
		- 
		\vk\pr
	\big) 
	\,,
\label{eq:Momentum_space_commutator_a} \\
	\left[
		\hb{\vp}
		,
		\hbdagg{\vp\pr}
	\right]
&=
	(2 \pi)^{3}
	\delta^{3}
	\big(
		\vp 
		- 
		\vp\pr
	\big) 
	\,.
\label{eq:Momentum_space_commutator_b}
\end{align}
We can then write the free Hamiltonian, $ \hH_{0} = \hH_{0,\Sy} + \hH_{0,\E} $, in the form
\begin{align}
	\hH_{0}
&=
	\int
	\frac{
		\dd^{3} k 
	}{ 
		(2 \pi)^{3}
	}
	\om_{k}
	\hadagg{\vk} 
	\ha{\vk}
	+
	\int
	\frac{
		\dd^{3} p 
	}{ 
		(2 \pi)^{3}
	}
	\Om_{p}
	\hbdagg{\vp} 
	\hb{\vp}
	\,,
\label{eq:Free_Hamiltonian}
\end{align}
where we have dropped the (infinite) zero-point energy term and note that it can be removed systematically by shifting the free Lagrangian density by a constant. For the interaction Hamiltonian, we choose to explicitly write the system field in terms of creation and annihilation operators. As shown in \cref{sec:SFME}, this allows us to easily transition from the Redfield master equation to the Davies master equation by isolating the time dependence of system operators in the interaction picture. Thus, we write $ \hH_{\intac} $ as
\begin{align}
    \hH_{\intac}
&=	
    \la
    \sum_{\al = 1}^{2}
    \int_{\vk}
    \hL{\vk, \al}
    \hO_{-\vk}
    \,,
\label{eq:Interaction_Hamiltonian_Fourier}
\end{align}
where we have defined the dimensionless system integral $ \int_{\vk} \equiv \int \frac{ \dd^{3} k }{ (2 \pi)^{3} } \frac{ 1 }{ \om_{k}^{3} } $, the dimensionless system operators
\begin{align}
    \hL{\vk, 1}
=
    \om_{k}^{3/2}
    \ha{\vk}
    \,,
\hspace{18pt}
    \hL{\vk, 2}
=
    \om_{k}^{3/2}
    \hadagg{-\vk}
    \,,
\end{align}
and the environment operator 
\begin{align}
\label{eq:Environment_operator}
    \hO_{-\vk}
&\equiv
    \frac{ \om_{k} }{ \sqrt{2} n! }
    \int
    \d^{3} x
    \hchi^{n}(\vx)
    e^{i \vk \cdot \vx}
    \,
\end{align}
corresponding to the system mode $ \vk $. In the interaction picture, $ \hL{\vk,\al} $ inherit the time dependence of the creation and annihilation operators,
\begin{align}
    \hL{\vk, \al}(t)
=
	e^{
		i
		(-1)^{\al}
		\om_{k}
		\left(
			t
			-
			t_{0}
		\right)
	}
    \hL{\vk, \al}
	\,,
\end{align}
where $t_0$ is the initial time at which all pictures coincide. We simply used the $t$ dependence to distinguish operators in the interaction picture from those in the Schr\"{o}dinger picture above and use the same simplified notation below as well. We find later that the $ \hL{} $ operators appear in the master equation in the same way as the Lindblad operators of a GKLS equation. Since it is not always possible to write the master equation in GKLS form, however, we refer to the $ \hL{} $ operators as the dimensionless eigenoperators of the system.

%%%%%%%%%%%%%%%%%%%%%%%%%%%%%%%%%%%%%%%%%%%%%%%%%%

%--------------------------
\section{Master equation construction}
\label{sec:me}
%--------------------------

In this section, we derive the Redfield equation following standard texts such as Refs.\ \cite{carmichael1999statistical,Breuer:2002pc,Schlosshauer2007}. We review the derivation in detail not only to set up our notation but also to highlight subtleties in the validity of various approximations for QFTs as compared to nonrelativistic systems. As usual, we start by assuming that the system and environment are initially uncorrelated and that the interaction is turned on at the time $ t_{0} $. The initial state of the composite system can then be written without loss of generality as a tensor product,
\begin{align}
	\hsig(t_{0})
=
	\hrho(t_{0})
	\otimes
	\hrho_{\E}
	\,,
\label{eq:Initial_joint_density_operator}
\end{align}
where $ \hrho_{\E} \equiv \hrho_{\E}(t_{0}) $. This can be time evolved in the standard way, using the unitary time-evolution operator $\hU(t,t_{0})$,
\begin{align}
	\hsig(t) 
= 
	\hU(t,t_{0}) 
	\hsig(t_{0}) 
	\hU\dagg(t,t_{0}) 
	\,.
\label{eq:Time_evoled_joint_density_operator}
\end{align}
For the composite system Hamiltonian $\hH = \hH_{0} + \hH_{\intac}$, $\hU(t,t_{0})$ is in turn given by
\begin{align}
	\hU(t,t_{0})
=
	e^{
		-
		i 
		\hH_{0} 
		(
			t
			-
			t_{0}
		)
	}
	T 
	e^{
		-
		i 
		\int_{t_{0}}^{t} 
		\hH_{\ipic}(t_{1}) 
		\dd t_{1}
	} 
	\,,
\label{eq:Time_evolution_operator}
\end{align}
where $ T $ indicates time ordering and we have defined the interaction Hamiltonian in the interaction picture: $ \hH_{\ipic}(t) \equiv e^{ i \hH_{0} ( t - t_{0}) } \hH_{\intac}(t_{0}) e^{ - i \hH_{0} ( t - t_{0}) } $, where $ \hH_{\intac}(t_{0}) $ is simply $ \hH_{\intac} $. Now, taking a time derivative of \cref{eq:Time_evoled_joint_density_operator} yields the von Neumann equation, which we write in the interaction picture as  
\begin{align}
	\frac{\dd}{\dd t} 
	\hsigi(t)
=
	- i
	\left[
		\hH_{\ipic}(t)
		,
		\hsigi(t)
	\right] 
	.
\label{eq:von_Neumann_equation}
\end{align}
We are interested in writing dynamical expressions that are valid to second order in the coupling without treating the density operator itself as a perturbative object. To this end, we first integrate \cref{eq:von_Neumann_equation} from $ t_{0} $ to $ t $ to find the formal solution for the density operator,
\begin{align}
	\hsigi(t)
&=
	\hsig(t_{0})
	-
	i
	\int_{t_{0}}^{t} \dd t_{1}
	\left[
		\hH_{\ipic}(t_{1})
		, 
		\hsigi(t_{1})
	\right]	
	.
\end{align}
We then substitute this result for $ \hsigi(t) $ on the right-hand side of \cref{eq:von_Neumann_equation} to obtain
\begin{align}
	\frac{\dd}{\dd t} 
	\hsigi(t)
&=
	- 
	i
	\left[
		\hH_{\ipic}(t)
		, 
		\hsig(t_{0})
	\right]
\nn
&\hspace{11.5pt}
	-
	\int_{t_{0}}^{t} \dd t_{1}
	\left[
		\hH_{\ipic}(t)
		, 
		\left[
			\hH_{\ipic}(t_{1})
			, 
			\hsigi(t_{1})
		\right]	
	\right]
	.
\label{eq:Integral_equation_for_sigma}
\end{align}
Note that although the second term contains two factors of the interaction Hamiltonian, we have not made any approximation; hence, the expression above is valid for {\it any} interaction strength at {\it all} orders of $\la$. It is easily verified that replacing $ \hsigi(t_1) $ on the right-hand side with $\hsig(t_{0})$ would exactly recover the standard second-order perturbation theory expression, but we will not make this approximation here.

We now perform a partial trace over the environment degrees of freedom in \cref{eq:Integral_equation_for_sigma} to obtain the master equation for the reduced density operator $ \hrhoi(t) \equiv \Tr_{\E} \hsigi(t) $,
\begin{align}
	\frac{\dd}{\dd t} 
	\hrhoi(t)
&=
	- 
	i
	\left[
		\hH_{\eff}(t)
		,
		\hrho(t_{0})
	\right]
\nn
&\hspace{11.5pt}
	-
	\int_{t_{0}}^{t} \dd t_{1}
	\Tr_{\E}
	\left[\hH_{\ipic}(t), \left[\hH_{\ipic}(t_{1}), \hsig_{\ipic}(t_{1})\right]\right]
	,
\label{eq:FormalSolutionForRho}
\end{align}
where we have used \cref{eq:Initial_joint_density_operator} to rewrite the first commutator on the right-hand side in terms of an effective Hamiltonian $ \hH_{\eff}(t) \equiv \Tr_{\E} \big\{ \hH_{\ipic}(t) \hrho_{\E} \big\} $. It is common to assume that the environment operators coupling to the system have zero mean in the state $ \hrho_{\E} $, so that $ \hH_{\eff}(t) = 0 $. Although this is not always true and depends on the choice of interaction and initial state, it can generally be imposed by including a mean environment field term in the system Hamiltonian~\cite{carmichael1999statistical}. We choose to explicitly keep the effective Hamiltonian term but find that it does not contribute to the equal-time two-point function of the reduced system in any of the cases considered in this work.

%--------------------------
\subsection{Redfield master equation}
\label{sec:RedfieldME}
%--------------------------

As mentioned earlier,~\cref{eq:FormalSolutionForRho} still describes the exact dynamics of the composite system; however, to make further progress, we now restrict to the perturbative regime and make two standard approximations, namely, the Born and Markov approximations. We would first like to reduce~\cref{eq:FormalSolutionForRho} to an integrodifferential equation for the reduced density operator $ \hrhoi(t) $. For this, it is useful to define a Zwanzig projection operator $ \hCP $ on the space of density operators for the composite system such that 
\begin{align}
	\hCP
	\hsigi(t)
&=
	\Tr_{\E}
	\left\{
		\hsigi(t)
	\right\}
	\otimes
	\hrho_{\r}
\nn
&=
	\hrhoi(t)
	\otimes
	\hrho_{\r} \, ,
\label{eq:DefProjectionOperator}
\end{align}
where $ \hrho_{\r} $ is a time-independent reference state for the environment. We construct a complementary operator $ \hCQ $ through the relationship $ \id = \hCP + \hCQ $, which allows us to decompose the full density operator as 
\begin{align}
	\hsigi(t)
&=
	\hCP
	\hsigi(t)
	+
	\hCQ
	\hsigi(t)
\nn
&=
	\hrhoi(t)
	\otimes
	\hrho_{\r}
	+
	\hCQ
	\hsigi(t)
	\,.
\label{eq:hsigPQ}
\end{align}
In the literature on projection operators, $ \hrhoi(t) \otimes \hrho_{\r} $ and $ \hCQ \hsigi(t) $ are referred to as the {\it relevant} and {\it irrelevant} parts, respectively~\cite{Giulini:1996nw,Breuer:2002pc}.  

Using \cref{eq:hsigPQ} in \cref{eq:FormalSolutionForRho} and choosing the reference state to be the initial state of the environment gives
\begin{widetext}
\begin{align}
    \frac{\dd}{\dd t} 
    \hrhoi(t)
&=
    - 
    i
    \left[
        \hH_{\eff}(t)
        ,
        \hrho(t_{0})
    \right]
    -
    \int_{t_{0}}^{t} 
    \dd t_{1}
    \Tr_{\E}
    \left\{
        \left[
            \hH_{\ipic}(t)
            , 
            \left[
                \hH_{\ipic}(t_{1})
                ,
                \hrhoi(t_{1})
                \otimes
                \hrho_{\E}
            \right]
        \right]
        +
        \left[
            \hH_{\ipic}(t)
            , 
            \left[
                \hH_{\ipic}(t_{1})
                ,
                \hCQ
                \hsigi(t_{1})
            \right]
        \right]
    \right\}
	.
\end{align}
\end{widetext}
We have come a bit closer to the goal of reducing all instances of $ \hsigi(t_{1}) $ to $ \hrhoi(t_{1}) $ and $ \hrho_{\E} $, with the exception of the irrelevant part of $ \hsigi(t_{1}) $, which brings us to our first approximation.

We now restrict to the weak-coupling regime and expand the term containing the irrelevant part of $ \hsigi(t_{1}) $ to second order in $ \la $. Since there are already two factors of $ \la $ (one from each $ \hH_{\ipic} $), we only need to make the expansion $ \hCQ \hsigi(t_{1}) = \hCQ \hsigi(t_{0}) + {\cal O}(\la) $. It is clear from the definitions in \cref{eq:Initial_joint_density_operator,eq:DefProjectionOperator}, however, that the projection operator leaves the initial density operator unchanged, from which we conclude that $ \hCQ \hsigi(t_{0}) = 0 $. Therefore, in the weak-coupling regime, we can write the master equation as
\begin{align}
	& \frac{\dd}{\dd t} 
	\hrhoi(t)
=
	- 
	i
	\left[
		\hH_{\eff}(t)
		,
		\hrho(t_{0})
	\right]
\nn
&\hspace{13.5pt}
	-
	\int_{t_{0}}^{t} 
	\dd t_{1}
	\Tr_{\E}
	\left[
		\hH_{\ipic}(t)
		, 
		\left[
			\hH_{\ipic}(t_{1})
			,
			\hrhoi(t_{1}) \otimes
			\hrho_{\E}
		\right]
	\right]
	.
\label{eq:CommutatorBornME}
\end{align}
This approximation is called the Born approximation and amounts to factorizing the composite system density operator as
\begin{align}
	\hsigi(t_{1})
\approx	
	\hrhoi(t_{1})
	\otimes
	\hrho_{\E}
	\,,
\label{eq:BornApproximation}
\end{align}
so that the density operator of the environment is almost stationary in the interaction picture. The Born approximation is also often presented as the assumption that the environment is sufficiently large and the interaction sufficiently weak so that environment excitations induced by the system are negligible compared to its free dynamics. Since weak coupling was sufficient in the above analysis, however, we make the Born approximation without any requirement on the number of degrees of freedom in the environment. We also note that although \cref{eq:CommutatorBornME} is only valid to second order in perturbation theory, the Born approximation is not equivalent to truncating the Dyson series expansion since $ \hrhoi(t_{1}) $ itself has not been expanded in a perturbation series. In fact, as shown in \cref{sec:TwoPointCorrelator}, leaving $ \hrhoi(t_{1}) $ intact allows one to obtain a second-order resummation for the equal-time two-point function of the reduced system.

We can write the master equation in \cref{eq:CommutatorBornME} in a more convenient form by expanding out the nested commutator on the right-hand side and using the explicit form of the interaction Hamiltonian from \cref{eq:Interaction_Hamiltonian_Fourier}. This gives
\begin{widetext}
    \begin{align}
        \frac{\dd}{\dd t} 
        \hrhoi(t)
    &=
        - 
        i
        \left[
            \hH_{\eff}(t)
            ,
            \hrho(t_{0})
        \right]
        -
        \la^{2}
        \sum_{\al, \be}
        \int_{\vk} 
		\frac{
			1
		}{
			2
			\om_{k}
		}
        \int_{t_{0}}^{t} \dd t_{1}
        \left\{
            \mathcal{C}_{k}(t,t_{1})
            \left[
                \hLdagg{\vk, \al}(t)
                ,
                \hL{\vk, \be}(t_{1})
                \hrhoi(t_{1})
            \right]
            +
            \HC
        \right\} ,
    \label{eq:BornME}
    \end{align}
\end{widetext}
where H.c. indicates the Hermitian conjugate, and we have moved all system operators out of the environment trace, defining the environment correlation function $ \mathcal{C}_{k}(t,t_{1}) $ through the relation~\footnote{We will see in \cref{sec:phichi,sec:phichi2} that for both interactions we consider, this environment correlator only depends on the magnitude $ k $.}
\begin{align}
   \Tr_{\E}
   \left\{
   \hO_{-\vk\pr}(t)
   \hO_{-\vk}(t_{1})
   \hrho_{\E}
   \right\}
=
	\frac{
		\om_{k}^{2}
	}{
		2
	}
	\mathcal{C}_{k}(t,t_{1})
	(2 \pi)^{3}
	\delta^{3}
	\big(
		\vk
		+
		\vk\pr
	\big)
   \, ,
\label{eq:EnvironementCorrelation}
\end{align}
with the $ \om_{k}^{2}/{2} $ separated out for later convenience. We can see from \cref{eq:BornME} that the Born approximation not only allows us to write an integrodifferential equation for $ \hrhoi $ but also to collate the effect of the environment into time-dependent correlation functions that appear as coefficients on system operators. 

Although we have made tidy progress by making the Born approximation, the time integral over $ \hrhoi(t_{1}) $ in \cref{eq:BornME} implies that we are still accounting for correlations that are not time-local due to the interaction with the environment. In the standard picture of open systems, the environment is typically a reservoir with correlation functions that decay rapidly compared to the timescale over which the system evolves appreciably, naturally suppressing time-nonlocal system correlations~\cite{Schlosshauer2007}~\footnote{Also see ref.\ \cite{Prudhoe:2022pte} for an effective time-local master equation construction and ref.\ \cite{Agarwal:2023lid} for an effective field theory-inspired approach.}. This lack of {\it memory} in the environment justifies the Markov approximation, which amounts to the replacement $ \hrhoi(t_{1}) \to \hrhoi(t) $ and is interpreted as neglecting coherent feedback from the environment. 

Before proceeding, we first argue that the Markov approximation is also justified in a different case, in particular, when the environment correlation function $ \mathcal{C}_{k}(t,t_{1}) $ is purely oscillatory and the characteristic timescale $ \tau_{\E} $ of $ \mathcal{C}_{k}(t,t_{1}) $ is much shorter than the characteristic timescale of the system $ \tau_{\Sy} $. We can see this by considering the $ t_{1} $ integral in \cref{eq:BornME},
\begin{align}
	\int_{t_{0}}^{t} 
        \dd t_{1}
	\mathcal{C}_{k}(t,t_{1}) 
	e^{ 
            i
            (-1)^{\be} 
            \om_{k} 
            \left( 
                t_{1} 
                - 
                t_{0} 
            \right) 
        } 
	\hrhoi(t_{1})
	\,,
\label{eq:BeforeByPartsMarkovApprox}
\end{align}
the exponential factor being the time dependence of $ \hL{\vk, \be}(t_{1}) $. We integrate by parts to find that~\footnote{Since we have used definite integrals in the integration by parts, there are terms that depend on the initial time $ t_{0} $. It is straightforward to show, however, that these terms cancel identically.}
\begin{align}
&
	\int_{t_{0}}^{t} \dd t_{1}
	\mathcal{C}_{k}(t,t_{1}) 
	e^{ i (-1)^{\be} \om_{k} \left( t_{1} - t_{0} \right) } 
	\hrhoi(t_{1})
\nn
&\hspace{11.5pt}
=
	\int_{t_{0}}^{t}
	\dd t_{1}
	\mathcal{C}_{k}(t,t_{1}) 
	e^{ i (-1)^{\be} \om_{k} \left( t_{1} - t_{0} \right) } 
	\hrhoi(t)
\nn
&\hspace{23pt}
	-
	\int_{t_{0}}^{t} \dd t_{1}
	\mathcal{K}_{k,\be}(t,t_{1}) 
	\frac{\dd}{\dd t_{1}}
	\hrhoi(t_{1})
	\,,
\label{eq:ByPartsMarkovApprox}
\end{align}
where we have defined
\begin{align}
	\mathcal{K}_{k,\be}(t,t_{1}) 
=
	\int_{t_{0}}^{t_{1}}
	\dd t_{2}
	\mathcal{C}_{k}(t,t_{2}) 
	e^{ i (-1)^{\be} \om_{k} \left( t_{2} - t_{0} \right) } 
	\,.
\label{eq:Kkbeta}
\end{align}
The first term on the right-hand side of \cref{eq:ByPartsMarkovApprox} is precisely the Markov approximation, so we are left with the task of showing that the second term is negligible. If $ \mathcal{K}_{k,\be}(t,t_{1}) $ is a purely sinusoidal function with a period of $ \tau_{\star} $, then we partition the domain of integration as 
\begin{align}
	(t_{0}, t) 
= 
	\bigcup_{n = 0}^{N-1}
	(t_{0} + n \tau_{\star}, t_{0} + (n+1) \tau_{\star}) 
	\cup 
	(t_{0} + N \tau_{\star}, t)
	\,,
\end{align}
where $ N = \lfloor (t-t_{0})/ \tau_{\star} \rfloor $, and rewrite the integral in the second term on the right-hand side of \cref{eq:ByPartsMarkovApprox} as $\int_{t_{0}}^{t} \dd t_{1} = \sum_{n = 0}^{N-1} \int_{t_{0} + n \tau_{\star}}^{t_{0} + (n+1) \tau_{\star}} \dd t_{1} + \int_{t_{0} + N \tau_{\star}}^{t} \dd t_{1}$. Now if $ \hrhoi(t_{1}) $ varies over a characteristic timescale $ \tau_{\Sy} \gg \tau_{\star} $, then we can move the derivative of $ \hrhoi(t_{1}) $ outside of the integrals as
\begin{align}
&
    \int_{t_{0}}^{t} \dd t_{1}
    \mathcal{K}_{k,\be}(t,t_{1}) 
    \frac{\dd}{\dd t_{1}}
    \hrhoi(t_{1})
\nn
&=
    \sum_{n = 0}^{N-1}
    \frac{\dd}{\dd t_{1}}
    \hrhoi(t_{1})
    \Big|_{t_{0} + (n+1) \tau_{\star}}
    \int_{t_{0} + n \tau_{\star}}^{t_{0} + (n+1) \tau_{\star}} \dd t_{1}
    \mathcal{K}_{k,\be}(t,t_{1}) 
\nn
&\hspace{11.5pt}
    +
    \frac{\dd}{\dd t_{1}}
    \hrhoi(t_{1})
    \Big|_{t}
    \int_{t_{0} + N \tau_{\star}}^{t} \dd t_{1}
    \mathcal{K}_{k,\be}(t,t_{1}) 
	\,.
\end{align}
The integrals under the summation now vanish identically, and we are only left with the second term on the right-hand side. Furthermore, since $ N \tau_{\star} \to t - t_{0} $ as $ \tau_{\star} \to 0 $, we find that  $ t - (t_{0} + N \tau_{\star}) \approx 0 $ for sufficiently small $ \tau_{\star} $, and so the second term is also negligible if the environment timescale is much shorter than the system timescale. We can, therefore, make the Markov approximation, replacing $ \hrhoi(t_{1}) \to \hrhoi(t) $ in \cref{eq:BeforeByPartsMarkovApprox}.

In either of the two cases discussed above, whether the environment correlation function is purely oscillatory with a short enough timescale (which we will find for $ \la \hphi \hchi $) or decays sufficiently fast (for $ \la \hphi \hchi^{2} $), the replacement $ \hrhoi(t_{1}) \to \hrhoi(t) $ is justified and leaves us with a time-local master equation called the Redfield equation,
\begin{align}
&	\frac{\dd}{\dd t}
	\hrhoi(t)
=
	- 
	i
	\left[
		\hH_{\eff}(t)
		,
		\hrho(t_{0})
	\right]
\nn
&\hspace{11.5pt}
	-
	\int_{t_{0}}^{t} \dd t_{1}
	\Tr_{\E}
	\left[
		\hH_{\ipic}(t)
		, 
            \hH_{\ipic}(t_{1})
            \hrhoi(t)
            \hrho_{\E}
	\right] 
        + \HC
        \, ,
\label{eq:RedfieldEquation}
\end{align}
where we have suppressed the $ \otimes $ between the density operators and restored the interaction Hamiltonian to highlight the general form. As discussed in the introduction, it is well known that the Redfield equation cannot be put into GKLS form in general and is thus not guaranteed to yield a completely positive map. The standard procedure to ensure positivity is to further take the Markovian limit and make the RWA, and the resulting equation is called the Davies equation. In the following two subsections, we rewrite the Redfield equation in a form that allows for a straightforward transition to the Davies equation so that we can easily compare our results in \cref{sec:TwoPointCorrelator} under different approximations.

%--------------------------
\subsection{Markovian limit}
\label{sec:MarkovianLimit}
%--------------------------

When the justification for the Markov approximation is the rapid decay of environment correlation functions, it is standard procedure to make a further simplification to the Redfield equation. Specifically, the initial state dependence can be removed by taking the Markovian limit~\cite{Breuer:2002pc}, the general procedure for which is as follows. Suppose we are interested in the integral of some function $f(t_1)$ from the initial time $ t_{0} $ to a later time $ t $. We can make a change of variables such that the integration is over the time elapsed since $ t_{0} $ as
\begin{align}
	I(t)
=
	\int_{t_{0}}^{t}
	\dd t_{1}
	f(t_{1}) 
\stackrel{ t_{1} \to t - t\pr }{ = }
	\int_{0}^{t-t_{0}}
	\dd t\pr
	f(t - t\pr) 
	\,.
\end{align}
Now, if the function $ f(t-t\pr) $ decays to zero over the domain of integration, the value of $I(t)$ does not change on increasing the upper limit of the integral. The initial time can, therefore, be removed by letting $ t - t_{0} \to \infty $, yielding the Markovian limit of $ I(t) $, 
\begin{align}
	I^{M}(t)
=
	\int_{0}^{\infty}
	\dd t_{1}
	f(t - t_{1}) \,,
\end{align}
and removing the memory of the system from the dynamics. Environments for which the Markovian limit is permissible are referred to as Markovian environments, and the reduced dynamics are said to be Markovian.

Rather than taking the Markovian limit at this stage, we will continue with the upper limit on the integral as $ t - t_{0} $ so that the Markovian limit can be obtained simply by letting $ t - t_{0} \to \infty $ in the $ t_{1} $ integral. We thus rewrite the Redfield equation in \cref{eq:RedfieldEquation} as 
\begin{align}
&	\frac{\dd}{\dd t}
    \hrhoi(t)
=
	- 
	i
	\left[
		\hH_{\eff}(t)
		,
		\hrho(t_{0})
	\right]
\nn
&\hspace{11.5pt}
	-
	\int_{t_{1}}
	\Tr_{\E}
	\left[
		\hH_{\ipic}(t)
		,
		\hH_{\ipic}(t - t_{1})
		\hrhoi(t)
		\hrho_{\E}
	\right]
	+
	\HC \, ,
\label{eq:Born_Markov_master_equation}
\end{align}
where $ \int_{t_{1}} \equiv \int_{0}^{t - t_{0}} \dd t_{1} $.

%--------------------------
\subsection{Standard form of the master equation}
\label{sec:SFME}
%--------------------------

While it is common to consider an environment in thermal equilibrium, here we initialize the environment in the vacuum state instead. We find that this suffices to justify the Markovian limit in the case of the nonlinear interaction considered in Sec. \ref{sec:phichi2}. With this choice of initial state, we now expand the environment correlations that appear in \cref{eq:Born_Markov_master_equation} using Wick's contractions. Since the resulting two-point correlators conserve momentum and are invariant under time translations, we can write \cref{eq:Born_Markov_master_equation} as
\begin{align}
&	\frac{\dd}{\dd t}
	\hrhoi(t)
=
	- 
	i
	\left[
		\hH_{\eff}(t)
		,
		\hrho(t_{0})
	\right]
\nn 
&\hspace{12.5pt}
	-
	\sum_{\al,\be}
	\int_{\vk}
	\Fk{\al \be}(t)
	\Gamk{\be}(t)
	\left[
		\hLdagg{\vk, \al}
		,
		\hL{\vk, \be}
		\hrhoi(t)
	\right]
	+
	\HC \, ,
\label{eq:Standard_form_1}
\end{align}
with the coefficients
\begin{align}
	\Fk{\al \be}(t)
&\equiv
	e^{
		-
		i
		\left(
			(-1)^{\al}
			-
			(-1)^{\be}
		\right)
		\om_{k}
		\left(
			t
			-
			t_{0}
		\right)
	}
	\,,
\label{eq:F_al_be} 
\\
    \Gamk{\be}(t)
&\equiv
	\la^{2}
	\int_{t_{1}}
	\frac{
		e^{
			-
			i
			(-1)^{\be}
			\om_{k}
			t_{1}
		}
	}{
		2
		\om_{k}
	}
	\mathcal{C}_{k}(t_{1})
	\,,
\label{eq:Gamma_al_be}
\end{align}
where $ \Fk{\al \be} $ is Hermitian in the sense that it satisfies the relation $\Fk{\al \be} = \Fk{\be \al}\conj$. The environment correlation function takes the explicit form~\footnote{Since we are working at $ O(\la^{2}) $, only tree-level diagrams contribute to the environment correlation function. This will be convenient when we renormalize the $ \la \phi \chi^{2} $ theory in \cref{sec:phichi2}.}
\begin{align}
    \mathcal{C}_{k}(t_{1})
=
    \frac{
        8 \pi^{3}
    }{
        n!
    }
	\int
    \delta^{3}
    \big(
        \vk
        +
        \textstyle{\sum_{i}}
        \vp_{i}
    \big)
	\prod_{j=1}^{n}
	\frac{
        \d^{3} p_{j}
	}{
		(2 \pi)^{3}
	}
	\frac{
        e^{ - i \Om_{p_{j}} t_{1} } 
	}{
        2
        \Om_{p_{j}}
	}
	\,,
\label{eq:ExplicitEnvCorr}
\end{align}
where we have dropped equal-time bubble diagrams that appear for $ n = 2 $, since they can be canceled by adding appropriate renormalization counterterms, as we show explicitly for the $n = 2$ case in \cref{sec:phichi2}, and used time translation invariance to write $ \mathcal{C}_{k} $ with a single time argument. As noted before, $ \mathcal{C}_{k} $ only depends on the magnitude of $ k $ since the $ \vp_{j} $ integrals can always be evaluated relative to the direction of $ \vk $. It is also worth noting that everything under the summation in \cref{eq:Standard_form_1} is dimensionless except for $ \Gamk{\be} $, which is a rate with mass dimension one. Finally, including the Hermitian conjugate term explicitly, we can write \cref{eq:Standard_form_1} in the standard form
\begin{widetext}
\begin{align}
    \frac{\dd}{\dd t}
    \hrhoi(t)
&=
    - 
    i
    \left[
        \hH_{\eff}(t)
        ,
        \hrho(t_{0})
    \right]
    -
    i
    \sum_{\al,\be}
    \int_{\vk}
    \Fk{\al \be}(t)
    \Sk{\al \be}(t)
    \left[
            \hLdagg{\vk, \al}
            \hL{\vk, \be}
            ,
            \hrhoi(t)
    \right]
\nn
&\hspace{11.5pt}
    +
    \sum_{\al,\be}
    \int_{\vk}
    \Fk{\al \be}(t)
    \gamk{\al \be}(t)
    \left[
            \hL{\vk, \be}
            \hrhoi(t)
            \hLdagg{\vk, \al}
            -
            \frac{
                1
            }{
                2
            }
            \left\{
                \hLdagg{\vk, \al}
                \hL{\vk, \be}
                ,
                \hrhoi(t)
            \right\}
    \right]
	,
\label{eq:Standard_form_2}
\end{align}
\end{widetext}
where we have defined
\begin{align}
	\gamk{\al \be}(t)
&\equiv
	\Gamk{\be}(t)
	+
	\Gamk{\al}\conj(t)
	\, ,
\label{eq:gammaAlBe} \\
	\Sk{\al \be}(t)
&\equiv
	\frac{
		1
	}{
		2
		i
	}
	\left[
		\Gamk{\be}(t)
		-
		\Gamk{\al}\conj(t)
	\right]
	,
\label{eq:SAlBe}
\end{align}
both of which are also Hermitian matrices that satisfy $ \gamk{\al \be} = \gamk{\be \al}\conj $ and $ \Sk{\al \be} = \Sk{\be \al}\conj $.

In \cref{eq:Standard_form_1,eq:Standard_form_2}, $ \Fk{\al \be} $ is the interaction picture time dependence of the dimensionless system operators $ \hL{\vk, \al} $, which we have separated out so that we may readily make the RWA. The motivation for the RWA is more evident for a generic master equation where this time dependence is $ \exp[- i (\om' - \om) t] $ and there are summations over both Bohr frequencies $ \om $ and $ \om' $, which can be positive or negative. When the difference of the Bohr frequencies is sufficiently large, this exponential rapidly oscillates on timescales over which the system dissipates and the corresponding terms average to zero. The RWA neglects these {\it off-diagonal} contributions, leaving only terms with $ \om' = \om $. Of course, this is not the case for systems with dense level spacing, further implying that the RWA is a poor approximation for systems with a continuous spectrum~\cite{Trushechkin2021,Mozgunov_2020}. This observation naturally signals that the RWA is not likely to be valid for a field theory where $ \om_{k} $ represents a continuum of Bohr frequencies. One subtlety in the present case is that momentum conservation has modified the above argument so that the magnitudes of $ \om $ and $ \om' $ are the same. Indeed, making the RWA amounts to replacing $ \Fk{\al \be} \to \delta_{\al \be} $ that neglects all terms proportional to $ \exp[\pm i 2 \om_{k} (t - t_{0})] $, requiring $ 2 \om_{k} $ to be large for all $ k $. This requirement is distinctly problematic in the massless limit, where $ \om_{k} = k $ can be arbitrarily small, and is not clearly resolved in the massive case. We will, therefore, not make this approximation in general but consider how it affects our results only from a pedagogical standpoint.

While we could continue to work in the interaction picture, we now choose to transform to the Schr\"odinger picture. This introduces a factor of $ \Fk{\al \be}\conj(t) $ that multiplies $ \Fk{\al \be}(t) $ to yield unity and a commutator of $ \hrho(t) $ with the free Hamiltonian of the system. This commutator can be combined with the second term of \cref{eq:Standard_form_2} to define a modified system Hamiltonian,
\begin{align}
	\hH_{\Sy}(t)
\equiv
	\hH_{0, \Sy}
	+
	\sum_{\al,\be}
	\int_{\vk}
	\Sk{\al \be}(t)
	\hLdagg{\vk, \al}
	\hL{\vk, \be}
	\,,
\label{eq:ShiftHamiltonian}
\end{align}
where, for $ \al = \be $, the second term on the right-hand side is identified as an environment-induced shift in energy levels of the system and is then referred to as the Lamb shift. Transforming back to the Schr\"odinger picture turns out to be mathematically convenient as it circumvents the need to handle time-dependent coefficients appearing in the Markovian limit, but we comment on the interaction picture formulation in the next section.

Using the fundamental relationship $ \hrhoi(t) = e^{ i \hH_{0, \Sy} ( t - t_{0}) } \hrho(t) e^{ - i \hH_{0, \Sy} ( t - t_{0}) } $, we now write the standard form of the Redfield equation in the Schr\"odinger picture as 
\begin{widetext}
\begin{align}
    \frac{\dd}{\dd t}
    \hrho(t)
&=
    - 
    i
    \left[
        \hH_{\eff}
        ,
        \hrho_{0}(t)
    \right]
    -
    i
    \left[
        \hH_{\Sy}(t)
        ,
        \hrho(t)
    \right]
    +
    \sum_{\al,\be}
    \int_{\vk}
    \gamk{\al \be}(t)
    \left[
        \hL{\vk, \be}
        \hrho(t)
        \hLdagg{\vk, \al}
        -
        \frac{
            1
        }{
            2
        }
        \left\{
            \hLdagg{\vk, \al}
            \hL{\vk, \be}
            ,
            \hrho(t)
        \right\}
    \right]
    ,
\label{eq:Standard_form}
\end{align}
\end{widetext}
where we have defined $ \hrho_{0}(t) \equiv e^{ - i \hH_{0, \Sy} ( t - t_{0}) } \hrho(t_{0}) e^{ i \hH_{0, \Sy} ( t - t_{0}) } $ and used $ \hrho_{\E} = \ket{0_{\E}} \bra{0_{\E}} $ to reduce the Schr\"odinger picture effective Hamiltonian to $ \hH_{\eff} = \bra{0_{\E}} \hH_{\intac} \ket{0_{\E}} $. The last term in eq.\ (\ref{eq:Standard_form}) is non-Hamiltonian and describes the decoherence of the system due to the environment. Equation (\ref{eq:Standard_form}) is the final master equation we will use in the remainder of the paper. We note that the procedure to make the RWA is not evident anymore since the time dependence of the rapidly oscillating terms has been absorbed into the states. The approximation can, nevertheless, be made by simply dropping the off-diagonal ($ \al \neq \be $) terms.

%%%%%%%%%%%%%%%%%%%%%%%%%%%%%%%%%%%%%%%%%%%%%%%%%%
%--------------------------
\section{Equal-time two-point function}
\label{sec:TwoPointCorrelator}
%--------------------------

In principle, one can solve the master equation in \cref{eq:Standard_form} for $\hrho$ and use it to investigate the dynamics of the reduced system. For two-level systems, for example, this is typically done by first writing the density operator in Bloch decomposition, then using the master equation to obtain coupled differential equations for components of the Bloch vector, and finally solving the resulting equations to obtain $\hrho$ at any time $t$ \cite{Breuer:2002pc}. Especially for systems with infinite-dimensional Hilbert spaces, such as QFTs, this is challenging as it amounts to determining all matrix elements of $ \hrho $. For such systems, we can instead use the master equation to directly solve for {\it system observables} such as equal-time correlation functions \cite{Boyanovsky:2015xoa}. In this section, we thus use \cref{eq:Standard_form} to write down coupled differential equations in correlation functions of $ \ha{\vk} $ and $ \hadagg{\vk} $. We further demonstrate that this correlation function approach allows for a direct comparison of how various approximations manifest in the reduced dynamics.

We start with the definition of the equal-time two-point correlation function for $ \hphi_{\vk} $, written in the Heisenberg picture as
$	
    \langle
        \hphi_{\vk}(t)
        \hphi_{\vk\pr}(t)
    \rangle
=
    \Tr
    \big\{
        \hphi_{\vk}(t)
        \hphi_{\vk\pr}(t)
        \hsig(t_{0})
    \big\}
    ,
$ 
where $ \hsig(t_{0}) $ is again the initial density operator for the composite system. Since we would ultimately like to use the Schr\"odinger picture master equation, \cref{eq:Standard_form}, we move the time dependence from the field operators to the density operator,
\begin{align}
	\Tr
	\left\{
		\hphi_{\vk}
		\hphi_{\vk\pr}
		\hsig(t)
	\right\}
=
	\CG_{k}(t)
	(2 \pi)^{3}
	\delta^{3}
	\big(
		\vk
		+
		\vk\pr
	\big) \,
	.
\label{eq:TwoPoint}
\end{align}
We now move the trace over the environment past $ \hphi_{\vk} $ and $ \hphi_{\vk\pr} $, use the definition of the reduced density operator $ \hrho(t) = \Tr_{\E} \hsig(t) $, and expand the left-hand side in terms of correlation functions of $ \ha{\vk} $ and $ \hadagg{\vk} $. Since each correlation function must conserve momentum, we define the functions $ \xi_{k} = \xi_{k,1} + i \xi_{k,2} $ and $ \xi_{k,3} $ such that
\begin{align}
	\Tr_{\Sy}
	\left\{
		\ha{\vk}
		\ha{\vk\pr}
        \hrho(t)
	\right\}
&=
	\xi_{k}(t)
	(2 \pi)^{3}
	\delta^{3}
	\big(
		\vk
		+
		\vk\pr
	\big) \,
	,
\label{eq:aa_Correlation} \\
	\Tr_{\Sy}
	\left\{
        \left[
    		\ha{\vk}
    		\hadagg{-\vk\pr}
    		+
    		\hadagg{-\vk}
    		\ha{\vk\pr}
        \right]
        \hrho(t)
	\right\}
&=
	\xi_{k,3}(t)
	(2 \pi)^{3}
	\delta^{3}
	\big(
		\vk
		+
		\vk\pr
	\big) \,
	,
\label{eq:aadagg_Correlations}
\end{align}
in terms of which $\CG_{k}$ is given by
\begin{align}
	\CG_{k}(t)
&=
	\frac{
		1
	}{
		2 
		\om_{k}
	}
	\left[
		2
		\xi_{k,1}(t)
		+
		\xi_{k,3}(t)
	\right]
    .
\label{eq:Gk}
\end{align}
We now multiply \cref{eq:Standard_form} by appropriate combinations of creation and annihilation operators and obtain a set of coupled first-order differential equations for $ \xi_{k} $ and $ \xi_{k,3} $. The contribution of the $ \hH_{\eff} $ term vanishes identically since it can always be written as the expectation of three system operators. Thus, the coefficients that appear in the coupled equations are combinations of the functions $ \Sk{\al \be} $ and $ \gamk{\al \be} $. Suppressing the $ k $ subscripts on all variables, we find that
\begin{align}
	\dot{\xi}(t)
&=
	-
	\left[
		\gamd{-}(t)
		+
		i
		2
		\omS(t)
	\right]
	\xi(t)
	-
	i
	2
	\Sd{\rm o}(t)
	\xi_{3}(t)
	-
	\gamd{\rm o}(t)
	\,,
\label{eq:Xi_diff_eq}
\\
	\dot{\xi}_{3}(t)
&=
	- 
	\gamd{-}(t)
	\xi_{3}(t)
	+
	8
	\Im
	\left[
		\Sd{\rm o}(t)
		\xi\conj(t)
	\right]
	+
	\gamd{+}(t)
	\,,
\label{eq:Xi3_diff_eq}
\end{align}
where we have defined the functions $ \gamd{\rm o}(t) \equiv \gamk{1 2}(t) $, $ \Sd{\rm o}(t) \equiv \Sk{1 2}(t) $, $ \gamd{\pm}(t) \equiv \gamk{1 1}(t) \pm \gamk{2 2}(t) $, and $ \omS(t) \equiv \om_{k} + \Sk{1 1}(t) + \Sk{2 2}(t) $, with the subscript "$ \text{o} $" denoting "off diagonal". Once an initial density operator is chosen, \cref{eq:Xi_diff_eq,eq:Xi3_diff_eq} can be solved using initial conditions obtained from \cref{eq:aa_Correlation,eq:aadagg_Correlations}. We denote the resulting two-point function obtained by using these solutions for $ \xi(t) $ and $ \xi_{3}(t) $ in \cref{eq:Gk} with $ \CG_{k}^{\text{R}}(t) $.

We note that although we have chosen to formulate differential equations for the system creation and annihilation operators in the Schr\"odinger picture, this is not necessary~\footnote{This qualitative equivalence between quantum master equations expressed in Schr{\"o}dinger and interaction pictures does not hold when the free evolution of the system itself is nonunitary \cite{Thorbeck2024}.}. Similar strategies have been employed in the interaction picture, where operators are time-dependent, and using the field basis, where $ \hL{\vk} $ are not eigenoperators of the free Hamiltonian~\cite{Boyanovsky:2015xoa,Martin:2018lin,Colas:2022hlq}. In these cases, the time dependence requires additional care|for example, if \cref{eq:Standard_form_2} is used in place of \cref{eq:Standard_form}, then the $ \xi $ variables become {\it mixed picture correlators} of Schr\"odinger picture operators with the interaction picture density operator. The two-point function can then be constructed by transferring the free time evolution from $ \hrho(t) $ to $ \hphi_{\vk} $ in \cref{eq:TwoPoint} and expanding the two-point function in terms of the mixed picture correlators. The main differences are that $ \omS(t) \to \Sk{11}(t) + \Sk{22}(t) $ and the off-diagonal coefficients are multiplied by a factor of $ \Fk{1 2}(t) $.

%--------------------------
\subsection*{Approximations}
%--------------------------

The coupled differential equations (\ref{eq:Xi_diff_eq}) and (\ref{eq:Xi3_diff_eq}) obtained from the Redfield equation are well-suited for numerical solutions. Finding analytical solutions, on the other hand, is not feasible in general but may be possible after making further simplifications. We consider four simplified cases below: the Markovian limit, the RWA, the Markovian limit under the RWA, and the standard Dyson series perturbation theory.

%--------------------------
\subsubsection{Markovian limit}
%--------------------------

Under the Markovian limit, we let $ t - t_{0} \to \infty $ in $ \Gamk{\be} $, removing the time dependence from $ \gamk{\al \be} $ and $ \Sk{\al \be} $. Since there is no scope for ambiguity, we will indicate this limit by simply dropping the time argument. In this case, the coupled equations (\ref{eq:Xi_diff_eq}) and (\ref{eq:Xi3_diff_eq}) become 
\begin{align}
	\dot{\xi}(t)
&=
	-
	\left(
		\gamd{-}
		+
		i
		2
		\omS
	\right)
	\xi(t)
	-
	i
	2
	\Sd{\rm o}
	\xi_{3}(t)
	-
	\gamd{\rm o}
	\,,
\label{eq:Xi_diff_eq_Markovian} \\
	\dot{\xi}_{3}(t)
&=
	- 
	\gamd{-}
	\xi_{3}(t)
	+
	8
	\Im
	\left[
		\Sd{\rm o}
		\xi\conj(t)
	\right]
	+
	\gamd{+}
	\,.
\label{eq:Xi3_diff_eq_Markovian}
\end{align}
The Markovian limit is a substantial simplification and leads to a set of coupled, first-order autonomous differential equations that can be solved by using, for example, the Laplace transform, as detailed in \Cref{app:MarkovianLimitSolution}. We denote the resulting two-point function in the Markovian limit, obtained by using these solutions for $ \xi(t) $ and $ \xi_{3}(t) $ in \cref{eq:Gk}, with $ \CG_{k}^{\text{M}}(t) $.

%--------------------------
\subsubsection{RWA}
%--------------------------

Under the RWA, we drop the off-diagonal terms in \cref{eq:Xi_diff_eq,eq:Xi3_diff_eq}, resulting in the set of {\it decoupled} equations,
\begin{align}
	\dot{\xi}(t)
&=
	-
	\left[
		\gamd{-}(t)
		+
		i
		2
		\omS(t)
	\right]
	\xi(t)
	\,,
\label{eq:RWAXi} \\
	\dot{\xi}_{3}(t)
&=
	- 
	\gamd{-}(t)
	\xi_{3}(t)
	+
	\gamd{+}(t)
	\,.
\label{eq:RWAXi3}
\end{align}
The $ \xi $ equation is solved by simple separation of variables and the $ \xi_{3} $ equation can be solved with an integrating factor; the solutions are 
\begin{align}
&\hspace{4pt} \xi(t)
=
	\xi(t_{0})
	e^{
		-
		\int_{t_{0}}^{t}
		\dd t_{1}
		\left[
			\gamd{-}(t_{1})
			+
			i
			2
			\omS(t_{1})
		\right]
	}
	\,,
\\
&	\xi_{3}(t)
=
	\xi_{3}(t_{0})
	e^{
		-
		\int_{t_{0}}^{t}
		\dd t_{1}
		\gamd{-}(t_{1})
	}
\nn
&\hspace{11.5pt}
	+
	e^{
		-
		\int_{t_{0}}^{t}
		\dd t_{1}
		\gamd{-}(t_{1})
	}
	\int_{t_{0}}^{t}
	\dd t_{1}
	e^{
		\int_{t_{0}}^{t_{1}}
		\dd t_{2}
		\gamd{-}(t_{2})
	}
	\gamd{+}(t_{1})
	\,.
\end{align}
We denote the resulting two-point function under the RWA, obtained by using these solutions for $ \xi(t) $ and $ \xi_{3}(t) $ in \cref{eq:Gk}, with $ \CG_{k}^{\text{RWA}}(t) $.

%--------------------------
\subsubsection{RWA and Markovian limit}
%--------------------------

Performing the RWA and taking the Markovian limit together effectively reduces the Redfield equation to the Davies equation. In this case, the coefficients in eqs. (\ref{eq:RWAXi}) and (\ref{eq:RWAXi3}) become time-independent and their solutions reduce to 
\begin{align}
	\xi(t)
&=
	\xi(t_{0})
	e^{
		-
		\left(
			\gamd{-}
			+
			i
			2
			\omS
		\right)
		\left(
			t
			-
			t_{0}
		\right)
	}
	\,,
\\
	\xi_{3}(t)
&=
	\xi_{3}(t_{0})
	e^{
		-
		\gamd{-}
		\left(
			t
			-
			t_{0}
		\right)
	}
	+
	\frac{
		\gamd{+}
	}{
		\gamd{-}
	}
	\left[
		1
		-
		e^{
			-
			\gamd{-}
			\left(
				t
				-
				t_{0}
			\right)
		}
	\right]
	.
\end{align}
In both examples considered in the following sections, where we assume that the system starts in the vacuum of the free theory, we find that $ \gamd{+} = \gamd{-} $ and that these solutions reduce to their initial values: $ \xi(t) = 0 $ and $ \xi_{3}(t) = 1 $. Therefore, the Davies master equation does not capture the open system dynamics that we are interested in, and we do not consider it further in the paper.

%--------------------------
\subsubsection{Dyson series perturbation theory}
%--------------------------

As discussed in \cref{sec:RedfieldME}, the Redfield equation matches the standard Dyson series at second order in $ \la $ when $ \hrhoi(t_{1}) $ is expanded to zeroth order on the right-hand side of \cref{eq:CommutatorBornME}, replacing $ \hrhoi(t_{1}) $ there with $\hat{\rho}(t_0)$. This also decouples \cref{eq:Xi_diff_eq,eq:Xi3_diff_eq}, resulting in the equations 
\begin{align}
	\dot{\xi}(t)
&=
    -
    i
    2
    \om_{k}
    \xi(t)
    -
    \left[
        \gamd{-}(t)
        +
        i
        2
        \tilde{\om}_{\rm S}(t)
    \right]
    \xi(t_{0})
\nn
&\hspace{11.5pt}
    -
    i
    2
    \Sd{\rm o}(t)
    \xi_{3}(t_{0})
    -
    \gamd{\rm o}(t)
    \,,
\label{eq:Xi_diff_eq_PT}
\\
    \dot{\xi}_{3}(t)
&=
    - 
    \gamd{-}(t)
    \xi_{3}(t_{0})
    +
    8
    \Im
    \left[
        \Sd{\rm o}(t)
        \xi\conj(t_{0})
    \right]
    +
    \gamd{+}(t)
    \,,
\label{eq:Xi3_diff_eq_PT}
\end{align}
where $ \tilde{\om}_{\rm S}(t) = \Sk{1 1}(t) + \Sk{2 2}(t) $. These expressions can be obtained directly from \cref{eq:Xi_diff_eq,eq:Xi3_diff_eq} by replacing all $ \xi(t) $ and $ \xi_{3}(t) $ on the right-hand side with their respective initial conditions, except for the $ \om_{k} \xi(t) $ term that originates from transforming the derivative on the left-hand side of \cref{eq:Standard_form_2} back to the Schr\"odinger picture and should, therefore, not be replaced. We denote the resulting two-point function in perturbation theory, obtained by using the solutions for $ \xi(t) $ and $ \xi_{3}(t) $ in \cref{eq:Gk}, with $ \CG_{k}^{\text{PT}}(t) $. Note that $ \CG_{k}^{\text{PT}}(t) $ is essentially the second-order loop correction to the two-point function.
%%%%%%%%%%%%%%%%%%%%%%%%%%%%%%%%%%%%%%%%%%%%%%%%%%

%--------------------------
\section{\texorpdfstring{$\lambda\phi\chi$}{lpc} interaction}
\label{sec:phichi}
%--------------------------

The simplest way to couple the system and environment fields is through a bilinear coupling. This corresponds to setting $ n = 1 $ in \cref{eq:Phi_Chi_Action,eq:Interaction_Hamiltonian}, so that the interaction Hamiltonian is given by
\begin{align}
	\hH_{\intac}
=	
	\la
	\int
	\dd^{3}
	x
	\hphi(\vx)
	\hchi(\vx)
	\,.
\label{eq:Interaction_Hamiltonian_phi_chi}
\end{align}
The resulting theory is {\it exactly} solvable since the full Hamiltonian remains quadratic. The $\lambda\phi\chi$ interaction thus provides an excellent benchmark for the master equation result under various approximations. The exact calculation is detailed in \Cref{app:exactsoln} and, in particular, the exact solution for the two-point function, denoted $\CG_{k}^{\text{exact}}(t)$, is shown in \cref{eq:Gkexact}. We note that the exact solution breaks down for a large enough $\lambda$ because the potential is unbounded from below and that the constraint on $\la$ can be avoided by instead choosing the interaction to be $ \frac{1}{2} \la \left( \phi - \chi \right)^{2} $. We continue to work with the $ \la \phi \chi $ interaction, however, as it remains well behaved in the perturbative regime that we are interested in here.

%------------------------------------
\begin{figure}[!t]
    \centering
    \includegraphics[width=\columnwidth]{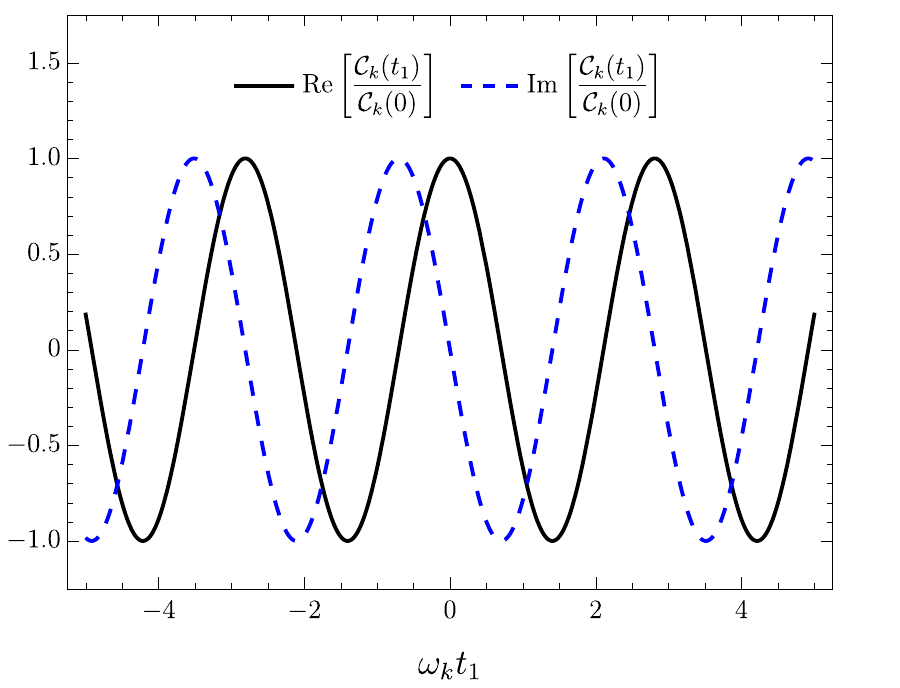} 
	\caption{The environment correlation as a function of elapsed time $ t_{1} $ in a $ \la \phi \chi $ interacting theory, computed for $ M = 3 m $ and $ k = m $, and normalized to unity at $ t_{1} = 0 $.}
\label{fig:phi_chi_env_corr} 
\end{figure}
%------------------------------------

We now discuss the master equation solution for the $ \la \phi \chi $ interaction. Consider first the correlation function $ \mathcal{C}_{k}(t_{1}) $, given by \cref{eq:ExplicitEnvCorr}, that appears in the $ \Gamk{\be}(t) $ coefficients. With $ n = 1 $, it becomes
\begin{align}
    \mathcal{C}_{k}(t_{1})
=
    \frac{
        e^{ - i \Om_{k} t_{1} } 
    }{
        2
        \Om_{k}
    }
    \,,
\label{eq:phi_chi_env_corr}
\end{align}
where evaluating the single $ \vp $ integral over the momentum-conserving delta function forces the system and environment momenta to be equal. The Redfield equation coefficients are now constructed by using the above expression for $ \mathcal{C}_{k}(t_{1}) $ in \cref{eq:Gamma_al_be}, using the resulting $ \Gamk{\be} $ in \cref{eq:gammaAlBe,eq:SAlBe}, and finally using the resulting $\gamk{\al \be}$ and $\Sk{\al \be}$ in the definitions of $ \gamd{\rm o} $, $ \Sd{\rm o} $, $ \gamd{\pm} $, and $ \omS $ after \cref{eq:Xi3_diff_eq}. We show the coefficients explicitly in \Cref{app:phichicoeff} for completeness.

It is clear from \cref{eq:phi_chi_env_corr} that the environment correlation $ \mathcal{C}_{k}(t_{1}) $ for this interaction is purely oscillatory, as shown in \cref{fig:phi_chi_env_corr} for clarity. Given our discussion in \cref{sec:MarkovianLimit}, we should thus not expect the Markovian limit to capture the dynamics in this case accurately. Another way to see that the dynamics here must be non-Markovian is to notice that each mode in the system is coupled to exactly one mode in the environment; thus, the composite system behaves like a collection of pairwise coupled oscillators. Consequently, there will be substantial memory in the evolution of the system density operator that should not be neglected. The Markovian limit can, nevertheless, be imposed by setting the upper limit of the time integral in $ \Gamk{\be}(t) $ to $ \infty $, yielding the time-independent function of $ k $,
\begin{align}
	\Gamk{\be}
&=
	\la^{2}
	\int_{0}^{\infty} \dd t_{1}
	\frac{
		e^{ - i (-1)^{\be} \om_{k} t_{1} }
	}{
		4 \om_{k} \Om_{k}
	}
	e^{ - i \Om_{k} t_{1} } 
	\,,
\label{eq:Gamma_be_phi_chi_Markovian}
\end{align}
which can be simplified by resolving the right-hand side into a sum of Fourier cosine and sine transforms.

%------------------------------------
\begin{figure}[!t] 
\centering
	\includegraphics[width=\columnwidth]{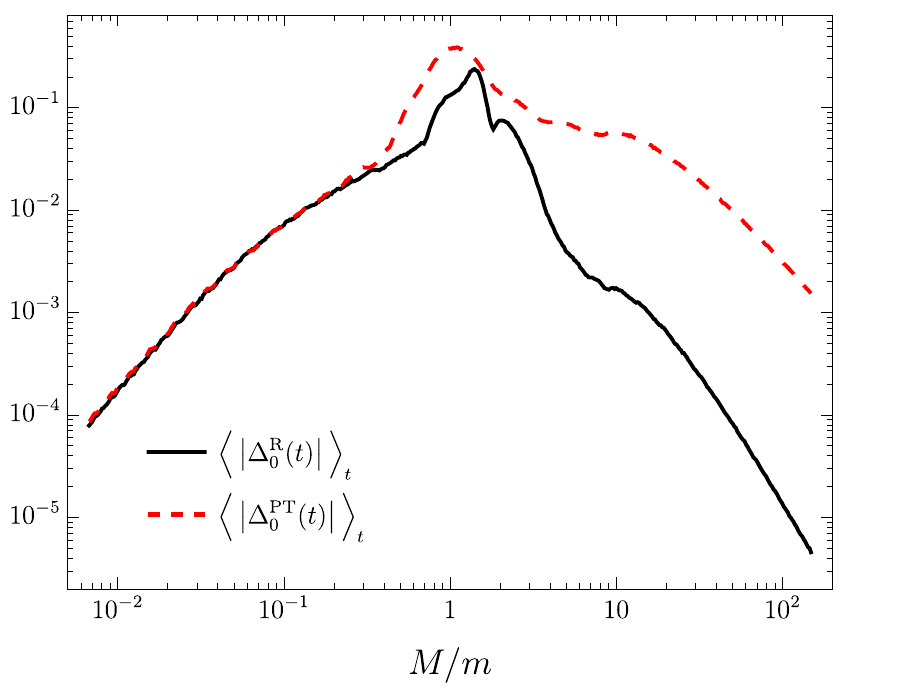} 
	\caption{The time-averaged absolute error in the two-point function calculated using the Redfield equation and perturbation theory as a function of $ M/m $ in a $ \la \phi \chi $ interacting theory, for $ \la = M m /2 $ and $ k = 0 $, and averaged over the time interval $ \sqrt{2 \la} t \in [0,10] $.}
\label{fig:MassComparison} 
\end{figure}
%------------------------------------

Although we do not expect the Markovian limit to be valid since the correlation function does not decay in time, the Markov approximation used in constructing our master equation can still hold for this interaction for a large environment mass. In \cref{sec:RedfieldME}, we concluded that the Markov approximation is justifiable for rapidly oscillating environment correlators so long as the characteristic timescale $ \tau_{\Sy} $ of the system is much longer than the period $ \tau_{\star}$ of the kernel $ \mathcal{K}_{k,\be} $. While $ \tau_{\Sy} \sim \om_{k}^{-1} $, we can obtain $\tau_{\star}$ by setting $ \be = 1 $ in \cref{eq:Kkbeta}, so as to obtain the longest relevant timescale for this interaction, which gives $ \tau_{\star} \sim \abs{\Om_{k} - \om_{k}}^{-1} $. We thus require that $ \abs{\Om_{k} - \om_{k}} \gg \om_{k} $. While this condition breaks down for $ k \to \infty $, the effect of the interaction is suppressed in this limit since $ \Gamk{\beta}(t) $ is itself suppressed. For $ k \to 0 $, on the other hand, this condition translates into $ \abs{M - m} \gg m $. We next show that the Redfield equation-based solution is more accurate than the perturbation theory-based solution for $M \gg m$, suggesting that the Markov approximation is indeed a good approximation in this limit.

To distinguish the validity of the Markov approximation from that of perturbation theory, we consider the relative error of the solution obtained using the Redfield equation, 
\begin{align}
	\Delta_{k}^{\text{R}}(t)
=
	\frac{
		\CG_{k}^{\text{R}}(t)
		-
		\CG_{k}^{\text{exact}}(t)
	}{
		\CG_{k}^{\text{exact}}(t)
	}
	\,,
\label{eq:DeltaR}
\end{align}
and that obtained using Dyson series perturbation theory, $\Delta_{k}^{\text{PT}}(t)$. We solve for $\Delta_{k}^{\text{R}}(t)$ and $\Delta_{k}^{\text{PT}}(t)$ numerically, using \cref{eq:Gkexact} for $\CG_{k}^{\text{exact}}(t)$. In \cref{fig:MassComparison}, we plot the time-averaged zero-mode absolute errors $ \langle |\Delta_{0}^{\text{R}}(t)| \rangle_{t} $ and $ \langle |\Delta_{0}^{\text{PT}}(t)| \rangle_{t} $ as a function of $ M/m $, upon fixing the interaction strength, $ \la = M m / 2 = \text{constant} $~\footnote{We never have to explicitly choose a value for $ M m $ since the explicit dependence on $ M $ and $ m $ appears as their ratio for this choice of parameters.}, and averaging over the time interval $ \sqrt{2 \la} t \in [0,10] $. As expected, the absolute error decreases as the mass scales separate for the Redfield and perturbation theory solutions due to a corresponding decrease in $\Gamk{\be}$. For $M > m$, however, the error in the Redfield solution decreases more rapidly than in perturbation theory, indicating that the Redfield solution offers an improvement over perturbation theory in this regime due to the improving validity of the Markov approximation.

%------------------------------------
\begin{figure}[!t]
    \centering
    \includegraphics[width=\columnwidth]{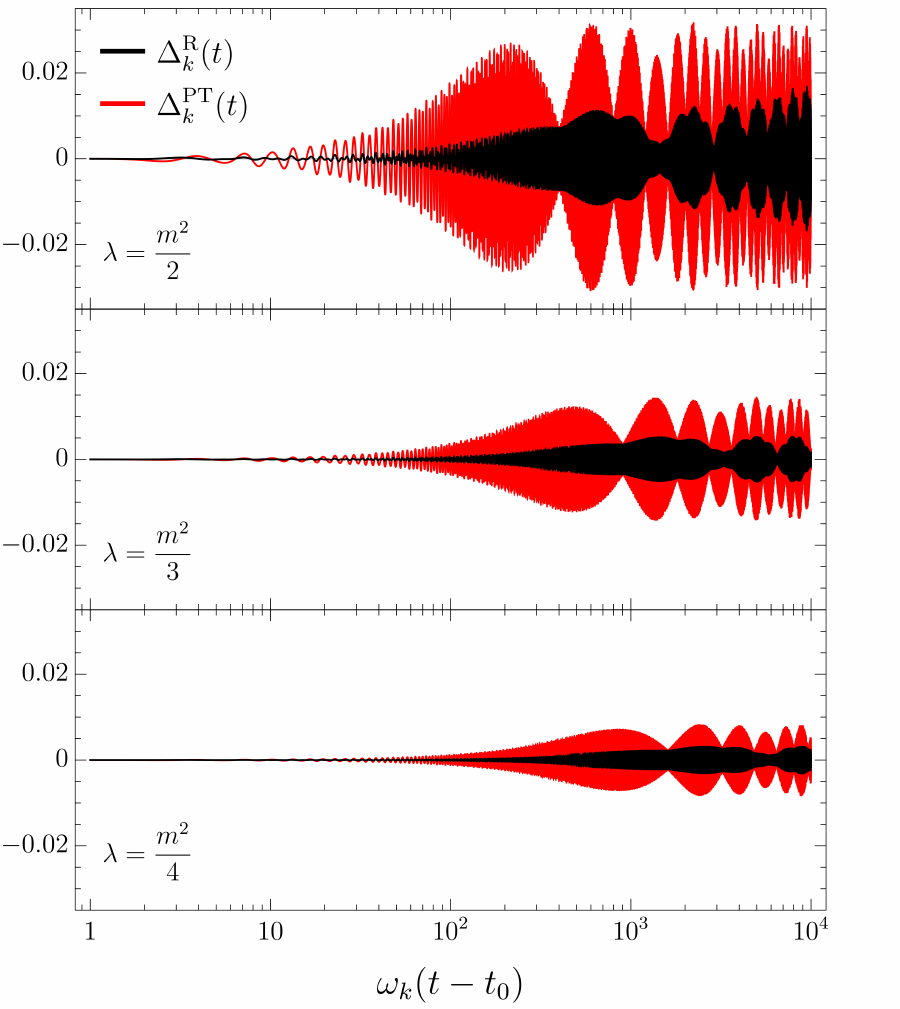} 
    \caption{The relative error in the two-point function calculated using the Redfield equation and perturbation theory as a function of time in a $ \la \phi \chi $ interacting theory, for three relatively large values of $ \la $, $ k = m $, and $ M = 3 m $. Expectedly, the error is suppressed for smaller values of $ \la $.}
 \label{fig:VaryTheCoupling} 
\end{figure}
%------------------------------------

In \cref{fig:VaryTheCoupling}, we compare the relative errors of the two solutions further for a given $M > m$ and for different choices of interaction strength $\la$. We find that the relative error is smaller for the Redfield solution and, while both solutions break down for large interaction strength and at late times as expected, the breakdown of the Redfield solution is {\it slower} than the perturbation theory one. This suggests that solving the coupled differential equations (\ref{eq:Xi_diff_eq}) and (\ref{eq:Xi3_diff_eq}) obtained from the Redfield equation provides a perturbative resummation that offers an improvement over the Dyson series perturbation theory result or, equivalently, the second-order loop correction. Understanding the set of diagrams that the Redfield solution resums|for example, whether it resums all one-particle irreducible (1PI) diagrams at second order in $\la$ or other diagrams|would be an interesting direction to pursue in future work. 

\begin{figure}[ht]
    \centering
    \includegraphics[width=\columnwidth]{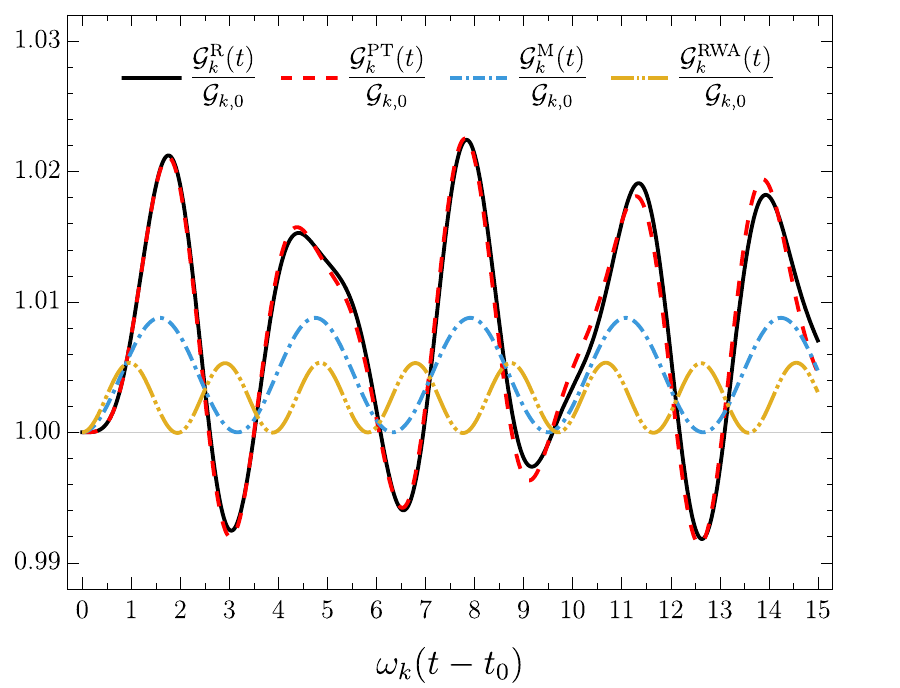}
    \caption{The two-point function calculated using the Redfield equation, perturbation theory, the Redfield equation in the Markovian limit, and the Redfield equation under the RWA as a function of time in a $ \la \phi \chi $ interacting theory, for $ \la = m^{2}/2 $, $ k = m $, and $ M = 3 m $, and normalized by the free theory solution, $ \CG_{k,0} = (2 \om_{k})^{-1} $.}
\label{fig:phi_chi_Approximations} 
\end{figure}

Finally, we compare the Redfield and perturbation theory solutions to those obtained in the Markovian limit and under the RWA in \cref{fig:phi_chi_Approximations}. As expected, neither approximation coincides with the Redfield dynamics, which capture the exact dynamics on these timescales, in a meaningful way. As mentioned earlier, we do not expect the Markovian limit to be valid since the environment correlation is purely oscillatory, and we do not expect the RWA to be a good approximation for systems with a continuous spectrum.

%------------------------------------
%------------------------------------

%%%%%%%%%%%%%%%%%%%%%%%%%%%%%%%%%%%%%%%%%%%%%%%%%%

%--------------------------
\section{\texorpdfstring{$\lambda\phi\chi^2$}{lpc2} interaction}
\label{sec:phichi2}
%--------------------------

We next couple the system and environment fields nonlinearly, specifically setting $ n = 2 $ in \cref{eq:Phi_Chi_Action}, so that the interaction is given by $ \la\phi\chi^{2} $. Note that we now need to renormalize the theory since loop corrections to, for example, $\hphi$ correlations will contain UV divergent (environment) momentum integrals.
We, therefore, introduce counterterms to the Lagrangian, which we use to absorb the divergences in the system equal-time two-point correlation function for this interaction. To introduce counterterms, we return to \cref{eq:Phi_Chi_Action} with $ n = 2 $ and follow standard renormalization techniques~\cite{Peskin:1995ev,Srednicki:2007qs}|we first shift and rescale both fields as $ \phi \to Z_{\phi}^{1/2} (\phi - f_{\phi}) $ and $ \chi \to Z_{\chi}^{1/2} (\chi - f_{\chi}) $ to correctly define the one-point function and one-particle state, and then redefine $ m $, $ M $, and $ \lambda $ to recover the original Lagrangian density plus the following counterterm Lagrangian density,
\begin{align}
	\mathcal{L}_{\ct}
&=
    -
    \frac{
        1
    }{
        2
    }
    \delta_{\phi}
    (\partial \phi)^{2}
    -
    \frac{
        1
    }{
        2
    }
    \delta_{m}
    \phi^{2}
    +
    Y_{\phi}
    \phi
\nn
&\hspace{11.5pt}
    -
    \frac{
        1
    }{
        2
    }
    \delta_{\chi}
    (\partial \chi)^{2}
    -
    \frac{
        1
    }{
        2
    }
    \delta_{M}
    \chi^{2}
    +
    Y_{\chi}
    \chi
    -
    \frac{1}{2}
    \delta_{\lambda}
    \phi
    \chi^{2}
	\,,
\label{eq:Lct}
\end{align}
where $\delta_{\phi}$, $\delta_{\chi}$, $\delta_m$, $\delta_M$, $Y_{\phi}$, $Y_{\chi}$, and $\delta_{\lambda}$ are the usual counterterms required to cancel any UV divergences that depend on $ Z_{\phi} $, $ Z_{\chi} $, $ f_{\phi} $, $ f_{\chi} $, and the original (bare) Lagrangian parameters $ m $, $ M $, and $ \lambda $. Note that, in general, we may need to add counterterms in the initial state as well~\cite{Calzetta:2008,Baacke:1997zz,Baacke:1999ia,Collins:2005nu,Collins:2014qna,Chaykov:2022pwd}, but they are not needed for the interaction that we consider here.

In fact, we can set all $\chi$ counterterms in \cref{eq:Lct} to zero: $\delta_{\chi}$ and $\delta_M$ can be set to zero since environment correlation functions are calculated at tree level and $Y_{\chi}$ can be set to zero since the interaction is quadratic in $\chi$ and does not generate a one-point expectation value. Additionally, since this interaction is linear in $ \phi $, the divergence of $ \langle \hphi(\vx,t) \rangle $ is at first order and results from the contraction of the two factors of $ \hchi $ in the interaction Hamiltonian. Therefore, for this interaction and choice of environment state, we can drop the $ Y_{\phi} $ counterterm as well and, equivalently, normal order the environment operator in the interaction Hamiltonian~\cite{Boyanovsky:2015xoa}. This, in fact, sets $ \hH_{\eff}(t) = 0 $, as discussed after \cref{eq:FormalSolutionForRho}, and is also consistent with dropping equal-time bubble diagrams in the environment as discussed after \cref{eq:ExplicitEnvCorr}. 
Finally, since the vertex correction does not contribute at $ O(\la^{2}) $~\cite{Srednicki:2007qs}, we ignore the $\delta_{\lambda}$ term. With these simplifications, the quantized interaction Hamiltonian that we consider in this section becomes
\begin{align}
	\hH_{\intac}
&=
	\int \dd^{3}x
	\bigg[
		\frac{\delta_{\phi}}{2}
        \left(
            \hpi_{\phi}^{2}(\vx)
            +
            (\vec{\nabla} \hphi(\vx))^{2}
        \right)
		+
		\frac{\delta_{m}}{2}
		\hphi^{2}(\vx)
\nn
&\hspace{46pt}
        +
        \frac{
            \la
        }{
            2
        }
        \hphi(\vx)
        :\hchi^{2}(\vx):
    \bigg]
    \,.
\label{eq:Hctphichi2}
\end{align}
We now write the counterterm contribution in terms of the dimensionless eigenoperators of the system introduced in \cref{sec:setup}. This leads to a correction in the system's unitary dynamics, such that $\Sk{\al \be}(t)$ in \cref{eq:SAlBe} becomes
\begin{align}
	\Sk{\al \be}(t)
&\equiv
    \frac{
        1
    }{
        2
        i
    }
    \left[
        \Gamk{\be}(t)
        -
        \Gamk{\al}\conj(t)
    \right]
\nn
&\hspace{11.5pt}
    +
    \frac{
        1
    }{
        4 
        \om_{k}
    }
    \left[
        \delta_{\phi}
        \left(
           k^{2}
           +
           (-1)^{\al + \be}
           \om_{k}^{2}
        \right)
         +
         \delta_{m}
    \right]
	\,,
\end{align}
which enters the modified system Hamiltonian in \cref{eq:ShiftHamiltonian}, which in turn enters the master equation in \cref{eq:Standard_form}.

Let us now consider the environment correlation function for $ \la \phi \chi^{2} $. For $ n = 2 $, \cref{eq:ExplicitEnvCorr} retains one momentum integral, with the momentum-conserving delta function restricting one of the environment modes to be the sum of the other two modes in the interaction, hence
\begin{align}
    \mathcal{C}_{k}(t_{1})
&=
	\frac{
        1
	}{
		8
	}
	\int\frac{
		\d^{3} p
	}{
		(2 \pi)^{3}
	}
	\frac{
        e^{ 
            - 
            i 
            \left(
                \Om_{p} 
                +
                \Om_{\abs*{\vk + \vp}} 
            \right)
            t_{1} 
       } 
	}{
        \Om_{p}
        \Om_{\abs*{\vk + \vp}}
	}
	\,.
\label{eq:EnvCorr_phi_chi2}
\end{align}
We set $ M=0 $ to simplify the integral over $ \vp $ and find, as usual, that it is UV divergent. On regulating the integral by introducing a small imaginary piece to the time parameter~\footnote{We note that one could obtain the same final results using the dimensional regularization procedure of ref.~\cite{Chaykov:2022pwd}. We have instead used the $i\epsilon$ regularization scheme here as it is more instructive since the delta function piece of the environment correlation function emerges naturally.} such that $ t_{1} \to t_{1} - i \epsilon $ with $ \epsilon \in \R^{+} $, we obtain the $ \epsilon $-dependent environment correlation function,
\begin{align}
    \mathcal{C}_{k}(t_{1})
&=
    -
    \frac{
        i
    }{
        32
        \pi^{2}
    }
    \frac{
        e^{
            -
            i
            k
            t_{1}
        }
    }{
        t_{1}
        -
        i
        \epsilon
    }
\label{eq:PhiChi2EnvCorrIntegrated}
 \\
 &=
    -
    \frac{
        1
    }{
        32
        \pi
    }
    \delta(t_{1})
    -
    \frac{
        i
        e^{
            -
            i
            k
            t_{1}
        }
    }{
        32
        \pi^{2}
    }
    \mathcal{P}
    \left(
        \frac{
            1
        }{
            t_{1}
        }
    \right)
    \,,
\label{eq:PhiChi2EnvCorrIntegrateddelta}
\end{align}
where we have used the identity $ \lim_{\ep \to 0} (x \pm i \ep)^{-1} = \mp i \pi \delta(x) + \mathcal{P} (1/x) $. Although $\mathcal{C}_{k}(t_{1})$ does not diverge as $\epsilon \rightarrow 0$, the regularization scheme simply converts the momentum divergence to an initial time one, which becomes apparent when computing the principal value piece of the $ t_{1} $ integral. We can, nevertheless, infer a couple of important points from \cref{eq:EnvCorr_phi_chi2,eq:PhiChi2EnvCorrIntegrated}, which we highlight before continuing with our renormalization procedure.

\begin{figure}[t!]
    \centering
    \includegraphics[width=\columnwidth]{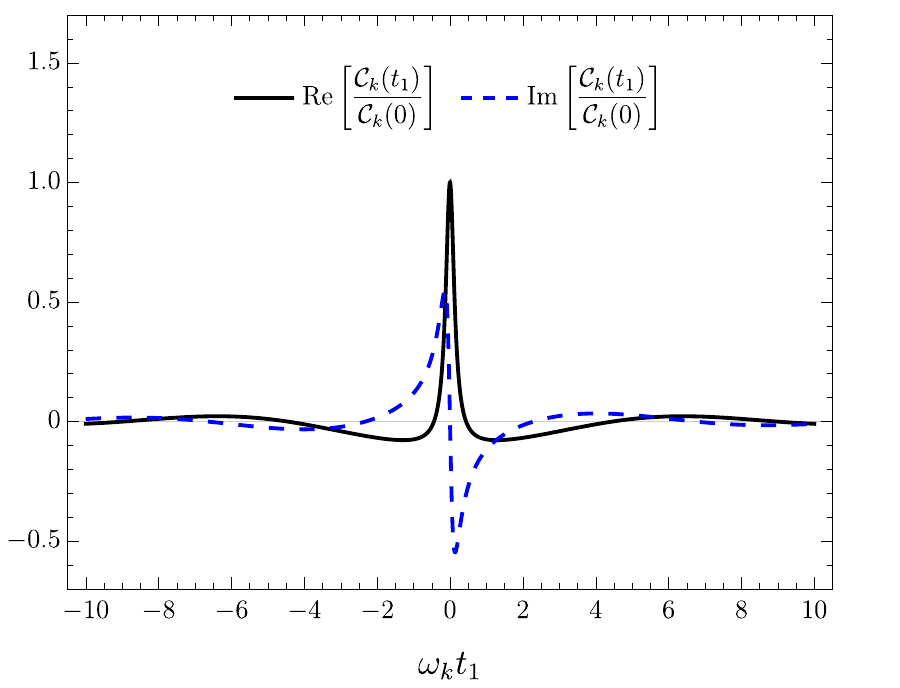}
	\caption{The environment correlation \cref{eq:PhiChi2EnvCorrIntegrated} as a function of elapsed time $ t_{1} $ in a $ \la \phi \chi^{2} $ interacting theory, computed for $ \ep = 1/10 $ and $ k = m $, and normalized to unity at $ t_{1} = 0 $.}
\label{fig:phi_chi2_env_corr}  
\end{figure}

First, we note that the momentum integral in \cref{eq:EnvCorr_phi_chi2} is a direct result of the nonlinearity of the interaction. In contrast to the $ \la \phi \chi $ theory, where a given system mode only couples to a single environment mode, in the $ \la \phi \chi^2 $ theory, a given system mode couples to {\it all} environment modes. In this sense, the environment can be considered {\it large}, analogous to the famous (Markovian) Caldeira-Leggett model of quantum Brownian motion~\cite{Caldeira:1981,Caldeira:1982iu}. Second, from \cref{eq:PhiChi2EnvCorrIntegrated} and \cref{fig:phi_chi2_env_corr}, we see that, even with the environment mass set to zero, $ \mathcal{C}_{k}(t_{1}) $ decays rapidly away from $t_1 = 0$, suppressing memory in the system density operator. In fact, from \cref{eq:PhiChi2EnvCorrIntegrateddelta} we see that for the special case $ k = 0 $, the real part of the environment correlation function becomes exactly a delta function. More generally, for $ k \neq 0 $ this memory suppression implies that both the Markov approximation and the Markovian limit should be valid in the late-time limit and, in turn, that the Redfield equation should capture the late-time behavior of the system, which is what we find below.

%------------------------------------
%------------------------------------

Let us now return to the renormalization procedure. Using \cref{eq:PhiChi2EnvCorrIntegrated} [or, equivalently, \cref{eq:PhiChi2EnvCorrIntegrateddelta}] in \cref{eq:Gamma_al_be}, we find that the time integral leads to a $ \ln \ep $ divergence that persists in $ \Sk{\al \be}(t) $. We impose the condition that all instances of $ \ep $ vanish in the master equation, choosing the counterterms to be
\begin{align}
    \delta_{\phi}   
=
    0
    \,,
    \hspace{20pt}
    \delta_{m}   
=
	-
    \frac{\la^{2}}{16 \pi^{2}}
    \ln \left( \ep \mu_{\rm r} \right)
    ,
\label{eq:FinalCounterterms}
\end{align}
where we have introduced the renormalization scale $ \mu_{\rm r} $. We show the final coefficients that enter the master equation for this interaction and with the above choice of counterterms in \Cref{app:phichi2coeff}. We note that while the coefficients in \cref{eq:So_phichi2,eq:omS_phichi2} diverge at the initial time, since $\text{Ci}(x)$ diverges at zero, correlation functions must remain finite at the initial time. A simple argument for this is that the perturbation theory solution continuously matches the initial conditions at the initial time, and the Redfield solution reduces to the perturbation theory one at early times. Also, as expected, the coefficients depend on $ \mu_{\rm r} $, which is typically chosen in a fixed loop order calculation such that the result has reduced dependence on it at the energy scales of interest. We discuss some subtleties related to choosing $ \mu_{\rm r} $ in the next section and simply choose a reasonable value for it in the analysis that follows.

%------------------------------------
\begin{figure*}[t!]
    \includegraphics[width=0.95\columnwidth]{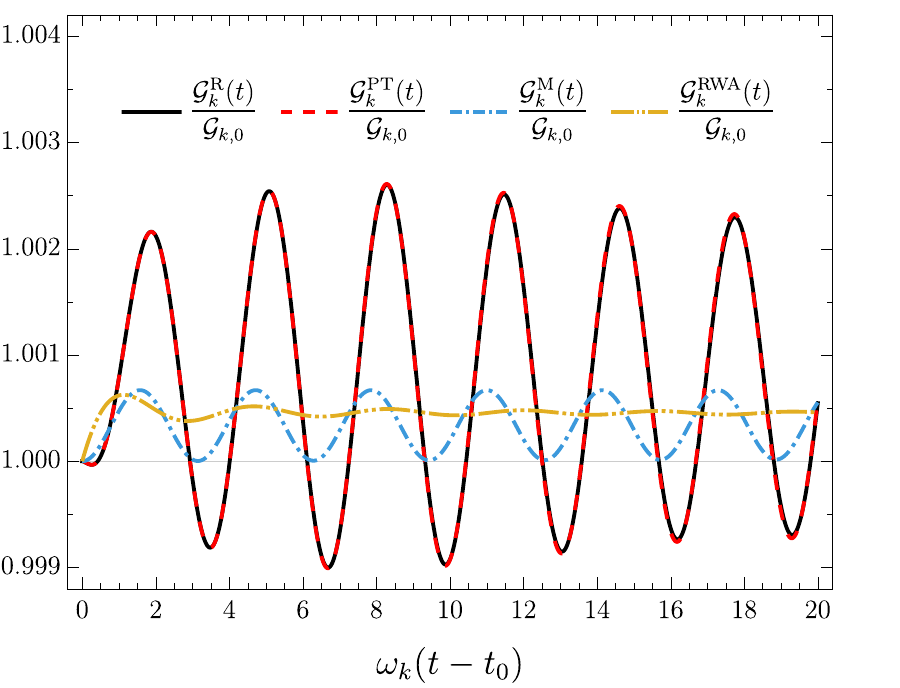} 
    \includegraphics[width=0.95\columnwidth]{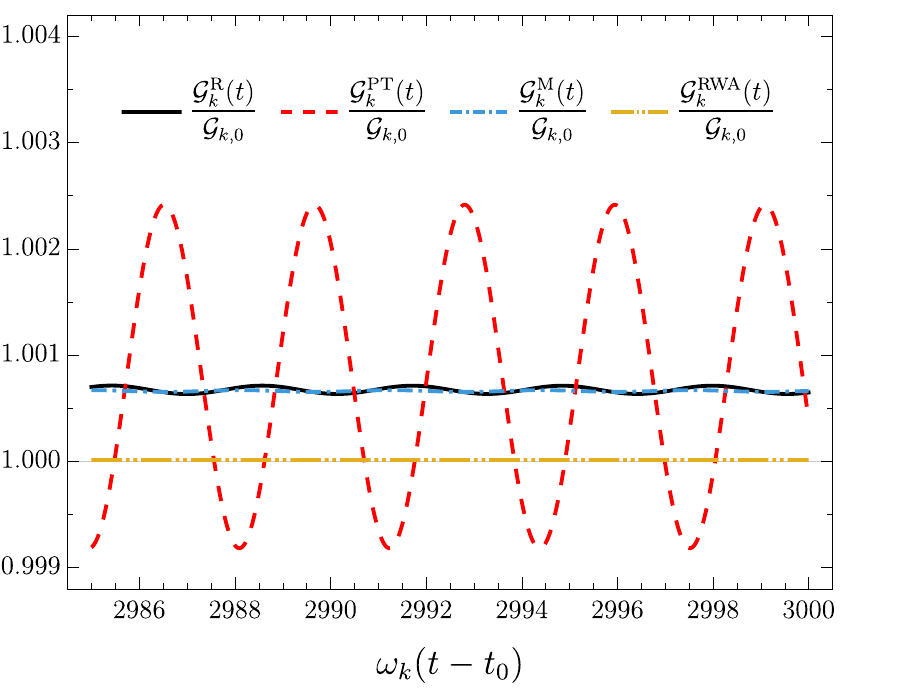}     
    \caption{The two-point function calculated using the Redfield equation, perturbation theory, the Redfield equation in the Markovian limit, and the Redfield equation under the RWA as a function of time in a $ \la \phi \chi^{2} $ interacting theory, for $ \la = m/2 $, $ k = m $, and $ \mu_{\rm r} = 10 m $, and normalized by the free theory solution, $ \CG_{k,0} = (2 \om_{k})^{-1} $. (Left) The two-point function at early times. (Right) The two-point function at sufficiently late times showing a relaxation of the Redfield solution as well as in the Markovian limit. We note that for the numerical Redfield solution, we impose initial conditions at $ \om_{k} (t-t_{0}) = 10^{-10} $ since some of the coefficients diverge as $ t \to t_{0} $.}
\label{fig:Results_phi_chi2}
\end{figure*}
%------------------------------------

We now compare the Redfield equation-based solution to the solutions obtained using perturbation theory, in the Markovian limit, and under the RWA in \cref{fig:Results_phi_chi2}. From the left panel of \cref{fig:Results_phi_chi2}, we see that the Redfield and perturbation theory solutions agree on short timescales. The Markovian limit and RWA, on the other hand, are both poor approximations at short timescales, although the solution in the Markovian limit retains more of the phase information. From the right panel of \cref{fig:Results_phi_chi2}, we see that the Redfield solution relaxes at late times, whereas the perturbation theory one continues to oscillate. This qualitative difference can be attributed to the Redfield equation representing a resummation of second-order dynamics. We further see that the Markovian limit is a good approximation at late times, since the solution relaxes to the same value as that obtained using the Redfield equation. This supports our expectation that the environment in the $\la \phi \chi^2$ theory is approximately Markovian. Finally, we see that the RWA is a poor approximation at late times as well, as the solution relaxes to the free theory one.

%%%%%%%%%%%%%%%%%%%%%%%%%%%%%%%%%%%%%%%%%%%%%%%%%%

%--------------------------
\section{Comments on the renormalization scale}
\label{sec:URU}
%--------------------------

We next show that the renormalization scale $ \mu_{\rm r} $, encountered in the previous section, can have a significant effect on the late-time solution obtained using the master equation, and certain choices of $ \mu_{\rm r} $ can lead to physically inconsistent results. For the $ \la \phi \chi^{2} $ theory, we can write an analytical expression for the late-time correlation in the Markovian limit where, as shown in the previous section, the solution agrees with the Redfield solution,
\begin{align}
    \CG_{k,0}^{-1}
    \CG_{k}^{\text{M}}(t \to \infty)
=
	\frac{
		\frac{
			\la^{2}
		}{
			16
			\pi^{2}
			\om_{k}^{2}
		}
		\ln
		\left(
			\frac{
				\om_{k}
				+
				k
			}{
				m
			}
		\right)
		-
		1
	}{
		\frac{
			\la^{2}
		}{
			16
			\pi^{2}
			\om_{k}^{2}
		}
		\ln
		\left(
			\frac{
				\mu_{\rm r}
			}{
				m
				e^{
					\gamma
				}
			}
		\right)
		-
		1
	}
	\,,
\end{align}
where $ \gamma $ is the Euler-Mascheroni constant. Using the Markovian limit solutions for $ \xi(t) $ and $ \xi_{3}(t) $, we can further construct the two-point function for the conjugate momentum and find that $ \lim_{t \to \infty} \langle \hpi_{\phi,\vk}(t) \hpi_{\phi,\vk\pr}(t) \rangle = \langle \hpi_{\phi,\vk}(t_{0}) \hpi_{\phi,\vk\pr}(t_{0}) \rangle $. The uncertainty principle in the late-time limit, therefore, reduces to
\begin{align}
    \lim_{t \to \infty}
    \Delta \phi_{\vk}(t)
    \Delta \pi_{\phi,\vk}(t)
=
    \frac{1}{2}
    \sqrt{
        \CG_{k,0}^{-1}
        \CG_{k}^{\text{M}}(t \to \infty)
    }
\geq
    \frac{1}{2}
\,.
\label{eq:UncertaintyRelation}
\end{align}
Imposing the inequality in \cref{eq:UncertaintyRelation} then leads to the following condition on $ \mu_{\rm r} $ for the theory to be physical,
\begin{align}
	\mu_{\rm r} 
>
	\left( 
		\om_{k} 
		+ 
		k 
	\right) 
	e^{\gamma} 
	\,.
\label{eq:mu_inequality}
\end{align}
The $ k $ dependence of $ \mu_{\rm r} $ implied by the above equation is not implausible as we may need to choose $ \mu_{\rm r} $ in such a way to remove any dependence on it at a given energy scale $k$. Nevertheless, the breakdown of uncertainty for an {\it incorrect choice} of $ \mu_{\rm r} $ is not encountered in perturbation theory. In fact, the perturbation theory solution does not break the uncertainty principle for either $ \la \phi \chi $ or $ \la \phi \chi^{2} $ and, in particular, for any $ \mu_{\rm r} > 0 $ for the latter. This suggests that the breakdown of uncertainty here is \textit{not} a feature of renormalization in QFT but is rather inherited from the Redfield equation construction. Interestingly, whereas the pathology of the Redfield description is typically diagnosed via positivity violation of the dynamical state of the system, here we diagnose this by demanding physical bounds on the relevant field correlators. While we leave a detailed study of this issue and potential connections between the unphysical predictions to future work, we briefly examine the Hamiltonian corrections, or the $ \Sk{\al \be}(t) $ coefficients, in the master equations for both interactions, as any $ \mu_{\rm r} $ dependence only appears in these terms.

We first consider $ \Sk{\al \be}(t) $, which can be interpreted as a Lamb-type shift in the system frequency spectrum, with negative eigenvalues \cite{Caldeira:1981,Caldeira:1982iu}. For the $\la \phi \chi$ interaction and considering the simpler case of the Davies equation, the eigenvalues can be calculated exactly and are given by
\begin{align}
	\Sk{11}
&=
	-
	\frac{
		\la^{2}
	}{
		4
		\om_{k}
		\Om_{k}
		\left(
			\Om_{k}
			-
			\om_{k}
		\right)
	}
	\,,
\\
	\Sk{22}
&=
	-
	\frac{
		\la^{2}
	}{
		4
		\om_{k}
		\Om_{k}
		\left(
			\Om_{k}
			+
			\om_{k}
		\right)
	}
	\,.
\end{align}
While the second eigenvalue is always negative, the first one is negative only for $ \Om_{k} > \om_{k} $, which is consistent with the requirement of $M > m$ for the Markov approximation to be valid. Violating this mass hierarchy, therefore, leads to an unphysical frequency shift and can lead to a broken uncertainty principle. For the more general case of the Redfield equation, on the other hand, we find that one eigenvalue of $ \Sk{\al \be} $ is positive and the other is negative. For parameter values that preserve the uncertainty principle, however, we find that the {\it trace} of $ \Sk{\al \be} $ is negative, at least at late times. We also find that the uncertainty principle is preserved when {\it all} unitary corrections, in both $ \Sk{\al \be} $ and $ \gamk{\al \be} $, are ignored.

We next consider the $ \Sk{\al \be}(t) $ coefficients for the $\la \phi \chi^2$ interaction. For the Davies equation, its eigenvalues are given by
\begin{align}
	\Sk{11}
&=
	-
	\frac{
		\la^{2}
	}{
		64
		\pi^{2}
		\om_{k}
	}
	\ln 
	\left[
		\frac{
			\mu_{\rm r}
		}{
			\left(
				\om_{k}
				-
				k
			\right)
			e^{\gamma}
		}
	\right] ,
\\
	\Sk{22}
&=
	-
	\frac{
		\la^{2}
	}{
		64
		\pi^{2}
		\om_{k}
	}
	\ln 
	\left[
		\frac{
			\mu_{\rm r}
		}{
			\left(
				\om_{k}
				+
				k
			\right)
			e^{\gamma}
		}
	\right] .
\end{align}
Demanding that both eigenvalues are negative gives us two conditions, with the second one being more restrictive and exactly matching the condition obtained from the uncertainty principle in \cref{eq:mu_inequality}. For the Redfield equation, on the other hand, we find the same qualitative behavior as mentioned in the previous paragraph for $\la \phi \chi$. We note that dropping all unitary corrections in $ \Sk{\al \be} $ and $ \gamk{\al \be} $ would now also drop all instances of $ \mu_{\rm r} $. These observations suggest that the broken uncertainty principle originates from the Lamb-type frequency shift terms, which are known to be problematic in general \cite{Agarwal1974}. Our procedure to set the counterterm appropriately to rectify the Lamb shift-induced violation of uncertainty is reminiscent of recent results that have shown how, by making suitable modifications to the system Hamiltonian, {\it fitter} Redfield descriptions can be obtained both in terms of accuracy of steady state predictions and describing transient dynamics \cite{Timofeev:2022tbl}.

%--------------------------
\section{Discussion}
\label{sec:disc}
%--------------------------

In summary, we showed that the Redfield master equation provides a {\it perturbative resummation} to calculate the non-Markovian dynamics of system observables. To this end, we first derived the Redfield master equation for the reduced density operator of a system field $\phi$ by tracing out an environment field $\chi$ at second order in $\lambda$ and under the standard Born and Markov approximations. We then used the master equation to set up coupled differential equations in two-point correlations of the creation and annihilation operators of $ \hphi_{\vk}(t) $ and discussed how the coupled equations simplify in various limits, including the Markovian limit, RWA, and standard perturbation theory expansion. We finally solved the resulting coupled equations to obtain the resummed equal-time two-point function of $ \hphi_{\vk}(t) $ in any of these limits. We used this approach for both bilinear ($ \la \phi \chi $) and nonlinear ($ \la \phi \chi^2 $) system-environment interactions and found that the evolution of the system field is approximately Markovian in the latter case.

In the case of the bilinear interaction, we argued that the Markovian limit must be invalid since (i) the environment correlation function does not decay in time and (ii) a single system mode only couples to a single environment mode. However, we showed that the Markov approximation, and thus the Redfield master equation construction, remains valid as long as the $\chi$ field is much heavier than the $\phi$ field. We next used the resulting Redfield equation to solve for the two-point function analytically/numerically in various limits and showed that the Markovian limit indeed fails to capture the late-time behavior. Finally, we showed that the Redfield equation-based solution offers a substantial improvement over the standard perturbation theory-based one, suggesting that the Redfield equation provides a perturbative resummation for the calculation of observables.

In the case of the nonlinear interaction, an immediate consequence was that the environment correlation function became UV divergent. We found, however, that the master equation approach admits standard renormalization techniques, such that adding appropriate counterterms in the unitary dynamics allowed us to cancel all UV divergences, while introducing a renormalization scale $ \mu_{\rm r} $ in the master equation. We next argued that the Markovian limit must be valid for this interaction since (i) the (regularized) environment correlation function decays rapidly in time and (ii) a single system mode couples to all environment modes. Choosing an arbitrary value of $ \mu_{\rm r} = 10 m $, we then solved the resulting Redfield equation for the two-point function analytically/numerically in various limits and showed that the Markovian limit indeed captures the late-time behavior of the full Redfield solution. Finally, we showed that the perturbation theory solution significantly differs from the Redfield solution at late times, again suggesting that the Redfield equation provides a perturbative resummation for the calculation of observables.

Finally, we highlight a few open questions that we have left to future work. First, while our results suggested that the Redfield solution provides a perturbative resummation, it is unclear which diagrams it resums|for example, whether it resums all 1PI diagrams at second order in $\lambda$ or other diagrams as well. It would be interesting to understand this structure and contrast it with other diagrammatic approaches, such as those of Refs. \cite{Burrage:2018pyg,ganguly2023,Hosseinabadi2024,Khan:2024zwr}. Second, while we found the master equation to be compatible with standard counterterm renormalization, it is not clear whether this would always be true. For example, it would be interesting to explore a more general, possibly time-dependent \cite{Burrage:2018pyg}, renormalization procedure and understand whether it could also help address positivity violation in the Redfield construction through an appropriate renormalization condition. Third, for the $ \la \phi \chi^2 $ interaction, we found that resummed correlations can break the uncertainty principle at late times for certain values of $\mu_{\rm r}$. We argued that this may be an artifact of an unphysical Lamb-type correction in the master equation construction, but deserves further investigation.

%%%%%%%%%%%%%%%%%%%%%%%%%%%%%%%%%%%%%%%%%%%%%%%%%%

\acknowledgments{It is a pleasure to thank Unnati Akhouri, Spasen Chaykov, Yi-Zen Chu, Sachin Jain, Christian K\"ading, Andrew Keefe, Sean Prudhoe, Hugo Ribeiro, and Sarah Shandera for useful discussions. B.B. and N.A. were supported by the Department of Energy under Awards No. DE-SC0019515 and No. DE-SC0020360. A.K. was supported by the National Science Foundation under Award No. DMR-2047357.}

%%%%%%%%%%%%%%%%%%%%%%%%%%%%%%%%%%%%%%%%%%%%%%%%%%

\appendix

\begin{widetext}
%--------------------------
\section{Solving the coupled differential equations in the Markovian limit}
\label{app:MarkovianLimitSolution}
%--------------------------

In this appendix, we solve for the functions $\xi(t)$ and $\xi_3(t)$, defined in \cref{eq:aa_Correlation,eq:aadagg_Correlations}, in the Markovian limit. We start with the coupled equations (\ref{eq:Xi_diff_eq_Markovian}) and (\ref{eq:Xi3_diff_eq_Markovian}), which we write again for convenience,
\begin{align}
	\dot{\xi}(t)
&=
	-
	\left(
		\gamd{-}
		+
		i
		2
		\omS
	\right)
	\xi(t)
	-
	i
	2
	\Sd{\rm o}
	\xi_{3}(t)
	-
	\gamd{\rm o}
	\,, \\
	\dot{\xi}_{3}(t)
&=
	- 
	\gamd{-}
	\xi_{3}(t)
	+
	8
	\Im
	\left[
		\Sd{\rm o}
		\xi\conj(t)
	\right]
	+
	\gamd{+}
	\,,
\end{align}
where $\gamma_o$, $S_o$, $\gamma_{\pm}$, and $\omS$ are all time-independent. In terms of the real and imaginary parts of $ \xi = \xi_{1} + i \xi_{2} $ and the time parameter $ T = t-t_{0} $, these can be written as the three equations,
\begin{align}
	\frac{
		\dd \xi_{1}
	}{
		\dd T
	}
&=
	-
	\gamd{-}
	\xi_{1}
	+
	2
	\omS
	\xi_{2}
	+
	2
	\Im
	\left[
		\Sd{\rm o}
	\right]
	\xi_{3}
	-
	\Re
	\left[
		\gamd{\rm o}
	\right]
	\,,
\\
	\frac{
		\dd \xi_{2}
	}{
		\dd T
	}
&=
	-
	\gamd{-}
	\xi_{2}
	-
	2
	\omS
	\xi_{1}
	-
	2
	\Re
	\left[
		\Sd{\rm o}
	\right]
	\xi_{3}
	-
	\Im
	\left[
		\gamd{\rm o}
	\right]
	\,,
\\
	\frac{
		\dd \xi_{3}
	}{
		\dd T
	}
&=
	- 
	\gamd{-}
	\xi_{3}
	+
	8
	\Im
	\left[
		\Sd{\rm o}
	\right]
	\xi_{1}
	-
	8
	\Re
	\left[
		\Sd{\rm o}
	\right]
	\xi_{2}
	+
	\gamd{+}
	\,.
\end{align}
We now simplify the above equations by making the change of variables $ \xi_{i}(T) = \eta_{i}(T) e^{- \gamd{-} T} $, which gives
\begin{align}
	\frac{
		\dd \eta_{1}
	}{
		\dd T
	}
	-
	2
	\omS
	\eta_{2}
	-
	2
	\Im
	\left[
		\Sd{\rm o}
	\right]
	\eta_{3}
&=
	-
	\Re
	\left[
		\gamd{\rm o}
	\right]
	e^{
		\gamd{-}
		T
	}
	\,,
\\
	2
	\omS
	\eta_{1}
	+
	\frac{
		\dd \eta_{2}
	}{
		\dd T
	}
	+
	2
	\Re
	\left[
		\Sd{\rm o}
	\right]
	\eta_{3}
&=
	-
	\Im
	\left[
		\gamd{\rm o}
	\right]
	e^{
		\gamd{-}
		T
	}
	\,,
\\
	-
	8
	\Im
	\left[
		\Sd{\rm o}
	\right]
	\eta_{1}
	+
	8
	\Re
	\left[
		\Sd{\rm o}
	\right]
	\eta_{2}
	+
	\frac{
		\dd \eta_{3}
	}{
		\dd T
	}
&=
	\gamd{+}
	e^{
		\gamd{-}
		T
	}
	\,.
\end{align}
We next take the Laplace transform to move from the $ T $ domain to the $ \varepsilon $ domain, making use of the property
\begin{align}
    \mathcal{L}_{T}
    \left\{
        \frac{\dd}{\dd T}
        f(T)
    \right\}
=
    \int_{0}^{\infty}
    \dd T
    \frac{\dd}{\dd T}
    f(T)
    e^{
        -
        \varepsilon
        T
    }
=
    \lim_{b \to \infty}
    \left[
        f(T)
        e^{
            -
            \varepsilon
            T
        }
    \right]_{0}^{b}
    +
    \varepsilon
    \int_{0}^{\infty}
    \dd T
    \left[
        f(T)
        e^{
            -
            \varepsilon
            T
        }
    \right]
=
    \varepsilon
    f(\varepsilon)
    -
    f(0)
    \,.
\end{align}
Then, the resulting equations can be written in matrix form as
\begin{align}
	\begin{bmatrix}
		\varepsilon
	&&
		-
		2
		\omS
	&&
		-
		2
		\Im
		\left[
			\Sd{\rm o}
		\right]
	\\
		2
		\omS
	&&
		\varepsilon
	&&
		2
		\Re
		\left[
			\Sd{\rm o}
		\right]
	\\
		-
		8
		\Im
		\left[
			\Sd{\rm o}
		\right]
	&&
		8
		\Re
		\left[
			\Sd{\rm o}
		\right]
	&&
		\varepsilon
	\end{bmatrix}
	\begin{bmatrix}
		\eta_{1}(\varepsilon)
		\\
		\eta_{2}(\varepsilon)
		\\
		\eta_{3}(\varepsilon)
	\end{bmatrix}
=
	\begin{bmatrix}
	\eta_{1}(0)
	-
	\frac{
		\Re
		\left[
			\gamd{\rm o}
		\right]
	}{
		\varepsilon
		-
		\gamd{-}
	}
	\\
	\eta_{2}(0)
	-
	\frac{
		\Im
		\left[
			\gamd{\rm o}
		\right]
	}{
		\varepsilon
		-
		\gamd{-}
	}
	\\
	\eta_{3}(0)
	+
	\frac{
		\gamd{+}
	}{
		\varepsilon
		-
		\gamd{-}
	}
	\end{bmatrix}
	\,,
\end{align}
and solutions for $ \xi_{i}(t) $ are then finally given by
\begin{align}
	\begin{bmatrix}
		\xi_{1}(t)
		\\
		\xi_{2}(t)
		\\
		\xi_{3}(t)
	\end{bmatrix}
=
	\mathcal{L}_{t - t_{0}}^{-1}
	\left\{
		\begin{bmatrix}
			\varepsilon
		&&
			-
			2
			\omS
		&&
			-
			2
			\Im
			\left[
				\Sd{\rm o}
			\right]
		\\
			2
			\omS
		&&
			\varepsilon
		&&
			2
			\Re
			\left[
				\Sd{\rm o}
			\right]
		\\
			-
			8
			\Im
			\left[
				\Sd{\rm o}
			\right]
		&&
			8
			\Re
			\left[
				\Sd{\rm o}
			\right]
		&&
			\varepsilon
		\end{bmatrix}^{-1}
		\begin{bmatrix}
		\eta_{1}(0)
		-
		\frac{
			\Re
			\left[
				\gamd{\rm o}
			\right]
		}{
			\varepsilon
			-
			\gamd{-}
		}
		\\
		\eta_{2}(0)
		-
		\frac{
			\Im
			\left[
				\gamd{\rm o}
			\right]
		}{
			\varepsilon
			-
			\gamd{-}
		}
		\\
		\eta_{3}(0)
		+
		\frac{
			\gamd{+}
		}{
			\varepsilon
			-
			\gamd{-}
		}
		\end{bmatrix}
	\right\}
	e^{
		-
		\gamd{-}
		\left(
			t
			-
			t_{0}
		\right)
	}
	\,,
\end{align}
where $ \mathcal{L}_{t-t_{0}}^{-1} $ denotes the inverse Laplace transform to the $ t - t_{0} $ domain. 

%--------------------------
\section{Exact solution for the equal-time two-point function in the \texorpdfstring{$ \la \phi \chi $}{lpc2} theory}
\label{app:exactsoln}
%--------------------------

In this appendix, we find the exact solution for the $ \la \phi \chi $ interacting theory. We start with the Heisenberg equations of motion for the system and environment annihilation operators, $\ha{\vk}$ and $\hb{\vk}$,
\begin{align}
	\frac{
		\dd 
	}{
		\dd t
	}
	\ha{\vk}(t)
&=
	-
	i
	\om_{k}
	\ha{\vk}(t)
	-
	i
	\frac{
		\la
	}{
		2 \sqrt{ \om_{k} \Om_{k} }
	}
	\left[
		\hb{\vk}(t) + \hbdagg{-\vk}(t)
	\right] ,
\label{eq:HEfora}
\\
	\frac{ \dd  }{ \dd t }
	\hb{\vk}(t)
&=
	-
	i
	\Om_{k}
	\hb{\vk}(t)
	-
	i
	\frac{
		\la
	}{
		2
		\sqrt{ \om_{k} \Om_{k} }
	}
	\left[
		\ha{\vk}(t) + \hadagg{-\vk}(t)
	\right] ,
\label{eq:HEforb}
\end{align}
and assume a general solution of the form
\begin{align}
	\ha{\vk}(t)
&=
	f_{0}(t) \ha{\vk} + f\conj_{1}(t) \hadagg{-\vk} + f_{2}(t) \hb{\vk} + f\conj_{3}(t) \hbdagg{-\vk}
	\,, \\
	\hb{\vk}(t)
&=
	h_{0}(t) \hb{\vk} + h\conj_{1}(t) \hbdagg{-\vk} + h_{2}(t) \ha{\vk} + h\conj_{3}(t) \hadagg{-\vk}
	\,,
\end{align}
where operators on the right-hand side are Schr\"odinger picture operators and the functions $ f_{i} $ and $ h_{i} $ are Bogoliubov coefficients that are yet to be determined. Substituting these equations into \cref{eq:HEfora,eq:HEforb} yields the set of coupled differential equations
\begin{align}
	\dot{f}_{0}(t)
	+
	i
	\om_{k}
	f_{0}(t)
	+
	i
	\frac{
		\la
	}{
		2 \sqrt{ \om_{k} \Om_{k} }
	}
	\left[
		h_{2}(t) + h_{3}(t)
	\right]
&=
	0 \, ,
\label{eq:f0_equation}
\\
	\dot{f}_{1}(t)
	-
	i
	\om_{k}
	f_{1}(t)
	-
	i
	\frac{
		\la
	}{
		2 \sqrt{ \om_{k} \Om_{k} }
	}
	\left[
		h_{2}(t) + h_{3}(t)
	\right]
&=
	0 \, ,
\label{eq:f1_equation}
\\
	\dot{h}_{2}(t)
	+
	i
	\Om_{k}
	h_{2}(t)
	+
	i
	\frac{
		\la
	}{
		2 \sqrt{ \om_{k} \Om_{k} }
	}
	\left[
		f_{0}(t) + f_{1}(t)
	\right]
&=
	0 \, ,
\label{eq:h2_equation}
\\
	\dot{h}_{3}(t)
	-	
	i
	\Om_{k}
	h_{3}(t)
	-
	i
	\frac{
		\la
	}{
		2 \sqrt{ \om_{k} \Om_{k} }
	}
	\left[
		f_{0}(t) + f_{1}(t)
	\right]
&=
	0 \, ,
\label{eq:h3_equation}
\end{align}
and a second set that can be obtained by making the two exchanges $ f_{i} \leftrightarrows h_{i} $ and $ \om_{k} \leftrightarrows \Om_{k} $ above. Equations (\ref{eq:f0_equation})-(\ref{eq:h3_equation}) can now be solved exactly using the same Laplace transform technique as in \Cref{app:MarkovianLimitSolution}. The equal-time two-point correlator for the system can then be written in terms of the functions $ f_{i} $ as 
\begin{align}
	\CG_{k}^{\text{exact}}(t)
&=
	\frac{
		1
	}{
		2 \om_{k}
	}
	\Big[
		\abs{f_{0}(t)}^{2}
		+
		\abs{f_{1}(t)}^{2}
		+
		\abs{f_{2}(t)}^{2}
		+	
		\abs{f_{3}(t)}^{2}
		+
		f_{0}(t)
		f\conj_{1}(t)
		+
		f\conj_{0}(t)
		f_{1}(t)
		+	
		f_{2}(t)
		f\conj_{3}(t)
		+	
		f\conj_{2}(t)
		f_{3}(t)
	\Big]
\end{align}
and is given by
\begin{align}
	\CG_{k}^{\text{exact}}(t)
&=
	\frac{
		1
	}{
		2 
		\om_{k}
	}
	\bigg\{
		1
		+
		\frac{
			\la^{2}
			\left(
				\Om_{k}
				-
				\om_{k}
			\right)
		}{
			2
			\Om_{k}
			\left( 
				\om_{+}^{2} 
				-
				\om_{-}^{2} 
			\right)^{2}
		}
		\bigg[
			\frac{
				2
				\la^{2}
				\left(
					3
					\Om_{k}
					-
					\om_{k}
				\right)
				+
				\Om_{k}^{2}
				\left(
					\Om_{k}
					-
					\om_{k}
				\right)^{2}
				\left(
					\Om_{k}
					+
					3
					\om_{k}
				\right)
			}{
				\left(
					\Om_{k}^{2}
					\om_{k}^{2}
					-
					\la^{2}
				\right)
				\left(
					\Om_{k}
					-
					\om_{k}
				\right)
			}
	\nn
	&\hspace{55.5pt}
			+
			\frac{
				\Om_{k}
				+
				\om_{k}
			}{
				\Om_{k}
				-
				\om_{k}
			}
			\left(
				\frac{
					\om_{+}^{2} 
					-
					\Om_{k}^{2}
				}{
					\om_{+}^{2} 
				}
				\cos
				\left[
					2
					\om_{+}
					\left(
						t
						-
						t_{0}
					\right)
				\right]
				-
				\frac{
					\om_{+}^{2} 
					-
					\om_{k}^{2}
				}{
					\om_{-}^{2} 
				}
				\cos
				\left[
					2
					\om_{-}
					\left(
						t
						-
						t_{0}
					\right)
				\right]
			\right)
	\nn
	&\hspace{55.5pt}
			+
			2
			\left(
				1
				+
				\frac{
					\Om_{k}
					\om_{k}
				}{
					\om_{+}
					\om_{-}
				}
			\right)
			\cos
			\left[
				\left(
					\om_{+}
					+
					\om_{-}
				\right)
				\left(
					t
					-
					t_{0}
				\right)
			\right]
			+
			2
			\left(
				1
				-
				\frac{
					\Om_{k}
					\om_{k}
				}{
					\om_{+}
					\om_{-}
				}
			\right)
			\cos
			\left[
				\left(
					\om_{+}
					-
					\om_{-}
				\right)
				\left(
					t
					-
					t_{0}
				\right)
			\right]
		\bigg]
	\bigg\}
	\,,
\label{eq:Gkexact}
\end{align}
where we have defined the frequencies
\begin{align}
	\om_{\pm}
&=
	\sqrt{
		\frac{
			\Om_{k}^{2}
			+
			\om_{k}^{2}
		}{
			2
		}
		\pm
		\sqrt{
			\la^{2}
			+
			\left(
				\frac{
					\Om_{k}^{2}
					-
					\om_{k}^{2} 
				}{
					2
				}
			\right)^{2}
		}
	}
	\,.
\end{align}
We note that the form of the frequencies $\om_{\pm}$ above suggests that the perturbative regime for this theory is characterized by the condition $ |\la| \ll |\Om_{k}^{2} - \om_{k}^{2}| = |M^{2} - m^{2}| $. Perturbation theory must, therefore, not be a good approximation when $m$ and $M$ are comparable, consistent with \cref{fig:MassComparison}.

%--------------------------
\section{Redfield equation coefficients for the \texorpdfstring{$ \la \phi \chi $}{lpc2} theory}
\label{app:phichicoeff}
%--------------------------

In this appendix, we display the coefficients of the Redfield equation for the $ \la \phi \chi $ interaction described in \cref{sec:phichi},
\begin{align}
    \gamd{\rm o}(t)
&=  
    \frac{
        i
        \la^{2}
    }{
        2
        \om_{k}
        \Om_{k}
        \left(
            \Om_{k}^{2}
            -
            \om_{k}^{2}
        \right)
    }
    \left[
        \om_{k}
        -
        \Big(
            \om_{k}
            \cos
            \left[
                \Om_{k}
                \left(
                    t
                    -
                    t_{0}
                \right)
            \right]
            +
            i
            \Om_{k}
            \sin
            \left[
                \Om_{k}
                \left(
                    t
                    -
                    t_{0}
                \right)
            \right]
        \Big)
        e^{
            -
            i
            \om_{k}
            \left(
                t
                -
                t_{0}
            \right)
        }
    \right]
    ,
\\
    \Sd{\rm o}(t)
&=
    -
    \frac{
        \la^{2}
    }{
        4
        \om_{k}
        \Om_{k}
        \left(
            \Om_{k}^{2}
            -
            \om_{k}^{2}
        \right)
    }
    \left[
        \Om_{k}
        -
        \Big(
            \Om_{k}
            \cos
            \left[
                \Om_{k}
                \left(
                    t
                    -
                    t_{0}
                \right)
            \right]
            +
            i
            \om_{k}
            \sin
            \left[
                \Om_{k}
                \left(
                    t
                    -
                    t_{0}
                \right)
            \right]
        \Big)
        e^{
            -
            i
            \om_{k}
            \left(
                t
                -
                t_{0}
            \right)
        }
    \right]
    ,
\\
    \gamd{+}(t)
&=
    \frac{
        \la^{2}
    }{
        \om_{k}
        \Om_{k}
        \left(
            \Om_{k}^{2}
            -
            \om_{k}^{2}
        \right)
    }
    \Big(
        \Om_{k}
        \cos
        \left[
            \om_{k}
            \left(
                t
                -
                t_{0}
            \right)
        \right]
        \sin
        \left[
            \Om_{k}
            \left(
                t
                -
                t_{0}
            \right)
        \right]
        -
        \om_{k}
        \cos
        \left[
            \Om_{k}
            \left(
                t
                -
                t_{0}
            \right)
        \right]
        \sin
        \left[
            \om_{k}
            \left(
                t
                -
                t_{0}
            \right)
        \right]
    \Big)
    \,,
\\
    \gamd{-}(t)
&=
    \frac{
        \la^{2}
    }{
        \om_{k}
        \Om_{k}
        \left(
            \Om_{k}^{2}
            -
            \om_{k}^{2}
        \right)
    }
    \Big(
        \om_{k}
        \cos
        \left[
            \om_{k}
            \left(
                t
                -
                t_{0}
            \right)
        \right]
        \sin
        \left[
            \Om_{k}
            \left(
                t
                -
                t_{0}
            \right)
        \right]
        -
        \Om_{k}
        \cos
        \left[
            \Om_{k}
            \left(
                t
                -
                t_{0}
            \right)
        \right]
        \sin
        \left[
            \om_{k}
            \left(
                t
                -
                t_{0}
            \right)
        \right]
    \Big)
    \,,
\\
    \omS(t)
&=
    \om_{k}
    -
    \frac{
        \la^{2}
    }{
        2
        \om_{k}
        \Om_{k}
        \left(
            \Om_{k}^{2}
            -
            \om_{k}^{2}
        \right)
    }
    \Big(
        \Om_{k}
        -
        \Om_{k}
        \cos
        \left[
            \om_{k}
            \left(
                t
                -
                t_{0}
            \right)
        \right]
        \cos
        \left[
            \Om_{k}
            \left(
                t
                -
                t_{0}
            \right)
        \right]
        -
        \om_{k}
        \sin
        \left[
            \Om_{k}
            \left(
                t
                -
                t_{0}
            \right)
        \right]
        \sin
        \left[
            \om_{k}
            \left(
                t
                -
                t_{0}
            \right)
        \right]
    \Big)
    \,.
\end{align}
For this interaction, the Markovian limit is only well defined {\it before} evaluating the time integral, as in \cref{eq:Gamma_be_phi_chi_Markovian}, so that the time integral can be written as a sum of Fourier cosine and sine transforms. The resulting coefficients can, however, also be obtained by dropping the oscillating terms in the expressions above.

%--------------------------
\section{Redfield equation coefficients for the \texorpdfstring{$ \la \phi \chi^{2} $}{lpc2} theory}
\label{app:phichi2coeff}
%--------------------------

In this appendix, we display the coefficients of the Redfield equation for the $ \la \phi \chi^{2} $ interaction described in \cref{sec:phichi2},
\begin{align}
    \gamd{\rm o}(t)
&=
    \frac{
        \la^{2}
    }{
        64
        \pi^{2}
        \om_{k}
    }
    \Big(
        \pi
        +
        \text{Si}
        \left[
            \left(
                \om_{k}
                -
                k
            \right)
            \left(
                t
                -
                t_{0}
            \right)
        \right]
        -
        \text{Si}
        \left[
            \left(
                \om_{k}
                +
                k
            \right)
            \left(
                t
                -
                t_{0}
            \right)
        \right]
    \Big)
\nn
&\hspace{11.5pt}
    +
    i
    \frac{
        \la^{2}
    }{
        64
        \pi^{2}
        \om_{k}
    }
    \Big(
        \text{Ci}
        \left[
            \left(
                \om_{k}
                -
                k
            \right)
            \left(
                t
                -
                t_{0}
            \right)
        \right]
        -
        \text{Ci}
        \left[
            \left(
                \om_{k}
                +
                k
            \right)
            \left(
                t
                -
                t_{0}
            \right)
        \right]
        +
        \ln
        \left[
            \frac{
                \om_{k}
                +
                k
            }{
                \om_{k}
                -
                k
            }
        \right]
    \Big)
    \,,
\\
    \Sd{\rm o}(t)
&=
    \frac{
        \la^{2}
    }{
        128
        \pi^{2}
        \om_{k}
    }
    \Big(
        2
        \gamma
        -
        2
        \ln
        \left[
            \frac{
                \mu_{\rm r}
            }{
                m
            }
        \right]
        -
        \text{Ci}
        \left[
            \left(
                \om_{k}
                -
                k
            \right)
            \left(
                t
                -
                t_{0}
            \right)
        \right]
        -
        \text{Ci}
        \left[
            \left(
                \om_{k}
                +
                k
            \right)
            \left(
                t
                -
                t_{0}
            \right)
        \right]
    \Big)
\label{eq:So_phichi2}
\nn
&\hspace{11.5pt}
    +
    i
    \frac{
        \la^{2}
    }{
        128
        \pi^{2}
        \om_{k}
    }
    \Big(
        \text{Si}
        \left[
            \left(
                \om_{k}
                -
                k
            \right)
            \left(
                t
                -
                t_{0}
            \right)
        \right]
        +
        \text{Si}
        \left[
            \left(
                \om_{k}
                +
                k
            \right)
            \left(
                t
                -
                t_{0}
            \right)
        \right]
    \Big)
    \,,
\\
    \gamd{+}(t)
&=
    \frac{
        \la^{2}
    }{
        32
        \pi^{2}
        \om_{k}
    }
    \Big(
        \pi
        +
        \text{Si}
        \left[
            \left(
                \om_{k}
                -
                k
            \right)
            \left(
                t
                -
                t_{0}
            \right)
        \right]
        -
        \text{Si}
        \left[
            \left(
                \om_{k}
                +
                k
            \right)
            \left(
                t
                -
                t_{0}
            \right)
        \right]
    \Big)
    \,,
\\
    \gamd{-}(t)
&=
    \frac{
        \la^{2}
    }{
        32
        \pi^{2}
        \om_{k}
    }
    \Big(
        \text{Si}
        \left[
            \left(
                \om_{k}
                -
                k
            \right)
            \left(
                t
                -
                t_{0}
            \right)
        \right]
        +
        \text{Si}
        \left[
            \left(
                \om_{k}
                +
                k
            \right)
            \left(
                t
                -
                t_{0}
            \right)
        \right]
    \Big)
    \,,
\\
    \omS(t)
&=
    \om_{k}
    +
    \frac{
        \la^{2}
    }{
        64
        \pi^{2}
        \om_{k}
    }
    \Big(
        2
        \gamma
        -
        2
        \ln
        \left[
            \frac{
                \mu_{\rm r}
            }{
                m
            }
        \right]
        -
        \text{Ci}
        \left[
            \left(
                \om_{k}
                -
                k
            \right)
            \left(
                t
                -
                t_{0}
            \right)
        \right]
        -
        \text{Ci}
        \left[
            \left(
                \om_{k}
                +
                k
            \right)
            \left(
                t
                -
                t_{0}
            \right)
        \right]
    \Big)
\label{eq:omS_phichi2}
    \,,
\end{align}
where we have used the standard definition for the sine and cosine integrals,
\begin{align}
    \text{Si}(z)
=
    \int_{0}^{z}
    \d t
    \frac{
        \sin(t)
    }{
        t
    }
    \,,
    \hspace{20pt}
    \text{Ci}(z)
=
    -
    \int_{z}^{\infty}
    \d t
    \frac{
        \cos(t)
    }{
        t
    }
    \,,
\end{align}
and the renormalization scale $ \mu_{\rm r} $ indicates where divergences were removed by choosing the counterterms in \cref{eq:FinalCounterterms}.  The Markovian limit is straightforward to take as each function has a well-defined limit as $ t - t_{0} \to \infty $.
\end{widetext}

\bibliography{references}

\begin{thebibliography}{91}%
\makeatletter
\providecommand \@ifxundefined [1]{%
 \@ifx{#1\undefined}
}%
\providecommand \@ifnum [1]{%
 \ifnum #1\expandafter \@firstoftwo
 \else \expandafter \@secondoftwo
 \fi
}%
\providecommand \@ifx [1]{%
 \ifx #1\expandafter \@firstoftwo
 \else \expandafter \@secondoftwo
 \fi
}%
\providecommand \natexlab [1]{#1}%
\providecommand \enquote  [1]{``#1''}%
\providecommand \bibnamefont  [1]{#1}%
\providecommand \bibfnamefont [1]{#1}%
\providecommand \citenamefont [1]{#1}%
\providecommand \href@noop [0]{\@secondoftwo}%
\providecommand \href [0]{\begingroup \@sanitize@url \@href}%
\providecommand \@href[1]{\@@startlink{#1}\@@href}%
\providecommand \@@href[1]{\endgroup#1\@@endlink}%
\providecommand \@sanitize@url [0]{\catcode `\\12\catcode `\$12\catcode
  `\&12\catcode `\#12\catcode `\^12\catcode `\_12\catcode `\%12\relax}%
\providecommand \@@startlink[1]{}%
\providecommand \@@endlink[0]{}%
\providecommand \url  [0]{\begingroup\@sanitize@url \@url }%
\providecommand \@url [1]{\endgroup\@href {#1}{\urlprefix }}%
\providecommand \urlprefix  [0]{URL }%
\providecommand \Eprint [0]{\href }%
\providecommand \doibase [0]{http://dx.doi.org/}%
\providecommand \selectlanguage [0]{\@gobble}%
\providecommand \bibinfo  [0]{\@secondoftwo}%
\providecommand \bibfield  [0]{\@secondoftwo}%
\providecommand \translation [1]{[#1]}%
\providecommand \BibitemOpen [0]{}%
\providecommand \bibitemStop [0]{}%
\providecommand \bibitemNoStop [0]{.\EOS\space}%
\providecommand \EOS [0]{\spacefactor3000\relax}%
\providecommand \BibitemShut  [1]{\csname bibitem#1\endcsname}%
\let\auto@bib@innerbib\@empty
%</preamble>
\bibitem [{\citenamefont {Gorini}\ \emph {et~al.}(1976)\citenamefont {Gorini},
  \citenamefont {Kossakowski},\ and\ \citenamefont
  {Sudarshan}}]{Gorini:1975nb}%
  \BibitemOpen
  \bibfield  {author} {\bibinfo {author} {\bibfnamefont {V.}~\bibnamefont
  {Gorini}}, \bibinfo {author} {\bibfnamefont {A.}~\bibnamefont {Kossakowski}},
  \ and\ \bibinfo {author} {\bibfnamefont {E.~C.~G.}\ \bibnamefont
  {Sudarshan}},\ }\bibfield  {title} {\enquote {\bibinfo {title} {{Completely
  positive dynamical semigroups of N level systems}},}\ }\href {\doibase
  10.1063/1.522979} {\bibfield  {journal} {\bibinfo  {journal} {J. Math.
  Phys.}\ }\textbf {\bibinfo {volume} {17}},\ \bibinfo {pages} {821} (\bibinfo
  {year} {1976})}\BibitemShut {NoStop}%
\bibitem [{\citenamefont {Lindblad}(1976)}]{Lindblad:1975ef}%
  \BibitemOpen
  \bibfield  {author} {\bibinfo {author} {\bibfnamefont {G.}~\bibnamefont
  {Lindblad}},\ }\bibfield  {title} {\enquote {\bibinfo {title} {{On the
  generators of quantum dynamical semigroups}},}\ }\href {\doibase
  10.1007/BF01608499} {\bibfield  {journal} {\bibinfo  {journal} {Commun. Math.
  Phys.}\ }\textbf {\bibinfo {volume} {48}},\ \bibinfo {pages} {119} (\bibinfo
  {year} {1976})}\BibitemShut {NoStop}%
\bibitem [{\citenamefont {Redfield}(1957)}]{Redfield1957}%
  \BibitemOpen
  \bibfield  {author} {\bibinfo {author} {\bibfnamefont {A.~G.}\ \bibnamefont
  {Redfield}},\ }\bibfield  {title} {\enquote {\bibinfo {title} {{On the theory
  of relaxation processes}},}\ }\href {\doibase 10.1147/rd.11.0019} {\bibfield
  {journal} {\bibinfo  {journal} {IBM Journal of Research and Development}\
  }\textbf {\bibinfo {volume} {1}},\ \bibinfo {pages} {19--31} (\bibinfo {year}
  {1957})}\BibitemShut {NoStop}%
\bibitem [{Note1()}]{Note1}%
  \BibitemOpen
  \bibinfo {note} {The Redfield equation is also referred to as the
  Born-Redfield equation \cite {Keefe:2024cia}.}\BibitemShut {Stop}%
\bibitem [{\citenamefont {Davies}(1974)}]{Davies_1974}%
  \BibitemOpen
  \bibfield  {author} {\bibinfo {author} {\bibfnamefont {E.~B.}\ \bibnamefont
  {Davies}},\ }\bibfield  {title} {\enquote {\bibinfo {title} {{Markovian
  master equations}},}\ }\href {\doibase 10.1007/BF01608389} {\bibfield
  {journal} {\bibinfo  {journal} {Commun. Math. Phys.}\ }\textbf {\bibinfo
  {volume} {39}},\ \bibinfo {pages} {91 -- 110} (\bibinfo {year}
  {1974})}\BibitemShut {NoStop}%
\bibitem [{\citenamefont {Davies}(1976)}]{Davies_1976}%
  \BibitemOpen
  \bibfield  {author} {\bibinfo {author} {\bibfnamefont {E.~B.}\ \bibnamefont
  {Davies}},\ }\bibfield  {title} {\enquote {\bibinfo {title} {{Markovian
  master equations. II.}}}\ }\href {\doibase 10.1007/BF01351898} {\bibfield
  {journal} {\bibinfo  {journal} {Math. Ann.}\ }\textbf {\bibinfo {volume}
  {219}},\ \bibinfo {pages} {147 -- 158} (\bibinfo {year} {1976})}\BibitemShut
  {NoStop}%
\bibitem [{\citenamefont {Gaspard}\ and\ \citenamefont
  {Nagaoka}(1999)}]{Gaspard:1999}%
  \BibitemOpen
  \bibfield  {author} {\bibinfo {author} {\bibfnamefont {P.}~\bibnamefont
  {Gaspard}}\ and\ \bibinfo {author} {\bibfnamefont {M.}~\bibnamefont
  {Nagaoka}},\ }\bibfield  {title} {\enquote {\bibinfo {title} {{Slippage of
  initial conditions for the Redfield master equation}},}\ }\href {\doibase
  10.1063/1.479867} {\bibfield  {journal} {\bibinfo  {journal} {The Journal of
  Chemical Physics}\ }\textbf {\bibinfo {volume} {111}},\ \bibinfo {pages}
  {5668--5675} (\bibinfo {year} {1999})}\BibitemShut {NoStop}%
\bibitem [{\citenamefont {Whitney}(2008)}]{Whitney:2008}%
  \BibitemOpen
  \bibfield  {author} {\bibinfo {author} {\bibfnamefont {R.}~\bibnamefont
  {Whitney}},\ }\bibfield  {title} {\enquote {\bibinfo {title} {{Staying
  positive: Going beyond Lindblad with perturbative master equations}},}\
  }\href {\doibase 10.1088/1751-8113/41/17/175304} {\bibfield  {journal}
  {\bibinfo  {journal} {Journal of Physics A: Mathematical and Theoretical}\
  }\textbf {\bibinfo {volume} {41}},\ \bibinfo {pages} {175304} (\bibinfo
  {year} {2008})},\ \Eprint {http://arxiv.org/abs/0711.0074} {arXiv:0711.0074
  [quant-ph]} \BibitemShut {NoStop}%
\bibitem [{\citenamefont {Schaller}\ and\ \citenamefont
  {Brandes}(2008)}]{Schaller:2008}%
  \BibitemOpen
  \bibfield  {author} {\bibinfo {author} {\bibfnamefont {G.}~\bibnamefont
  {Schaller}}\ and\ \bibinfo {author} {\bibfnamefont {T.}~\bibnamefont
  {Brandes}},\ }\bibfield  {title} {\enquote {\bibinfo {title} {{Preservation
  of positivity by dynamical coarse graining}},}\ }\href {\doibase
  10.1103/physreva.78.022106} {\bibfield  {journal} {\bibinfo  {journal} {Phys.
  Rev. A}\ }\textbf {\bibinfo {volume} {78}},\ \bibinfo {pages} {022106}
  (\bibinfo {year} {2008})},\ \Eprint {http://arxiv.org/abs/0804.2374}
  {arXiv:0804.2374 [quant-ph]} \BibitemShut {NoStop}%
\bibitem [{\citenamefont {Majenz}\ \emph {et~al.}(2013)\citenamefont {Majenz},
  \citenamefont {Albash}, \citenamefont {Breuer},\ and\ \citenamefont
  {Lidar}}]{Majenz:2013}%
  \BibitemOpen
  \bibfield  {author} {\bibinfo {author} {\bibfnamefont {C.}~\bibnamefont
  {Majenz}}, \bibinfo {author} {\bibfnamefont {T.}~\bibnamefont {Albash}},
  \bibinfo {author} {\bibfnamefont {H.-P.}\ \bibnamefont {Breuer}}, \ and\
  \bibinfo {author} {\bibfnamefont {D.~A.}\ \bibnamefont {Lidar}},\ }\bibfield
  {title} {\enquote {\bibinfo {title} {{Coarse graining can beat the
  rotating-wave approximation in quantum Markovian master equations}},}\ }\href
  {\doibase 10.1103/PhysRevA.88.012103} {\bibfield  {journal} {\bibinfo
  {journal} {Phys. Rev. A}\ }\textbf {\bibinfo {volume} {88}},\ \bibinfo
  {pages} {012103} (\bibinfo {year} {2013})},\ \Eprint
  {http://arxiv.org/abs/1303.6580} {arXiv:1303.6580 [quant-ph]} \BibitemShut
  {NoStop}%
\bibitem [{\citenamefont {Farina}\ and\ \citenamefont
  {Giovannetti}(2019)}]{Farina:2019}%
  \BibitemOpen
  \bibfield  {author} {\bibinfo {author} {\bibfnamefont {D.}~\bibnamefont
  {Farina}}\ and\ \bibinfo {author} {\bibfnamefont {V.}~\bibnamefont
  {Giovannetti}},\ }\bibfield  {title} {\enquote {\bibinfo {title}
  {{Open-quantum-system dynamics: Recovering positivity of the Redfield
  equation via the partial secular approximation}},}\ }\href {\doibase
  10.1103/PhysRevA.100.012107} {\bibfield  {journal} {\bibinfo  {journal}
  {Phys. Rev. A}\ }\textbf {\bibinfo {volume} {100}},\ \bibinfo {pages}
  {012107} (\bibinfo {year} {2019})},\ \Eprint
  {http://arxiv.org/abs/1903.07324} {arXiv:1903.07324 [quant-ph]} \BibitemShut
  {NoStop}%
\bibitem [{\citenamefont {Mozgunov}\ and\ \citenamefont
  {Lidar}(2020)}]{Mozgunov_2020}%
  \BibitemOpen
  \bibfield  {author} {\bibinfo {author} {\bibfnamefont {E.}~\bibnamefont
  {Mozgunov}}\ and\ \bibinfo {author} {\bibfnamefont {D.}~\bibnamefont
  {Lidar}},\ }\bibfield  {title} {\enquote {\bibinfo {title} {Completely
  positive master equation for arbitrary driving and small level spacing},}\
  }\href@noop {} {\bibfield  {journal} {\bibinfo  {journal} {Quantum}\ }\textbf
  {\bibinfo {volume} {4}},\ \bibinfo {pages} {227} (\bibinfo {year} {2020})},\
  \Eprint {http://arxiv.org/abs/1908.01095} {arXiv:1908.01095 [quant-ph]}
  \BibitemShut {NoStop}%
\bibitem [{\citenamefont {Davidovi{\'{c}}}(2020)}]{Davidovic:2020}%
  \BibitemOpen
  \bibfield  {author} {\bibinfo {author} {\bibfnamefont {D.}~\bibnamefont
  {Davidovi{\'{c}}}},\ }\bibfield  {title} {\enquote {\bibinfo {title}
  {{Completely {p}ositive, {s}imple, and {p}ossibly {h}ighly {a}ccurate
  {a}pproximation of the {R}edfield {e}quation}},}\ }\href {\doibase
  10.22331/q-2020-09-21-326} {\bibfield  {journal} {\bibinfo  {journal}
  {Quantum}\ }\textbf {\bibinfo {volume} {4}},\ \bibinfo {pages} {326}
  (\bibinfo {year} {2020})},\ \Eprint {http://arxiv.org/abs/2003.09063}
  {arXiv:2003.09063 [quant-ph]} \BibitemShut {NoStop}%
\bibitem [{\citenamefont {Nathan}\ and\ \citenamefont
  {Rudner}(2020)}]{Nathan:2020}%
  \BibitemOpen
  \bibfield  {author} {\bibinfo {author} {\bibfnamefont {F.}~\bibnamefont
  {Nathan}}\ and\ \bibinfo {author} {\bibfnamefont {M.~S.}\ \bibnamefont
  {Rudner}},\ }\bibfield  {title} {\enquote {\bibinfo {title} {{Universal
  Lindblad equation for open quantum systems}},}\ }\href {\doibase
  10.1103/PhysRevB.102.115109} {\bibfield  {journal} {\bibinfo  {journal}
  {Phys. Rev. B}\ }\textbf {\bibinfo {volume} {102}},\ \bibinfo {pages}
  {115109} (\bibinfo {year} {2020})},\ \Eprint
  {http://arxiv.org/abs/2004.01469} {arXiv:2004.01469 [cond-mat.mes-hall]}
  \BibitemShut {NoStop}%
\bibitem [{\citenamefont {Trushechkin}(2021)}]{Trushechkin2021}%
  \BibitemOpen
  \bibfield  {author} {\bibinfo {author} {\bibfnamefont {A.}~\bibnamefont
  {Trushechkin}},\ }\bibfield  {title} {\enquote {\bibinfo {title} {Unified
  gorini-kossakowski-lindblad-sudarshan quantum master equation beyond the
  secular approximation},}\ }\href {\doibase {10.1103/physreva.103.062226}}
  {\bibfield  {journal} {\bibinfo  {journal} {Phys. Rev. A}\ }\textbf {\bibinfo
  {volume} {103}},\ \bibinfo {pages} {062226} (\bibinfo {year} {2021})},\
  \Eprint {http://arxiv.org/abs/2103.12042} {arXiv:2103.12042 [quant-ph]}
  \BibitemShut {NoStop}%
\bibitem [{\citenamefont {Keefe}\ \emph {et~al.}(2025)\citenamefont {Keefe},
  \citenamefont {Agarwal},\ and\ \citenamefont {Kamal}}]{Keefe:2024cia}%
  \BibitemOpen
  \bibfield  {author} {\bibinfo {author} {\bibfnamefont {A.}~\bibnamefont
  {Keefe}}, \bibinfo {author} {\bibfnamefont {N.}~\bibnamefont {Agarwal}}, \
  and\ \bibinfo {author} {\bibfnamefont {A.}~\bibnamefont {Kamal}},\ }\bibfield
   {title} {\enquote {\bibinfo {title} {{Quantifying spectral signatures of
  non-Markovianity beyond the Born-Redfield master equation}},}\ }\href
  {\doibase 10.22331/q-2025-09-24-1863} {\bibfield  {journal} {\bibinfo
  {journal} {Quantum}\ }\textbf {\bibinfo {volume} {9}},\ \bibinfo {pages}
  {1863} (\bibinfo {year} {2025})},\ \Eprint {http://arxiv.org/abs/2405.01722}
  {arXiv:2405.01722 [quant-ph]} \BibitemShut {NoStop}%
\bibitem [{\citenamefont {Levy}\ and\ \citenamefont
  {Kosloff}(2014)}]{Levy2014}%
  \BibitemOpen
  \bibfield  {author} {\bibinfo {author} {\bibfnamefont {A.}~\bibnamefont
  {Levy}}\ and\ \bibinfo {author} {\bibfnamefont {R.}~\bibnamefont {Kosloff}},\
  }\bibfield  {title} {\enquote {\bibinfo {title} {The local approach to
  quantum transport may violate the second law of thermodynamics},}\ }\href
  {\doibase 10.1209/0295-5075/107/20004} {\bibfield  {journal} {\bibinfo
  {journal} {Europhys. Lett.}\ }\textbf {\bibinfo {volume} {107}},\ \bibinfo
  {pages} {20004} (\bibinfo {year} {2014})}\BibitemShut {NoStop}%
\bibitem [{\citenamefont {Lombardo}\ and\ \citenamefont
  {Mazzitelli}(1996)}]{Lombardo:1995fg}%
  \BibitemOpen
  \bibfield  {author} {\bibinfo {author} {\bibfnamefont {F.}~\bibnamefont
  {Lombardo}}\ and\ \bibinfo {author} {\bibfnamefont {F.~D.}\ \bibnamefont
  {Mazzitelli}},\ }\bibfield  {title} {\enquote {\bibinfo {title} {{Coarse
  graining and decoherence in quantum field theory}},}\ }\href {\doibase
  10.1103/PhysRevD.53.2001} {\bibfield  {journal} {\bibinfo  {journal} {Phys.
  Rev. D}\ }\textbf {\bibinfo {volume} {53}},\ \bibinfo {pages} {2001--2011}
  (\bibinfo {year} {1996})},\ \Eprint {http://arxiv.org/abs/hep-th/9508052}
  {arXiv:hep-th/9508052} \BibitemShut {NoStop}%
\bibitem [{\citenamefont {Koksma}\ \emph {et~al.}(2010)\citenamefont {Koksma},
  \citenamefont {Prokopec},\ and\ \citenamefont {Schmidt}}]{Koksma:2009wa}%
  \BibitemOpen
  \bibfield  {author} {\bibinfo {author} {\bibfnamefont {J.~F.}\ \bibnamefont
  {Koksma}}, \bibinfo {author} {\bibfnamefont {T.}~\bibnamefont {Prokopec}}, \
  and\ \bibinfo {author} {\bibfnamefont {M.~G.}\ \bibnamefont {Schmidt}},\
  }\bibfield  {title} {\enquote {\bibinfo {title} {{Decoherence in an
  interacting quantum field theory: The vacuum case}},}\ }\href {\doibase
  10.1103/PhysRevD.81.065030} {\bibfield  {journal} {\bibinfo  {journal} {Phys.
  Rev. D}\ }\textbf {\bibinfo {volume} {81}},\ \bibinfo {pages} {065030}
  (\bibinfo {year} {2010})},\ \Eprint {http://arxiv.org/abs/0910.5733}
  {arXiv:0910.5733 [hep-th]} \BibitemShut {NoStop}%
\bibitem [{\citenamefont {Koksma}\ \emph {et~al.}(2011)\citenamefont {Koksma},
  \citenamefont {Prokopec},\ and\ \citenamefont {Schmidt}}]{Koksma:2011dy}%
  \BibitemOpen
  \bibfield  {author} {\bibinfo {author} {\bibfnamefont {J.~F.}\ \bibnamefont
  {Koksma}}, \bibinfo {author} {\bibfnamefont {T.}~\bibnamefont {Prokopec}}, \
  and\ \bibinfo {author} {\bibfnamefont {M.~G.}\ \bibnamefont {Schmidt}},\
  }\bibfield  {title} {\enquote {\bibinfo {title} {{Decoherence in an
  interacting quantum field theory: Thermal case}},}\ }\href {\doibase
  10.1103/PhysRevD.83.085011} {\bibfield  {journal} {\bibinfo  {journal} {Phys.
  Rev. D}\ }\textbf {\bibinfo {volume} {83}},\ \bibinfo {pages} {085011}
  (\bibinfo {year} {2011})},\ \Eprint {http://arxiv.org/abs/1102.4713}
  {arXiv:1102.4713 [hep-th]} \BibitemShut {NoStop}%
\bibitem [{\citenamefont {Ag\'on}\ \emph {et~al.}(2018)\citenamefont {Ag\'on},
  \citenamefont {Balasubramanian}, \citenamefont {Kasko},\ and\ \citenamefont
  {Lawrence}}]{Agon:2014uxa}%
  \BibitemOpen
  \bibfield  {author} {\bibinfo {author} {\bibfnamefont {C.}~\bibnamefont
  {Ag\'on}}, \bibinfo {author} {\bibfnamefont {V.}~\bibnamefont
  {Balasubramanian}}, \bibinfo {author} {\bibfnamefont {S.}~\bibnamefont
  {Kasko}}, \ and\ \bibinfo {author} {\bibfnamefont {A.}~\bibnamefont
  {Lawrence}},\ }\bibfield  {title} {\enquote {\bibinfo {title} {{Coarse
  grained quantum dynamics}},}\ }\href {\doibase 10.1103/PhysRevD.98.025019}
  {\bibfield  {journal} {\bibinfo  {journal} {Phys. Rev. D}\ }\textbf {\bibinfo
  {volume} {98}},\ \bibinfo {pages} {025019} (\bibinfo {year} {2018})},\
  \Eprint {http://arxiv.org/abs/1412.3148} {arXiv:1412.3148 [hep-th]}
  \BibitemShut {NoStop}%
\bibitem [{\citenamefont
  {Boyanovsky}(2015{\natexlab{a}})}]{Boyanovsky:2015xoa}%
  \BibitemOpen
  \bibfield  {author} {\bibinfo {author} {\bibfnamefont {D.}~\bibnamefont
  {Boyanovsky}},\ }\bibfield  {title} {\enquote {\bibinfo {title} {{Effective
  field theory out of equilibrium: Brownian quantum fields}},}\ }\href
  {\doibase 10.1088/1367-2630/17/6/063017} {\bibfield  {journal} {\bibinfo
  {journal} {New J. Phys.}\ }\textbf {\bibinfo {volume} {17}},\ \bibinfo
  {pages} {063017} (\bibinfo {year} {2015}{\natexlab{a}})},\ \Eprint
  {http://arxiv.org/abs/1503.00156} {arXiv:1503.00156 [hep-ph]} \BibitemShut
  {NoStop}%
\bibitem [{\citenamefont {Ag\'on}\ and\ \citenamefont
  {Lawrence}(2018)}]{Agon:2017oia}%
  \BibitemOpen
  \bibfield  {author} {\bibinfo {author} {\bibfnamefont {C.}~\bibnamefont
  {Ag\'on}}\ and\ \bibinfo {author} {\bibfnamefont {A.}~\bibnamefont
  {Lawrence}},\ }\bibfield  {title} {\enquote {\bibinfo {title} {{Divergences
  in open quantum systems}},}\ }\href {\doibase 10.1007/JHEP04(2018)008}
  {\bibfield  {journal} {\bibinfo  {journal} {JHEP}\ }\textbf {\bibinfo
  {volume} {04}},\ \bibinfo {pages} {008} (\bibinfo {year} {2018})},\ \Eprint
  {http://arxiv.org/abs/1709.10095} {arXiv:1709.10095 [hep-th]} \BibitemShut
  {NoStop}%
\bibitem [{\citenamefont
  {Boyanovsky}(2018{\natexlab{a}})}]{Boyanovsky:2018fxl}%
  \BibitemOpen
  \bibfield  {author} {\bibinfo {author} {\bibfnamefont {D.}~\bibnamefont
  {Boyanovsky}},\ }\bibfield  {title} {\enquote {\bibinfo {title} {{Information
  loss in effective field theory: Entanglement and thermal entropies}},}\
  }\href {\doibase 10.1103/PhysRevD.97.065008} {\bibfield  {journal} {\bibinfo
  {journal} {Phys. Rev. D}\ }\textbf {\bibinfo {volume} {97}},\ \bibinfo
  {pages} {065008} (\bibinfo {year} {2018}{\natexlab{a}})},\ \Eprint
  {http://arxiv.org/abs/1801.06840} {arXiv:1801.06840 [hep-th]} \BibitemShut
  {NoStop}%
\bibitem [{\citenamefont {Burrage}\ \emph {et~al.}(2019)\citenamefont
  {Burrage}, \citenamefont {K\"ading}, \citenamefont {Millington},\ and\
  \citenamefont {Min\'a\v{r}}}]{Burrage:2018pyg}%
  \BibitemOpen
  \bibfield  {author} {\bibinfo {author} {\bibfnamefont {C.}~\bibnamefont
  {Burrage}}, \bibinfo {author} {\bibfnamefont {C.}~\bibnamefont {K\"ading}},
  \bibinfo {author} {\bibfnamefont {P.}~\bibnamefont {Millington}}, \ and\
  \bibinfo {author} {\bibfnamefont {J.}~\bibnamefont {Min\'a\v{r}}},\
  }\bibfield  {title} {\enquote {\bibinfo {title} {{Open quantum dynamics
  induced by light scalar fields}},}\ }\href {\doibase
  10.1103/PhysRevD.100.076003} {\bibfield  {journal} {\bibinfo  {journal}
  {Phys. Rev. D}\ }\textbf {\bibinfo {volume} {100}},\ \bibinfo {pages}
  {076003} (\bibinfo {year} {2019})},\ \Eprint
  {http://arxiv.org/abs/1812.08760} {arXiv:1812.08760 [hep-th]} \BibitemShut
  {NoStop}%
\bibitem [{\citenamefont {Banerjee}\ \emph {et~al.}(2023)\citenamefont
  {Banerjee}, \citenamefont {Choudhury}, \citenamefont {Chowdhury},
  \citenamefont {Knaute}, \citenamefont {Panda},\ and\ \citenamefont
  {Shirish}}]{Banerjee:2021lqu}%
  \BibitemOpen
  \bibfield  {author} {\bibinfo {author} {\bibfnamefont {S.}~\bibnamefont
  {Banerjee}}, \bibinfo {author} {\bibfnamefont {S.}~\bibnamefont {Choudhury}},
  \bibinfo {author} {\bibfnamefont {S.}~\bibnamefont {Chowdhury}}, \bibinfo
  {author} {\bibfnamefont {J.}~\bibnamefont {Knaute}}, \bibinfo {author}
  {\bibfnamefont {S.}~\bibnamefont {Panda}}, \ and\ \bibinfo {author}
  {\bibfnamefont {K.}~\bibnamefont {Shirish}},\ }\bibfield  {title} {\enquote
  {\bibinfo {title} {{Thermalization in quenched open quantum cosmology}},}\
  }\href {\doibase 10.1016/j.nuclphysb.2023.116368} {\bibfield  {journal}
  {\bibinfo  {journal} {Nucl. Phys. B}\ }\textbf {\bibinfo {volume} {996}},\
  \bibinfo {pages} {116368} (\bibinfo {year} {2023})},\ \Eprint
  {http://arxiv.org/abs/2104.10692} {arXiv:2104.10692 [hep-th]} \BibitemShut
  {NoStop}%
\bibitem [{\citenamefont {K\"ading}\ and\ \citenamefont
  {Pitschmann}(2023)}]{Kading:2022jjl}%
  \BibitemOpen
  \bibfield  {author} {\bibinfo {author} {\bibfnamefont {C.}~\bibnamefont
  {K\"ading}}\ and\ \bibinfo {author} {\bibfnamefont {M.}~\bibnamefont
  {Pitschmann}},\ }\bibfield  {title} {\enquote {\bibinfo {title} {{New method
  for directly computing reduced density matrices}},}\ }\href {\doibase
  10.1103/PhysRevD.107.016005} {\bibfield  {journal} {\bibinfo  {journal}
  {Phys. Rev. D}\ }\textbf {\bibinfo {volume} {107}},\ \bibinfo {pages}
  {016005} (\bibinfo {year} {2023})},\ \Eprint
  {http://arxiv.org/abs/2204.08829} {arXiv:2204.08829 [hep-th]} \BibitemShut
  {NoStop}%
\bibitem [{\citenamefont {K\"ading}\ and\ \citenamefont
  {Pitschmann}(2025)}]{Kading:2025cwg}%
  \BibitemOpen
  \bibfield  {author} {\bibinfo {author} {\bibfnamefont {C.}~\bibnamefont
  {K\"ading}}\ and\ \bibinfo {author} {\bibfnamefont {M.}~\bibnamefont
  {Pitschmann}},\ }\bibfield  {title} {\enquote {\bibinfo {title} {{Density
  matrices in quantum field theory: Non-Markovianity, path integrals and master
  equations}},}\ }\href@noop {} {\  (\bibinfo {year} {2025})},\ \Eprint
  {http://arxiv.org/abs/2503.08567} {arXiv:2503.08567 [hep-th]} \BibitemShut
  {NoStop}%
\bibitem [{\citenamefont {Lombardo}\ and\ \citenamefont
  {Nacir}(2005)}]{Lombardo:2005iz}%
  \BibitemOpen
  \bibfield  {author} {\bibinfo {author} {\bibfnamefont {F.~C.}\ \bibnamefont
  {Lombardo}}\ and\ \bibinfo {author} {\bibfnamefont {D.~L.}\ \bibnamefont
  {Nacir}},\ }\bibfield  {title} {\enquote {\bibinfo {title} {{Decoherence
  during inflation: The generation of classical inhomogeneities}},}\ }\href
  {\doibase 10.1103/PhysRevD.72.063506} {\bibfield  {journal} {\bibinfo
  {journal} {Phys. Rev. D}\ }\textbf {\bibinfo {volume} {72}},\ \bibinfo
  {pages} {063506} (\bibinfo {year} {2005})},\ \Eprint
  {http://arxiv.org/abs/gr-qc/0506051} {arXiv:gr-qc/0506051} \BibitemShut
  {NoStop}%
\bibitem [{\citenamefont
  {Boyanovsky}(2015{\natexlab{b}})}]{Boyanovsky:2015tba}%
  \BibitemOpen
  \bibfield  {author} {\bibinfo {author} {\bibfnamefont {D.}~\bibnamefont
  {Boyanovsky}},\ }\bibfield  {title} {\enquote {\bibinfo {title} {{Effective
  field theory during inflation: Reduced density matrix and its quantum master
  equation}},}\ }\href {\doibase 10.1103/PhysRevD.92.023527} {\bibfield
  {journal} {\bibinfo  {journal} {Phys. Rev. D}\ }\textbf {\bibinfo {volume}
  {92}},\ \bibinfo {pages} {023527} (\bibinfo {year} {2015}{\natexlab{b}})},\
  \Eprint {http://arxiv.org/abs/1506.07395} {arXiv:1506.07395 [astro-ph.CO]}
  \BibitemShut {NoStop}%
\bibitem [{\citenamefont {Burgess}\ \emph {et~al.}(2016)\citenamefont
  {Burgess}, \citenamefont {Holman},\ and\ \citenamefont
  {Tasinato}}]{Burgess:2015ajz}%
  \BibitemOpen
  \bibfield  {author} {\bibinfo {author} {\bibfnamefont {C.~P.}\ \bibnamefont
  {Burgess}}, \bibinfo {author} {\bibfnamefont {R.}~\bibnamefont {Holman}}, \
  and\ \bibinfo {author} {\bibfnamefont {G.}~\bibnamefont {Tasinato}},\
  }\bibfield  {title} {\enquote {\bibinfo {title} {{Open EFTs, IR effects \&
  late-time resummations: Systematic corrections in stochastic inflation}},}\
  }\href {\doibase 10.1007/JHEP01(2016)153} {\bibfield  {journal} {\bibinfo
  {journal} {JHEP}\ }\textbf {\bibinfo {volume} {01}},\ \bibinfo {pages} {153}
  (\bibinfo {year} {2016})},\ \Eprint {http://arxiv.org/abs/1512.00169}
  {arXiv:1512.00169 [gr-qc]} \BibitemShut {NoStop}%
\bibitem [{\citenamefont {Hollowood}\ and\ \citenamefont
  {McDonald}(2017)}]{Hollowood:2017bil}%
  \BibitemOpen
  \bibfield  {author} {\bibinfo {author} {\bibfnamefont {T.~J.}\ \bibnamefont
  {Hollowood}}\ and\ \bibinfo {author} {\bibfnamefont {J.~I.}\ \bibnamefont
  {McDonald}},\ }\bibfield  {title} {\enquote {\bibinfo {title} {{Decoherence,
  discord and the quantum master equation for cosmological perturbations}},}\
  }\href {\doibase 10.1103/PhysRevD.95.103521} {\bibfield  {journal} {\bibinfo
  {journal} {Phys. Rev. D}\ }\textbf {\bibinfo {volume} {95}},\ \bibinfo
  {pages} {103521} (\bibinfo {year} {2017})},\ \Eprint
  {http://arxiv.org/abs/1701.02235} {arXiv:1701.02235 [gr-qc]} \BibitemShut
  {NoStop}%
\bibitem [{\citenamefont {Shandera}\ \emph {et~al.}(2018)\citenamefont
  {Shandera}, \citenamefont {Agarwal},\ and\ \citenamefont
  {Kamal}}]{Shandera:2017qkg}%
  \BibitemOpen
  \bibfield  {author} {\bibinfo {author} {\bibfnamefont {S.}~\bibnamefont
  {Shandera}}, \bibinfo {author} {\bibfnamefont {N.}~\bibnamefont {Agarwal}}, \
  and\ \bibinfo {author} {\bibfnamefont {A.}~\bibnamefont {Kamal}},\ }\bibfield
   {title} {\enquote {\bibinfo {title} {{Open quantum cosmological system}},}\
  }\href {\doibase 10.1103/PhysRevD.98.083535} {\bibfield  {journal} {\bibinfo
  {journal} {Phys. Rev. D}\ }\textbf {\bibinfo {volume} {98}},\ \bibinfo
  {pages} {083535} (\bibinfo {year} {2018})},\ \Eprint
  {http://arxiv.org/abs/1708.00493} {arXiv:1708.00493 [hep-th]} \BibitemShut
  {NoStop}%
\bibitem [{\citenamefont
  {Boyanovsky}(2018{\natexlab{b}})}]{Boyanovsky:2018soy}%
  \BibitemOpen
  \bibfield  {author} {\bibinfo {author} {\bibfnamefont {D.}~\bibnamefont
  {Boyanovsky}},\ }\bibfield  {title} {\enquote {\bibinfo {title} {{Imprint of
  entanglement entropy in the power spectrum of inflationary fluctuations}},}\
  }\href {\doibase 10.1103/PhysRevD.98.023515} {\bibfield  {journal} {\bibinfo
  {journal} {Phys. Rev. D}\ }\textbf {\bibinfo {volume} {98}},\ \bibinfo
  {pages} {023515} (\bibinfo {year} {2018}{\natexlab{b}})},\ \Eprint
  {http://arxiv.org/abs/1804.07967} {arXiv:1804.07967 [astro-ph.CO]}
  \BibitemShut {NoStop}%
\bibitem [{\citenamefont {Akhtar}\ \emph {et~al.}(2020)\citenamefont {Akhtar},
  \citenamefont {Choudhury}, \citenamefont {Chowdhury}, \citenamefont
  {Goswami}, \citenamefont {Panda},\ and\ \citenamefont
  {Swain}}]{Akhtar:2019qdn}%
  \BibitemOpen
  \bibfield  {author} {\bibinfo {author} {\bibfnamefont {S.}~\bibnamefont
  {Akhtar}}, \bibinfo {author} {\bibfnamefont {S.}~\bibnamefont {Choudhury}},
  \bibinfo {author} {\bibfnamefont {S.}~\bibnamefont {Chowdhury}}, \bibinfo
  {author} {\bibfnamefont {D.}~\bibnamefont {Goswami}}, \bibinfo {author}
  {\bibfnamefont {S.}~\bibnamefont {Panda}}, \ and\ \bibinfo {author}
  {\bibfnamefont {A.}~\bibnamefont {Swain}},\ }\bibfield  {title} {\enquote
  {\bibinfo {title} {{Open quantum entanglement: A study of two atomic system
  in static patch of de Sitter space}},}\ }\href {\doibase
  10.1140/epjc/s10052-020-8302-2} {\bibfield  {journal} {\bibinfo  {journal}
  {Eur. Phys. J. C}\ }\textbf {\bibinfo {volume} {80}},\ \bibinfo {pages} {748}
  (\bibinfo {year} {2020})},\ \Eprint {http://arxiv.org/abs/1908.09929}
  {arXiv:1908.09929 [hep-th]} \BibitemShut {NoStop}%
\bibitem [{\citenamefont {Gong}\ and\ \citenamefont
  {Seo}(2019)}]{Gong:2019yyz}%
  \BibitemOpen
  \bibfield  {author} {\bibinfo {author} {\bibfnamefont {J.-O.}\ \bibnamefont
  {Gong}}\ and\ \bibinfo {author} {\bibfnamefont {M.-S.}\ \bibnamefont {Seo}},\
  }\bibfield  {title} {\enquote {\bibinfo {title} {{Quantum non-linear
  evolution of inflationary tensor perturbations}},}\ }\href {\doibase
  10.1007/JHEP05(2019)021} {\bibfield  {journal} {\bibinfo  {journal} {JHEP}\
  }\textbf {\bibinfo {volume} {05}},\ \bibinfo {pages} {021} (\bibinfo {year}
  {2019})},\ \Eprint {http://arxiv.org/abs/1903.12295} {arXiv:1903.12295
  [hep-th]} \BibitemShut {NoStop}%
\bibitem [{\citenamefont {Brahma}\ \emph {et~al.}(2020)\citenamefont {Brahma},
  \citenamefont {Alaryani},\ and\ \citenamefont
  {Brandenberger}}]{Brahma:2020zpk}%
  \BibitemOpen
  \bibfield  {author} {\bibinfo {author} {\bibfnamefont {S.}~\bibnamefont
  {Brahma}}, \bibinfo {author} {\bibfnamefont {O.}~\bibnamefont {Alaryani}}, \
  and\ \bibinfo {author} {\bibfnamefont {R.}~\bibnamefont {Brandenberger}},\
  }\bibfield  {title} {\enquote {\bibinfo {title} {{Entanglement entropy of
  cosmological perturbations}},}\ }\href {\doibase 10.1103/PhysRevD.102.043529}
  {\bibfield  {journal} {\bibinfo  {journal} {Phys. Rev. D}\ }\textbf {\bibinfo
  {volume} {102}},\ \bibinfo {pages} {043529} (\bibinfo {year} {2020})},\
  \Eprint {http://arxiv.org/abs/2005.09688} {arXiv:2005.09688 [hep-th]}
  \BibitemShut {NoStop}%
\bibitem [{\citenamefont {Banerjee}\ \emph {et~al.}(2022)\citenamefont
  {Banerjee}, \citenamefont {Choudhury}, \citenamefont {Chowdhury},
  \citenamefont {Das}, \citenamefont {Gupta}, \citenamefont {Panda},\ and\
  \citenamefont {Swain}}]{Banerjee:2020ljo}%
  \BibitemOpen
  \bibfield  {author} {\bibinfo {author} {\bibfnamefont {S.}~\bibnamefont
  {Banerjee}}, \bibinfo {author} {\bibfnamefont {S.}~\bibnamefont {Choudhury}},
  \bibinfo {author} {\bibfnamefont {S.}~\bibnamefont {Chowdhury}}, \bibinfo
  {author} {\bibfnamefont {R.~N.}\ \bibnamefont {Das}}, \bibinfo {author}
  {\bibfnamefont {N.}~\bibnamefont {Gupta}}, \bibinfo {author} {\bibfnamefont
  {S.}~\bibnamefont {Panda}}, \ and\ \bibinfo {author} {\bibfnamefont
  {A.}~\bibnamefont {Swain}},\ }\bibfield  {title} {\enquote {\bibinfo {title}
  {{Indirect detection of cosmological constant from interacting open quantum
  system}},}\ }\href {\doibase 10.1016/j.aop.2022.168941} {\bibfield  {journal}
  {\bibinfo  {journal} {Annals Phys.}\ }\textbf {\bibinfo {volume} {443}},\
  \bibinfo {pages} {168941} (\bibinfo {year} {2022})},\ \Eprint
  {http://arxiv.org/abs/2004.13058} {arXiv:2004.13058 [hep-th]} \BibitemShut
  {NoStop}%
\bibitem [{\citenamefont {Brahma}\ \emph {et~al.}(2022)\citenamefont {Brahma},
  \citenamefont {Berera},\ and\ \citenamefont
  {Calder\'on-Figueroa}}]{Brahma:2021mng}%
  \BibitemOpen
  \bibfield  {author} {\bibinfo {author} {\bibfnamefont {S.}~\bibnamefont
  {Brahma}}, \bibinfo {author} {\bibfnamefont {A.}~\bibnamefont {Berera}}, \
  and\ \bibinfo {author} {\bibfnamefont {J.}~\bibnamefont
  {Calder\'on-Figueroa}},\ }\bibfield  {title} {\enquote {\bibinfo {title}
  {{Universal signature of quantum entanglement across cosmological
  distances}},}\ }\href {\doibase 10.1088/1361-6382/aca066} {\bibfield
  {journal} {\bibinfo  {journal} {Class. Quant. Grav.}\ }\textbf {\bibinfo
  {volume} {39}},\ \bibinfo {pages} {245002} (\bibinfo {year} {2022})},\
  \Eprint {http://arxiv.org/abs/2107.06910} {arXiv:2107.06910 [hep-th]}
  \BibitemShut {NoStop}%
\bibitem [{\citenamefont {Hsiang}\ and\ \citenamefont
  {Hu}(2022)}]{Hsiang:2021kgh}%
  \BibitemOpen
  \bibfield  {author} {\bibinfo {author} {\bibfnamefont {J.-T.}\ \bibnamefont
  {Hsiang}}\ and\ \bibinfo {author} {\bibfnamefont {B.-L.}\ \bibnamefont
  {Hu}},\ }\bibfield  {title} {\enquote {\bibinfo {title} {{No intrinsic
  decoherence of inflationary cosmological perturbations}},}\ }\href {\doibase
  10.3390/universe8010027} {\bibfield  {journal} {\bibinfo  {journal}
  {Universe}\ }\textbf {\bibinfo {volume} {8}},\ \bibinfo {pages} {27}
  (\bibinfo {year} {2022})},\ \Eprint {http://arxiv.org/abs/2112.04092}
  {arXiv:2112.04092 [gr-qc]} \BibitemShut {NoStop}%
\bibitem [{\citenamefont {Colas}\ \emph {et~al.}(2022)\citenamefont {Colas},
  \citenamefont {Grain},\ and\ \citenamefont {Vennin}}]{Colas:2022hlq}%
  \BibitemOpen
  \bibfield  {author} {\bibinfo {author} {\bibfnamefont {T.}~\bibnamefont
  {Colas}}, \bibinfo {author} {\bibfnamefont {J.}~\bibnamefont {Grain}}, \ and\
  \bibinfo {author} {\bibfnamefont {V.}~\bibnamefont {Vennin}},\ }\bibfield
  {title} {\enquote {\bibinfo {title} {{Benchmarking the cosmological master
  equations}},}\ }\href {\doibase 10.1140/epjc/s10052-022-11047-9} {\bibfield
  {journal} {\bibinfo  {journal} {Eur. Phys. J. C}\ }\textbf {\bibinfo {volume}
  {82}},\ \bibinfo {pages} {1085} (\bibinfo {year} {2022})},\ \Eprint
  {http://arxiv.org/abs/2209.01929} {arXiv:2209.01929 [hep-th]} \BibitemShut
  {NoStop}%
\bibitem [{\citenamefont {Hammou}\ and\ \citenamefont
  {Bartolo}(2023)}]{DaddiHammou:2022itk}%
  \BibitemOpen
  \bibfield  {author} {\bibinfo {author} {\bibfnamefont {A.~D.}\ \bibnamefont
  {Hammou}}\ and\ \bibinfo {author} {\bibfnamefont {N.}~\bibnamefont
  {Bartolo}},\ }\bibfield  {title} {\enquote {\bibinfo {title} {{Cosmic
  decoherence: Primordial power spectra and non-Gaussianities}},}\ }\href
  {\doibase 10.1088/1475-7516/2023/04/055} {\bibfield  {journal} {\bibinfo
  {journal} {JCAP}\ }\textbf {\bibinfo {volume} {04}},\ \bibinfo {pages} {055}
  (\bibinfo {year} {2023})},\ \Eprint {http://arxiv.org/abs/2211.07598}
  {arXiv:2211.07598 [astro-ph.CO]} \BibitemShut {NoStop}%
\bibitem [{\citenamefont {Burgess}\ \emph {et~al.}(2023)\citenamefont
  {Burgess}, \citenamefont {Holman}, \citenamefont {Kaplanek}, \citenamefont
  {Martin},\ and\ \citenamefont {Vennin}}]{Burgess:2022nwu}%
  \BibitemOpen
  \bibfield  {author} {\bibinfo {author} {\bibfnamefont {C.~P.}\ \bibnamefont
  {Burgess}}, \bibinfo {author} {\bibfnamefont {R.}~\bibnamefont {Holman}},
  \bibinfo {author} {\bibfnamefont {G.}~\bibnamefont {Kaplanek}}, \bibinfo
  {author} {\bibfnamefont {J.}~\bibnamefont {Martin}}, \ and\ \bibinfo {author}
  {\bibfnamefont {V.}~\bibnamefont {Vennin}},\ }\bibfield  {title} {\enquote
  {\bibinfo {title} {{Minimal decoherence from inflation}},}\ }\href {\doibase
  10.1088/1475-7516/2023/07/022} {\bibfield  {journal} {\bibinfo  {journal}
  {JCAP}\ }\textbf {\bibinfo {volume} {07}},\ \bibinfo {pages} {022} (\bibinfo
  {year} {2023})},\ \Eprint {http://arxiv.org/abs/2211.11046} {arXiv:2211.11046
  [hep-th]} \BibitemShut {NoStop}%
\bibitem [{\citenamefont {Colas}\ \emph {et~al.}(2023)\citenamefont {Colas},
  \citenamefont {Grain},\ and\ \citenamefont {Vennin}}]{Colas:2022kfu}%
  \BibitemOpen
  \bibfield  {author} {\bibinfo {author} {\bibfnamefont {T.}~\bibnamefont
  {Colas}}, \bibinfo {author} {\bibfnamefont {J.}~\bibnamefont {Grain}}, \ and\
  \bibinfo {author} {\bibfnamefont {V.}~\bibnamefont {Vennin}},\ }\bibfield
  {title} {\enquote {\bibinfo {title} {{Quantum recoherence in the early
  Universe}},}\ }\href {\doibase 10.1209/0295-5075/acdd94} {\bibfield
  {journal} {\bibinfo  {journal} {Europhys. Lett.}\ }\textbf {\bibinfo {volume}
  {142}},\ \bibinfo {pages} {69002} (\bibinfo {year} {2023})},\ \Eprint
  {http://arxiv.org/abs/2212.09486} {arXiv:2212.09486 [gr-qc]} \BibitemShut
  {NoStop}%
\bibitem [{\citenamefont {Burgess}\ \emph {et~al.}(2024)\citenamefont
  {Burgess}, \citenamefont {Colas}, \citenamefont {Holman}, \citenamefont
  {Kaplanek},\ and\ \citenamefont {Vennin}}]{Burgess:2024eng}%
  \BibitemOpen
  \bibfield  {author} {\bibinfo {author} {\bibfnamefont {C.~P.}\ \bibnamefont
  {Burgess}}, \bibinfo {author} {\bibfnamefont {Thomas}\ \bibnamefont {Colas}},
  \bibinfo {author} {\bibfnamefont {R.}~\bibnamefont {Holman}}, \bibinfo
  {author} {\bibfnamefont {Greg}\ \bibnamefont {Kaplanek}}, \ and\ \bibinfo
  {author} {\bibfnamefont {Vincent}\ \bibnamefont {Vennin}},\ }\bibfield
  {title} {\enquote {\bibinfo {title} {{Cosmic purity lost: perturbative and
  resummed late-time inflationary decoherence}},}\ }\href {\doibase
  10.1088/1475-7516/2024/08/042} {\bibfield  {journal} {\bibinfo  {journal}
  {JCAP}\ }\textbf {\bibinfo {volume} {08}},\ \bibinfo {pages} {042} (\bibinfo
  {year} {2024})},\ \Eprint {http://arxiv.org/abs/2403.12240} {arXiv:2403.12240
  [gr-qc]} \BibitemShut {NoStop}%
\bibitem [{\citenamefont {Salcedo}\ \emph {et~al.}(2024)\citenamefont
  {Salcedo}, \citenamefont {Colas},\ and\ \citenamefont
  {Pajer}}]{Salcedo:2024smn}%
  \BibitemOpen
  \bibfield  {author} {\bibinfo {author} {\bibfnamefont {S.~A.}\ \bibnamefont
  {Salcedo}}, \bibinfo {author} {\bibfnamefont {T.}~\bibnamefont {Colas}}, \
  and\ \bibinfo {author} {\bibfnamefont {E.}~\bibnamefont {Pajer}},\ }\bibfield
   {title} {\enquote {\bibinfo {title} {{The open effective field theory of
  inflation}},}\ }\href {\doibase 10.1007/JHEP10(2024)248} {\bibfield
  {journal} {\bibinfo  {journal} {JHEP}\ }\textbf {\bibinfo {volume} {10}},\
  \bibinfo {pages} {248} (\bibinfo {year} {2024})},\ \Eprint
  {http://arxiv.org/abs/2404.15416} {arXiv:2404.15416 [hep-th]} \BibitemShut
  {NoStop}%
\bibitem [{\citenamefont {Brahma}\ \emph {et~al.}(2025)\citenamefont {Brahma},
  \citenamefont {Calder{\'o}n-Figueroa}, \citenamefont {Luo},\ and\
  \citenamefont {Seery}}]{Brahma:2024ycc}%
  \BibitemOpen
  \bibfield  {author} {\bibinfo {author} {\bibfnamefont {S.}~\bibnamefont
  {Brahma}}, \bibinfo {author} {\bibfnamefont {J.}~\bibnamefont
  {Calder{\'o}n-Figueroa}}, \bibinfo {author} {\bibfnamefont {X.}~\bibnamefont
  {Luo}}, \ and\ \bibinfo {author} {\bibfnamefont {D.}~\bibnamefont {Seery}},\
  }\bibfield  {title} {\enquote {\bibinfo {title} {{The special case of
  slow-roll attractors in de~Sitter: non-Markovian noise and evolution of
  entanglement entropy}},}\ }\href {\doibase 10.1088/1475-7516/2025/04/050}
  {\bibfield  {journal} {\bibinfo  {journal} {JCAP}\ }\textbf {\bibinfo
  {volume} {04}},\ \bibinfo {pages} {050} (\bibinfo {year} {2025})},\ \Eprint
  {http://arxiv.org/abs/2411.08632} {arXiv:2411.08632 [hep-th]} \BibitemShut
  {NoStop}%
\bibitem [{\citenamefont {Lopez}\ and\ \citenamefont
  {Bartolo}(2025)}]{Lopez:2025arw}%
  \BibitemOpen
  \bibfield  {author} {\bibinfo {author} {\bibfnamefont {F.}~\bibnamefont
  {Lopez}}\ and\ \bibinfo {author} {\bibfnamefont {N.}~\bibnamefont
  {Bartolo}},\ }\bibfield  {title} {\enquote {\bibinfo {title} {{Quantum
  signatures and decoherence during inflation from deep subhorizon
  perturbations}},}\ }\href@noop {} {\  (\bibinfo {year} {2025})},\ \Eprint
  {http://arxiv.org/abs/2503.23150} {arXiv:2503.23150 [astro-ph.CO]}
  \BibitemShut {NoStop}%
\bibitem [{\citenamefont {Li}(2025)}]{Li:2025azq}%
  \BibitemOpen
  \bibfield  {author} {\bibinfo {author} {\bibfnamefont {Y.-Z.}\ \bibnamefont
  {Li}},\ }\bibfield  {title} {\enquote {\bibinfo {title} {{Stochastic
  inflation as an open quantum system}},}\ }\href@noop {} {\  (\bibinfo {year}
  {2025})},\ \Eprint {http://arxiv.org/abs/2507.02070} {arXiv:2507.02070
  [hep-th]} \BibitemShut {NoStop}%
\bibitem [{\citenamefont {Cao}\ and\ \citenamefont
  {Boyanovsky}(2023)}]{Cao:2022kjn}%
  \BibitemOpen
  \bibfield  {author} {\bibinfo {author} {\bibfnamefont {S.}~\bibnamefont
  {Cao}}\ and\ \bibinfo {author} {\bibfnamefont {D.}~\bibnamefont
  {Boyanovsky}},\ }\bibfield  {title} {\enquote {\bibinfo {title}
  {{Nonequilibrium dynamics of axionlike particles: The quantum master
  equation}},}\ }\href {\doibase 10.1103/PhysRevD.107.063518} {\bibfield
  {journal} {\bibinfo  {journal} {Phys. Rev. D}\ }\textbf {\bibinfo {volume}
  {107}},\ \bibinfo {pages} {063518} (\bibinfo {year} {2023})},\ \Eprint
  {http://arxiv.org/abs/2212.05161} {arXiv:2212.05161 [astro-ph.CO]}
  \BibitemShut {NoStop}%
\bibitem [{\citenamefont {K\"ading}\ \emph {et~al.}(2023)\citenamefont
  {K\"ading}, \citenamefont {Pitschmann},\ and\ \citenamefont
  {Voith}}]{Kading:2023mdk}%
  \BibitemOpen
  \bibfield  {author} {\bibinfo {author} {\bibfnamefont {C.}~\bibnamefont
  {K\"ading}}, \bibinfo {author} {\bibfnamefont {M.}~\bibnamefont
  {Pitschmann}}, \ and\ \bibinfo {author} {\bibfnamefont {C.}~\bibnamefont
  {Voith}},\ }\bibfield  {title} {\enquote {\bibinfo {title} {{Dilaton-induced
  open quantum dynamics}},}\ }\href {\doibase 10.1140/epjc/s10052-023-11939-4}
  {\bibfield  {journal} {\bibinfo  {journal} {Eur. Phys. J. C}\ }\textbf
  {\bibinfo {volume} {83}},\ \bibinfo {pages} {767} (\bibinfo {year} {2023})},\
  \Eprint {http://arxiv.org/abs/2306.10896} {arXiv:2306.10896 [hep-ph]}
  \BibitemShut {NoStop}%
\bibitem [{\citenamefont {Bassi}\ \emph {et~al.}(2017)\citenamefont {Bassi},
  \citenamefont {Gro\ss{}ardt},\ and\ \citenamefont
  {Ulbricht}}]{Bassi:2017szd}%
  \BibitemOpen
  \bibfield  {author} {\bibinfo {author} {\bibfnamefont {A.}~\bibnamefont
  {Bassi}}, \bibinfo {author} {\bibfnamefont {A.}~\bibnamefont {Gro\ss{}ardt}},
  \ and\ \bibinfo {author} {\bibfnamefont {H.}~\bibnamefont {Ulbricht}},\
  }\bibfield  {title} {\enquote {\bibinfo {title} {{Gravitational
  decoherence}},}\ }\href {\doibase 10.1088/1361-6382/aa864f} {\bibfield
  {journal} {\bibinfo  {journal} {Class. Quant. Grav.}\ }\textbf {\bibinfo
  {volume} {34}},\ \bibinfo {pages} {193002} (\bibinfo {year} {2017})},\
  \Eprint {http://arxiv.org/abs/1706.05677} {arXiv:1706.05677 [quant-ph]}
  \BibitemShut {NoStop}%
\bibitem [{\citenamefont {Sharifian}\ \emph {et~al.}(2024)\citenamefont
  {Sharifian}, \citenamefont {Zarei}, \citenamefont {Abdi}, \citenamefont
  {Bartolo},\ and\ \citenamefont {Matarrese}}]{Sharifian:2023jem}%
  \BibitemOpen
  \bibfield  {author} {\bibinfo {author} {\bibfnamefont {M.}~\bibnamefont
  {Sharifian}}, \bibinfo {author} {\bibfnamefont {M.}~\bibnamefont {Zarei}},
  \bibinfo {author} {\bibfnamefont {M.}~\bibnamefont {Abdi}}, \bibinfo {author}
  {\bibfnamefont {N.}~\bibnamefont {Bartolo}}, \ and\ \bibinfo {author}
  {\bibfnamefont {S.}~\bibnamefont {Matarrese}},\ }\bibfield  {title} {\enquote
  {\bibinfo {title} {{Open quantum system approach to the gravitational
  decoherence of spin-1/2 particles}},}\ }\href {\doibase
  10.1103/PhysRevD.109.043510} {\bibfield  {journal} {\bibinfo  {journal}
  {Phys. Rev. D}\ }\textbf {\bibinfo {volume} {109}},\ \bibinfo {pages}
  {043510} (\bibinfo {year} {2024})},\ \Eprint
  {http://arxiv.org/abs/2309.07236} {arXiv:2309.07236 [gr-qc]} \BibitemShut
  {NoStop}%
\bibitem [{\citenamefont {Zarei}\ \emph {et~al.}(2021)\citenamefont {Zarei},
  \citenamefont {Bartolo}, \citenamefont {Bertacca}, \citenamefont
  {Ricciardone},\ and\ \citenamefont {Matarrese}}]{Zarei:2021dpb}%
  \BibitemOpen
  \bibfield  {author} {\bibinfo {author} {\bibfnamefont {M.}~\bibnamefont
  {Zarei}}, \bibinfo {author} {\bibfnamefont {N.}~\bibnamefont {Bartolo}},
  \bibinfo {author} {\bibfnamefont {D.}~\bibnamefont {Bertacca}}, \bibinfo
  {author} {\bibfnamefont {A.}~\bibnamefont {Ricciardone}}, \ and\ \bibinfo
  {author} {\bibfnamefont {S.}~\bibnamefont {Matarrese}},\ }\bibfield  {title}
  {\enquote {\bibinfo {title} {{Non-Markovian open quantum system approach to
  the early Universe: Damping of gravitational waves by matter}},}\ }\href
  {\doibase 10.1103/PhysRevD.104.083508} {\bibfield  {journal} {\bibinfo
  {journal} {Phys. Rev. D}\ }\textbf {\bibinfo {volume} {104}},\ \bibinfo
  {pages} {083508} (\bibinfo {year} {2021})},\ \Eprint
  {http://arxiv.org/abs/2104.04836} {arXiv:2104.04836 [astro-ph.CO]}
  \BibitemShut {NoStop}%
\bibitem [{\citenamefont {Kaplanek}\ and\ \citenamefont
  {Burgess}(2020{\natexlab{a}})}]{Kaplanek:2019dqu}%
  \BibitemOpen
  \bibfield  {author} {\bibinfo {author} {\bibfnamefont {G.}~\bibnamefont
  {Kaplanek}}\ and\ \bibinfo {author} {\bibfnamefont {C.~P.}\ \bibnamefont
  {Burgess}},\ }\bibfield  {title} {\enquote {\bibinfo {title} {{Hot
  accelerated qubits: Decoherence, thermalization, secular growth and reliable
  late-time predictions}},}\ }\href {\doibase 10.1007/JHEP03(2020)008}
  {\bibfield  {journal} {\bibinfo  {journal} {JHEP}\ }\textbf {\bibinfo
  {volume} {03}},\ \bibinfo {pages} {008} (\bibinfo {year}
  {2020}{\natexlab{a}})},\ \Eprint {http://arxiv.org/abs/1912.12951}
  {arXiv:1912.12951 [hep-th]} \BibitemShut {NoStop}%
\bibitem [{\citenamefont {Kaplanek}\ and\ \citenamefont
  {Burgess}(2020{\natexlab{b}})}]{Kaplanek:2019vzj}%
  \BibitemOpen
  \bibfield  {author} {\bibinfo {author} {\bibfnamefont {G.}~\bibnamefont
  {Kaplanek}}\ and\ \bibinfo {author} {\bibfnamefont {C.~P.}\ \bibnamefont
  {Burgess}},\ }\bibfield  {title} {\enquote {\bibinfo {title} {{Hot cosmic
  qubits: Late-time de Sitter evolution and critical slowing down}},}\ }\href
  {\doibase 10.1007/JHEP02(2020)053} {\bibfield  {journal} {\bibinfo  {journal}
  {JHEP}\ }\textbf {\bibinfo {volume} {02}},\ \bibinfo {pages} {053} (\bibinfo
  {year} {2020}{\natexlab{b}})},\ \Eprint {http://arxiv.org/abs/1912.12955}
  {arXiv:1912.12955 [hep-th]} \BibitemShut {NoStop}%
\bibitem [{\citenamefont {Kaplanek}\ and\ \citenamefont
  {Tjoa}(2023)}]{Kaplanek:2022xrr}%
  \BibitemOpen
  \bibfield  {author} {\bibinfo {author} {\bibfnamefont {G.}~\bibnamefont
  {Kaplanek}}\ and\ \bibinfo {author} {\bibfnamefont {E.}~\bibnamefont
  {Tjoa}},\ }\bibfield  {title} {\enquote {\bibinfo {title} {{Effective master
  equations for two accelerated qubits}},}\ }\href {\doibase
  10.1103/PhysRevA.107.012208} {\bibfield  {journal} {\bibinfo  {journal}
  {Phys. Rev. A}\ }\textbf {\bibinfo {volume} {107}},\ \bibinfo {pages}
  {012208} (\bibinfo {year} {2023})},\ \Eprint
  {http://arxiv.org/abs/2207.13750} {arXiv:2207.13750 [quant-ph]} \BibitemShut
  {NoStop}%
\bibitem [{\citenamefont {Jana}\ \emph {et~al.}(2020)\citenamefont {Jana},
  \citenamefont {Loganayagam},\ and\ \citenamefont {Rangamani}}]{Jana:2020vyx}%
  \BibitemOpen
  \bibfield  {author} {\bibinfo {author} {\bibfnamefont {C.}~\bibnamefont
  {Jana}}, \bibinfo {author} {\bibfnamefont {R.}~\bibnamefont {Loganayagam}}, \
  and\ \bibinfo {author} {\bibfnamefont {M.}~\bibnamefont {Rangamani}},\
  }\bibfield  {title} {\enquote {\bibinfo {title} {{Open quantum systems and
  Schwinger-Keldysh holograms}},}\ }\href {\doibase 10.1007/JHEP07(2020)242}
  {\bibfield  {journal} {\bibinfo  {journal} {JHEP}\ }\textbf {\bibinfo
  {volume} {07}},\ \bibinfo {pages} {242} (\bibinfo {year} {2020})},\ \Eprint
  {http://arxiv.org/abs/2004.02888} {arXiv:2004.02888 [hep-th]} \BibitemShut
  {NoStop}%
\bibitem [{\citenamefont {Loganayagam}\ \emph {et~al.}(2020)\citenamefont
  {Loganayagam}, \citenamefont {Ray},\ and\ \citenamefont
  {Sivakumar}}]{Loganayagam:2020eue}%
  \BibitemOpen
  \bibfield  {author} {\bibinfo {author} {\bibfnamefont {R.}~\bibnamefont
  {Loganayagam}}, \bibinfo {author} {\bibfnamefont {K.}~\bibnamefont {Ray}}, \
  and\ \bibinfo {author} {\bibfnamefont {A.}~\bibnamefont {Sivakumar}},\
  }\bibfield  {title} {\enquote {\bibinfo {title} {{Fermionic open EFT from
  holography}},}\ }\href@noop {} {\  (\bibinfo {year} {2020})},\ \Eprint
  {http://arxiv.org/abs/2011.07039} {arXiv:2011.07039 [hep-th]} \BibitemShut
  {NoStop}%
\bibitem [{\citenamefont {Loganayagam}\ \emph {et~al.}(2023)\citenamefont
  {Loganayagam}, \citenamefont {Rangamani},\ and\ \citenamefont
  {Virrueta}}]{Loganayagam:2022zmq}%
  \BibitemOpen
  \bibfield  {author} {\bibinfo {author} {\bibfnamefont {R.}~\bibnamefont
  {Loganayagam}}, \bibinfo {author} {\bibfnamefont {M.}~\bibnamefont
  {Rangamani}}, \ and\ \bibinfo {author} {\bibfnamefont {J.}~\bibnamefont
  {Virrueta}},\ }\bibfield  {title} {\enquote {\bibinfo {title} {{Holographic
  open quantum systems: toy models and analytic properties of thermal
  correlators}},}\ }\href {\doibase 10.1007/JHEP03(2023)153} {\bibfield
  {journal} {\bibinfo  {journal} {JHEP}\ }\textbf {\bibinfo {volume} {03}},\
  \bibinfo {pages} {153} (\bibinfo {year} {2023})},\ \Eprint
  {http://arxiv.org/abs/2211.07683} {arXiv:2211.07683 [hep-th]} \BibitemShut
  {NoStop}%
\bibitem [{\citenamefont {Pelliconi}\ and\ \citenamefont
  {Sonner}(2024)}]{Pelliconi:2023ojb}%
  \BibitemOpen
  \bibfield  {author} {\bibinfo {author} {\bibfnamefont {P.}~\bibnamefont
  {Pelliconi}}\ and\ \bibinfo {author} {\bibfnamefont {J.}~\bibnamefont
  {Sonner}},\ }\bibfield  {title} {\enquote {\bibinfo {title} {{The influence
  functional in open holography: Entanglement and R\'enyi entropies}},}\ }\href
  {\doibase 10.1007/JHEP06(2024)185} {\bibfield  {journal} {\bibinfo  {journal}
  {JHEP}\ }\textbf {\bibinfo {volume} {06}},\ \bibinfo {pages} {185} (\bibinfo
  {year} {2024})},\ \Eprint {http://arxiv.org/abs/2310.13047} {arXiv:2310.13047
  [hep-th]} \BibitemShut {NoStop}%
\bibitem [{\citenamefont {Srednicki}({2007})}]{Srednicki:2007qs}%
  \BibitemOpen
  \bibfield  {author} {\bibinfo {author} {\bibfnamefont {M.}~\bibnamefont
  {Srednicki}},\ }\href@noop {} {\emph {\bibinfo {title} {{Quantum field
  theory}}}}\ (\bibinfo  {publisher} {{Cambridge University Press}},\ \bibinfo
  {year} {{2007}})\BibitemShut {NoStop}%
\bibitem [{\citenamefont {Carmichael}({1999})}]{carmichael1999statistical}%
  \BibitemOpen
  \bibfield  {author} {\bibinfo {author} {\bibfnamefont {H.~J.}\ \bibnamefont
  {Carmichael}},\ }\href@noop {} {\emph {\bibinfo {title} {{Statistical methods
  in quantum optics 1: Master equations and Fokker-Planck equations}}}}\
  (\bibinfo  {publisher} {{Springer}},\ \bibinfo {year} {{1999}})\BibitemShut
  {NoStop}%
\bibitem [{\citenamefont {Breuer}\ and\ \citenamefont
  {Petruccione}({2002})}]{Breuer:2002pc}%
  \BibitemOpen
  \bibfield  {author} {\bibinfo {author} {\bibfnamefont {H.-P.}\ \bibnamefont
  {Breuer}}\ and\ \bibinfo {author} {\bibfnamefont {F.}~\bibnamefont
  {Petruccione}},\ }\href@noop {} {\emph {\bibinfo {title} {{The theory of open
  quantum systems}}}}\ (\bibinfo  {publisher} {{Oxford University Press}},\
  \bibinfo {year} {{2002}})\BibitemShut {NoStop}%
\bibitem [{\citenamefont {Schlosshauer}({2007})}]{Schlosshauer2007}%
  \BibitemOpen
  \bibfield  {author} {\bibinfo {author} {\bibfnamefont {M.~A.}\ \bibnamefont
  {Schlosshauer}},\ }\href@noop {} {\emph {\bibinfo {title} {{Decoherence and
  the quantum-to-classical transition}}}}\ (\bibinfo  {publisher}
  {{Springer-Verlag Berlin Heidelberg}},\ \bibinfo {year} {{2007}})\BibitemShut
  {NoStop}%
\bibitem [{\citenamefont {Joos}\ \emph {et~al.}(2003)\citenamefont {Joos},
  \citenamefont {Zeh}, \citenamefont {Kiefer}, \citenamefont {Giulini},
  \citenamefont {Kupsch},\ and\ \citenamefont {Stamatescu}}]{Giulini:1996nw}%
  \BibitemOpen
  \bibfield  {author} {\bibinfo {author} {\bibfnamefont {E.}~\bibnamefont
  {Joos}}, \bibinfo {author} {\bibfnamefont {H.~D.}\ \bibnamefont {Zeh}},
  \bibinfo {author} {\bibfnamefont {C.}~\bibnamefont {Kiefer}}, \bibinfo
  {author} {\bibfnamefont {D.}~\bibnamefont {Giulini}}, \bibinfo {author}
  {\bibfnamefont {J.}~\bibnamefont {Kupsch}}, \ and\ \bibinfo {author}
  {\bibfnamefont {I.~O.}\ \bibnamefont {Stamatescu}},\ }\href@noop {} {\emph
  {\bibinfo {title} {{Decoherence and the appearance of a classical world in
  quantum theory}}}}\ (\bibinfo  {publisher} {{Springer}},\ \bibinfo {year}
  {2003})\BibitemShut {NoStop}%
\bibitem [{Note2()}]{Note2}%
  \BibitemOpen
  \bibinfo {note} {We will see in \protect \cref {sec:phichi,sec:phichi2} that
  for both interactions we consider, this environment correlator only depends
  on the magnitude $ k $.}\BibitemShut {Stop}%
\bibitem [{Note3()}]{Note3}%
  \BibitemOpen
  \bibinfo {note} {Also see ref.\ \cite {Prudhoe:2022pte} for an effective
  time-local master equation construction and ref.\ \cite {Agarwal:2023lid} for
  an effective field theory-inspired approach.}\BibitemShut {Stop}%
\bibitem [{Note4()}]{Note4}%
  \BibitemOpen
  \bibinfo {note} {Since we have used definite integrals in the integration by
  parts, there are terms that depend on the initial time $ t_{0} $. It is
  straightforward to show, however, that these terms cancel
  identically.}\BibitemShut {Stop}%
\bibitem [{Note5()}]{Note5}%
  \BibitemOpen
  \bibinfo {note} {Since we are working at $ O(\lambda ^{2}) $, only tree-level
  diagrams contribute to the environment correlation function. This will be
  convenient when we renormalize the $ \lambda \phi \chi ^{2} $ theory in
  \protect \cref {sec:phichi2}.}\BibitemShut {Stop}%
\bibitem [{Note6()}]{Note6}%
  \BibitemOpen
  \bibinfo {note} {This qualitative equivalence between quantum master
  equations expressed in Schr{\"o}dinger and interaction pictures does not hold
  when the free evolution of the system itself is nonunitary \cite
  {Thorbeck2024}.}\BibitemShut {Stop}%
\bibitem [{\citenamefont {Martin}\ and\ \citenamefont
  {Vennin}(2018)}]{Martin:2018lin}%
  \BibitemOpen
  \bibfield  {author} {\bibinfo {author} {\bibfnamefont {J.}~\bibnamefont
  {Martin}}\ and\ \bibinfo {author} {\bibfnamefont {V.}~\bibnamefont
  {Vennin}},\ }\bibfield  {title} {\enquote {\bibinfo {title}
  {{Non-Gaussianities from quantum decoherence during inflation}},}\ }\href
  {\doibase 10.1088/1475-7516/2018/06/037} {\bibfield  {journal} {\bibinfo
  {journal} {JCAP}\ }\textbf {\bibinfo {volume} {06}},\ \bibinfo {pages} {037}
  (\bibinfo {year} {2018})},\ \Eprint {http://arxiv.org/abs/1805.05609}
  {arXiv:1805.05609 [astro-ph.CO]} \BibitemShut {NoStop}%
\bibitem [{Note7()}]{Note7}%
  \BibitemOpen
  \bibinfo {note} {We never have to explicitly choose a value for $ M m $ since
  the explicit dependence on $ M $ and $ m $ appears as their ratio for this
  choice of parameters.}\BibitemShut {Stop}%
\bibitem [{\citenamefont {Peskin}\ and\ \citenamefont
  {Schroeder}({1995})}]{Peskin:1995ev}%
  \BibitemOpen
  \bibfield  {author} {\bibinfo {author} {\bibfnamefont {M.~E.}\ \bibnamefont
  {Peskin}}\ and\ \bibinfo {author} {\bibfnamefont {D.~V.}\ \bibnamefont
  {Schroeder}},\ }\href@noop {} {\emph {\bibinfo {title} {{An introduction to
  quantum field theory}}}}\ (\bibinfo  {publisher} {{Addison-Wesley}},\
  \bibinfo {address} {{Reading, USA}},\ \bibinfo {year} {{1995}})\BibitemShut
  {NoStop}%
\bibitem [{\citenamefont {Calzetta}\ and\ \citenamefont
  {Hu}(2008)}]{Calzetta:2008}%
  \BibitemOpen
  \bibfield  {author} {\bibinfo {author} {\bibfnamefont {E.~A.}\ \bibnamefont
  {Calzetta}}\ and\ \bibinfo {author} {\bibfnamefont {B.-L.~B.}\ \bibnamefont
  {Hu}},\ }\href@noop {} {\emph {\bibinfo {title} {{Nonequilibrium quantum
  field theory}}}}\ (\bibinfo  {publisher} {\textnormal{Cambridge University
  Press}},\ \bibinfo {year} {2008})\BibitemShut {NoStop}%
\bibitem [{\citenamefont {Baacke}\ \emph {et~al.}(1998)\citenamefont {Baacke},
  \citenamefont {Heitmann},\ and\ \citenamefont {Patzold}}]{Baacke:1997zz}%
  \BibitemOpen
  \bibfield  {author} {\bibinfo {author} {\bibfnamefont {J.}~\bibnamefont
  {Baacke}}, \bibinfo {author} {\bibfnamefont {K.}~\bibnamefont {Heitmann}}, \
  and\ \bibinfo {author} {\bibfnamefont {C.}~\bibnamefont {Patzold}},\
  }\bibfield  {title} {\enquote {\bibinfo {title} {{On the choice of initial
  states in nonequilibrium dynamics}},}\ }\href {\doibase
  10.1103/PhysRevD.57.6398} {\bibfield  {journal} {\bibinfo  {journal} {Phys.
  Rev. D}\ }\textbf {\bibinfo {volume} {57}},\ \bibinfo {pages} {6398--6405}
  (\bibinfo {year} {1998})},\ \Eprint {http://arxiv.org/abs/hep-th/9711144}
  {arXiv:hep-th/9711144} \BibitemShut {NoStop}%
\bibitem [{\citenamefont {Baacke}\ \emph {et~al.}(2001)\citenamefont {Baacke},
  \citenamefont {Boyanovsky},\ and\ \citenamefont {de~Vega}}]{Baacke:1999ia}%
  \BibitemOpen
  \bibfield  {author} {\bibinfo {author} {\bibfnamefont {J.}~\bibnamefont
  {Baacke}}, \bibinfo {author} {\bibfnamefont {D.}~\bibnamefont {Boyanovsky}},
  \ and\ \bibinfo {author} {\bibfnamefont {H.~J.}\ \bibnamefont {de~Vega}},\
  }\bibfield  {title} {\enquote {\bibinfo {title} {{Initial time singularities
  in nonequilibrium evolution of condensates and their resolution in the
  linearized approximation}},}\ }\href {\doibase 10.1103/PhysRevD.63.045023}
  {\bibfield  {journal} {\bibinfo  {journal} {Phys. Rev. D}\ }\textbf {\bibinfo
  {volume} {63}},\ \bibinfo {pages} {045023} (\bibinfo {year} {2001})},\
  \Eprint {http://arxiv.org/abs/hep-ph/9907337} {arXiv:hep-ph/9907337}
  \BibitemShut {NoStop}%
\bibitem [{\citenamefont {Collins}\ and\ \citenamefont
  {Holman}(2005)}]{Collins:2005nu}%
  \BibitemOpen
  \bibfield  {author} {\bibinfo {author} {\bibfnamefont {H.}~\bibnamefont
  {Collins}}\ and\ \bibinfo {author} {\bibfnamefont {R.}~\bibnamefont
  {Holman}},\ }\bibfield  {title} {\enquote {\bibinfo {title} {{Renormalization
  of initial conditions and the trans-Planckian problem of inflation}},}\
  }\href {\doibase 10.1103/PhysRevD.71.085009} {\bibfield  {journal} {\bibinfo
  {journal} {Phys. Rev. D}\ }\textbf {\bibinfo {volume} {71}},\ \bibinfo
  {pages} {085009} (\bibinfo {year} {2005})},\ \Eprint
  {http://arxiv.org/abs/hep-th/0501158} {arXiv:hep-th/0501158} \BibitemShut
  {NoStop}%
\bibitem [{\citenamefont {Collins}\ \emph {et~al.}(2014)\citenamefont
  {Collins}, \citenamefont {Holman},\ and\ \citenamefont
  {Vardanyan}}]{Collins:2014qna}%
  \BibitemOpen
  \bibfield  {author} {\bibinfo {author} {\bibfnamefont {H.}~\bibnamefont
  {Collins}}, \bibinfo {author} {\bibfnamefont {R.}~\bibnamefont {Holman}}, \
  and\ \bibinfo {author} {\bibfnamefont {T.}~\bibnamefont {Vardanyan}},\
  }\bibfield  {title} {\enquote {\bibinfo {title} {{Renormalizing an initial
  state}},}\ }\href {\doibase 10.1007/JHEP10(2014)124} {\bibfield  {journal}
  {\bibinfo  {journal} {JHEP}\ }\textbf {\bibinfo {volume} {10}},\ \bibinfo
  {pages} {124} (\bibinfo {year} {2014})},\ \Eprint
  {http://arxiv.org/abs/1408.4801} {arXiv:1408.4801 [hep-th]} \BibitemShut
  {NoStop}%
\bibitem [{\citenamefont {Chaykov}\ \emph {et~al.}(2023)\citenamefont
  {Chaykov}, \citenamefont {Agarwal}, \citenamefont {Bahrami},\ and\
  \citenamefont {Holman}}]{Chaykov:2022pwd}%
  \BibitemOpen
  \bibfield  {author} {\bibinfo {author} {\bibfnamefont {S.}~\bibnamefont
  {Chaykov}}, \bibinfo {author} {\bibfnamefont {N.}~\bibnamefont {Agarwal}},
  \bibinfo {author} {\bibfnamefont {S.}~\bibnamefont {Bahrami}}, \ and\
  \bibinfo {author} {\bibfnamefont {R.}~\bibnamefont {Holman}},\ }\bibfield
  {title} {\enquote {\bibinfo {title} {{Loop corrections in Minkowski spacetime
  away from equilibrium. Part II. Finite-time results}},}\ }\href {\doibase
  10.1007/JHEP02(2023)094} {\bibfield  {journal} {\bibinfo  {journal} {JHEP}\
  }\textbf {\bibinfo {volume} {02}},\ \bibinfo {pages} {094} (\bibinfo {year}
  {2023})},\ \Eprint {http://arxiv.org/abs/2206.11289} {arXiv:2206.11289
  [hep-th]} \BibitemShut {NoStop}%
\bibitem [{Note8()}]{Note8}%
  \BibitemOpen
  \bibinfo {note} {We note that one could obtain the same final results using
  the dimensional regularization procedure of ref.~\cite {Chaykov:2022pwd}. We
  have instead used the $i\epsilon $ regularization scheme here as it is more
  instructive since the delta function piece of the environment correlation
  function emerges naturally.}\BibitemShut {Stop}%
\bibitem [{\citenamefont {Caldeira}\ and\ \citenamefont
  {Leggett}(1981)}]{Caldeira:1981}%
  \BibitemOpen
  \bibfield  {author} {\bibinfo {author} {\bibfnamefont {A.~O.}\ \bibnamefont
  {Caldeira}}\ and\ \bibinfo {author} {\bibfnamefont {A.~J.}\ \bibnamefont
  {Leggett}},\ }\bibfield  {title} {\enquote {\bibinfo {title} {{Influence of
  dissipation on quantum tunneling in macroscopic systems}},}\ }\href {\doibase
  10.1103/PhysRevLett.46.211} {\bibfield  {journal} {\bibinfo  {journal} {Phys.
  Rev. Lett.}\ }\textbf {\bibinfo {volume} {46}},\ \bibinfo {pages} {211}
  (\bibinfo {year} {1981})}\BibitemShut {NoStop}%
\bibitem [{\citenamefont {Caldeira}\ and\ \citenamefont
  {Leggett}(1983)}]{Caldeira:1982iu}%
  \BibitemOpen
  \bibfield  {author} {\bibinfo {author} {\bibfnamefont {A.~O.}\ \bibnamefont
  {Caldeira}}\ and\ \bibinfo {author} {\bibfnamefont {A.~J.}\ \bibnamefont
  {Leggett}},\ }\bibfield  {title} {\enquote {\bibinfo {title} {{Path integral
  approach to quantum Brownian motion}},}\ }\href {\doibase
  10.1016/0378-4371(83)90013-4} {\bibfield  {journal} {\bibinfo  {journal}
  {Physica A}\ }\textbf {\bibinfo {volume} {121}},\ \bibinfo {pages} {587--616}
  (\bibinfo {year} {1983})}\BibitemShut {NoStop}%
\bibitem [{\citenamefont {Agarwal}({1974})}]{Agarwal1974}%
  \BibitemOpen
  \bibfield  {author} {\bibinfo {author} {\bibfnamefont {G.~S.}\ \bibnamefont
  {Agarwal}},\ }\enquote {\bibinfo {title} {{Quantum statistical theories of
  spontaneous emission and their relation to other approaches}},}\ in\ \href
  {\doibase 10.1007/BFb0042382} {\emph {\bibinfo {booktitle} {{Quantum
  Optics}}}}\ (\bibinfo  {publisher} {{Springer Berlin Heidelberg}},\ \bibinfo
  {year} {{1974}})\ pp.\ \bibinfo {pages} {1--128}\BibitemShut {NoStop}%
\bibitem [{\citenamefont {Timofeev}\ and\ \citenamefont
  {Trushechkin}(2022)}]{Timofeev:2022tbl}%
  \BibitemOpen
  \bibfield  {author} {\bibinfo {author} {\bibfnamefont {G.}~\bibnamefont
  {Timofeev}}\ and\ \bibinfo {author} {\bibfnamefont {A.}~\bibnamefont
  {Trushechkin}},\ }\bibfield  {title} {\enquote {\bibinfo {title}
  {{Hamiltonian of mean force in the weak-coupling and high-temperature
  approximations and refined quantum master equations}},}\ }\href {\doibase
  10.1142/S0217751X22430217} {\bibfield  {journal} {\bibinfo  {journal} {Int.
  J. Mod. Phys. A}\ }\textbf {\bibinfo {volume} {37}},\ \bibinfo {pages}
  {2243021} (\bibinfo {year} {2022})},\ \Eprint
  {http://arxiv.org/abs/2204.00599} {arXiv:2204.00599 [quant-ph]} \BibitemShut
  {NoStop}%
\bibitem [{\citenamefont {Ganguly}\ and\ \citenamefont
  {Agarwalla}(2023)}]{ganguly2023}%
  \BibitemOpen
  \bibfield  {author} {\bibinfo {author} {\bibfnamefont {K.}~\bibnamefont
  {Ganguly}}\ and\ \bibinfo {author} {\bibfnamefont {B.~K.}\ \bibnamefont
  {Agarwalla}},\ }\bibfield  {title} {\enquote {\bibinfo {title} {Study of
  non-equilibrium {G}reen's functions beyond {B}orn approximation in open
  quantum systems},}\ }\href@noop {} {\  (\bibinfo {year} {2023})},\ \Eprint
  {http://arxiv.org/abs/2309.03776} {arXiv:2309.03776} \BibitemShut {NoStop}%
\bibitem [{\citenamefont {Hosseinabadi}\ \emph {et~al.}(2024)\citenamefont
  {Hosseinabadi}, \citenamefont {Chang},\ and\ \citenamefont
  {Marino}}]{Hosseinabadi2024}%
  \BibitemOpen
  \bibfield  {author} {\bibinfo {author} {\bibfnamefont {H.}~\bibnamefont
  {Hosseinabadi}}, \bibinfo {author} {\bibfnamefont {D.~E.}\ \bibnamefont
  {Chang}}, \ and\ \bibinfo {author} {\bibfnamefont {J.}~\bibnamefont
  {Marino}},\ }\bibfield  {title} {\enquote {\bibinfo {title} {Far from
  equilibrium field theory for strongly coupled light and matter: {D}ynamics of
  frustrated multimode cavity {QED}},}\ }\href {\doibase
  10.1103/PhysRevResearch.6.043314} {\bibfield  {journal} {\bibinfo  {journal}
  {Phys. Rev. Res.}\ }\textbf {\bibinfo {volume} {6}},\ \bibinfo {pages}
  {043314} (\bibinfo {year} {2024})},\ \Eprint
  {http://arxiv.org/abs/2312.11624} {arXiv:2312.11624 [cond-mat]} \BibitemShut
  {NoStop}%
\bibitem [{\citenamefont {Khan}\ \emph {et~al.}(2025)\citenamefont {Khan},
  \citenamefont {Rathore},\ and\ \citenamefont {Jain}}]{Khan:2024zwr}%
  \BibitemOpen
  \bibfield  {author} {\bibinfo {author} {\bibfnamefont {S.}~\bibnamefont
  {Khan}}, \bibinfo {author} {\bibfnamefont {L.~S.}\ \bibnamefont {Rathore}}, \
  and\ \bibinfo {author} {\bibfnamefont {S.}~\bibnamefont {Jain}},\ }\bibfield
  {title} {\enquote {\bibinfo {title} {{Steady-state correlation function
  beyond the standard weak-coupling limit and consistency with the
  Kubo-Martin-Schwinger relation}},}\ }\href {\doibase
  10.1103/PhysRevA.111.032214} {\bibfield  {journal} {\bibinfo  {journal}
  {Phys. Rev. A}\ }\textbf {\bibinfo {volume} {111}},\ \bibinfo {pages}
  {032214} (\bibinfo {year} {2025})},\ \Eprint
  {http://arxiv.org/abs/2401.16488} {arXiv:2401.16488 [quant-ph]} \BibitemShut
  {NoStop}%
\bibitem [{\citenamefont {Prudhoe}\ and\ \citenamefont
  {Shandera}(2023)}]{Prudhoe:2022pte}%
  \BibitemOpen
  \bibfield  {author} {\bibinfo {author} {\bibfnamefont {S.}~\bibnamefont
  {Prudhoe}}\ and\ \bibinfo {author} {\bibfnamefont {S.}~\bibnamefont
  {Shandera}},\ }\bibfield  {title} {\enquote {\bibinfo {title} {{Classifying
  the non-time-local and entangling dynamics of an open qubit system}},}\
  }\href {\doibase 10.1007/JHEP02(2023)007} {\bibfield  {journal} {\bibinfo
  {journal} {JHEP}\ }\textbf {\bibinfo {volume} {02}},\ \bibinfo {pages} {007}
  (\bibinfo {year} {2023})},\ \Eprint {http://arxiv.org/abs/2201.07080}
  {arXiv:2201.07080 [quant-ph]} \BibitemShut {NoStop}%
\bibitem [{\citenamefont {Agarwal}\ and\ \citenamefont
  {Chu}(2024)}]{Agarwal:2023lid}%
  \BibitemOpen
  \bibfield  {author} {\bibinfo {author} {\bibfnamefont {N.}~\bibnamefont
  {Agarwal}}\ and\ \bibinfo {author} {\bibfnamefont {Y.-Z.}\ \bibnamefont
  {Chu}},\ }\bibfield  {title} {\enquote {\bibinfo {title} {{Initial value
  formulation of a quantum damped harmonic oscillator}},}\ }\href {\doibase
  10.1103/PhysRevResearch.6.023113} {\bibfield  {journal} {\bibinfo  {journal}
  {Phys. Rev. Res.}\ }\textbf {\bibinfo {volume} {6}},\ \bibinfo {pages}
  {023113} (\bibinfo {year} {2024})},\ \Eprint
  {http://arxiv.org/abs/2303.04829} {arXiv:2303.04829 [hep-th]} \BibitemShut
  {NoStop}%
\bibitem [{\citenamefont {Thorbeck}\ \emph {et~al.}(2024)\citenamefont
  {Thorbeck}, \citenamefont {Xiao}, \citenamefont {Kamal},\ and\ \citenamefont
  {Govia}}]{Thorbeck2024}%
  \BibitemOpen
  \bibfield  {author} {\bibinfo {author} {\bibfnamefont {T.}~\bibnamefont
  {Thorbeck}}, \bibinfo {author} {\bibfnamefont {Z.}~\bibnamefont {Xiao}},
  \bibinfo {author} {\bibfnamefont {A.}~\bibnamefont {Kamal}}, \ and\ \bibinfo
  {author} {\bibfnamefont {L.~C.~G.}\ \bibnamefont {Govia}},\ }\bibfield
  {title} {\enquote {\bibinfo {title} {{Readout-induced suppression and
  enhancement of superconducting qubit lifetimes}},}\ }\href {\doibase
  10.1103/PhysRevLett.132.090602} {\bibfield  {journal} {\bibinfo  {journal}
  {Phys. Rev. Lett.}\ }\textbf {\bibinfo {volume} {132}},\ \bibinfo {pages}
  {090602} (\bibinfo {year} {2024})},\ \Eprint
  {http://arxiv.org/abs/2305.10508} {arXiv:2305.10508 [quant-ph]} \BibitemShut
  {NoStop}%
\end{thebibliography}%

\end{document}